\documentclass[12pt,letterpaper]{article}
\pdfoutput=1
\usepackage{epsfig}
\usepackage{amsmath}
\usepackage{amssymb}
\usepackage{amsthm}
\usepackage{indentfirst}
\usepackage{xspace}
\usepackage{multirow}
\usepackage{hyperref}
\usepackage{xcolor}
\definecolor{darkred}{rgb}{0.5,0.15,0.15}
\hypersetup{colorlinks=true,urlcolor=darkred,linkcolor=darkred,citecolor=darkred}
\usepackage[letterpaper,margin=1in,headheight=15pt]{geometry}
\usepackage{mathpazo}
\usepackage{authblk}
\usepackage{empheq}
\usepackage{feynmp}
\usepackage{graphicx}
\usepackage[matrix,arrow]{xy}
\usepackage{young}
\usepackage[vcentermath]{youngtab}
\usepackage{slashed}
\usepackage{tikz-cd}
\usepackage{tikz}
\usetikzlibrary{positioning}
\usetikzlibrary{shapes.geometric, arrows}
\tikzstyle{rect} = [rectangle,rounded corners,minimum width=3cm, minimum height=1cm, text centered, draw=black]
\tikzstyle{arrow} = [thick,->,>=stealth]
\usepackage{stackrel}

\usepackage{array,ragged2e}
\usepackage[sorting=ydnt,bibstyle=authoryear-comp,labelyear=false,defernumbers=true,maxnames=20,firstinits=true, uniquename=init,dashed=false,backend=bibtex]{biblatex}

\DeclareNameAlias{sortname}{first-last}

\DeclareFieldFormat{url}{\url{#1}}
\DeclareFieldFormat[article]{pages}{#1}
\DeclareFieldFormat[inproceedings]{pages}{\lowercase{pp.}#1}
\DeclareFieldFormat[incollection]{pages}{\lowercase{pp.}#1}
\DeclareFieldFormat[article]{volume}{\textbf{#1}}
\DeclareFieldFormat[article]{number}{(#1)}
\DeclareFieldFormat[article]{title}{\MakeCapital{#1}}
\DeclareFieldFormat[inproceedings]{title}{#1}
\DeclareFieldFormat{shorthandwidth}{#1}

\DeclareBibliographyDriver{article}{%
  \usebibmacro{bibindex}%
  \usebibmacro{begentry}%
  \usebibmacro{author/editor}%
  \setunit{\labelnamepunct}\newblock
  \MakeSentenceCase{\usebibmacro{title}}%
  \newunit
  \printlist{language}%
  \newunit\newblock
  \usebibmacro{byauthor}%
  \newunit\newblock
  \usebibmacro{byeditor+others}%
  \newunit\newblock
  \printfield{version}%
  \newunit\newblock
  \usebibmacro{journal+issuetitle}%
  \newunit\newblock
  \printfield{note}%
  \printfield{pages}
  \newunit\newblock
  \usebibmacro{eprint}
  \newunit\newblock
  \printfield{addendum}%
  \newunit\newblock
  \usebibmacro{pageref}%
  \usebibmacro{finentry}}

\renewbibmacro*{journal+issuetitle}{%
  \usebibmacro{journal}%
  \setunit*{\addspace}%
  \iffieldundef{series}
    {}
    {\newunit
     \printfield{series}%
     \setunit{\addspace}}%
  \printfield{volume}%
  \printfield{number}%
  \setunit{\addcomma\space}%
  \printfield{eid}%
  \setunit{\addspace}%
  \usebibmacro{issue+date}%
  \newunit\newblock
  \usebibmacro{issue}%
  \newunit}

\def\makebibcategory#1#2{\DeclareBibliographyCategory{#1}\defbibheading{#1}{\section*{#2}}}
\makebibcategory{books}{Books}
\makebibcategory{papers}{Refereed research papers}
\makebibcategory{chapters}{Book chapters}
\makebibcategory{conferences}{Papers in conference proceedings}
\makebibcategory{techreports}{Unpublished working papers}
\makebibcategory{bookreviews}{Book reviews}
\makebibcategory{editorials}{Editorials}
\makebibcategory{phd}{PhD thesis}
\makebibcategory{subpapers}{Submitted papers}
\makebibcategory{curpapers}{Current projects}

\renewcommand*{\bibitem}{\addtocounter{papers}{1}\item \mbox{}\hskip-0.85cm\hbox to 0.85cm{\hfill\arabic{papers}.~~}}
\defbibenvironment{bibliography}
{\list{}
  {\setlength{\leftmargin}{\bibhang}%
   \setlength{\itemsep}{\bibitemsep}%
   \setlength{\parsep}{\bibparsep}}}
{\endlist}
{\bibitem}

\newcounter{papers}
\newcounter{sumpapers}

\numberwithin{equation}{section}

\newcommand{\CN}{\mathcal{N}}
\newcommand{\be}{\begin{equation}}
\newcommand{\ee}{\end{equation}}

\newcommand*\widefbox[1]{\fbox{\hspace{2em}#1\hspace{2em}}}

\newcommand{\ba}{\begin{array}}
\newcommand{\ea}{\end{array}}
\newcommand{\bi}{\begin{itemize}}
\newcommand{\ei}{\end{itemize}}
 \newcommand{\ben}{\begin{enumerate}}
\newcommand{\een}{\end{enumerate}}
\newcommand{\bean}{\begin{eqnarray*}}
\newcommand{\eean}{\end{eqnarray*}}
\newcommand{\eref}[1]{(\ref{#1})}

\newcommand{\fa}{{\mathfrak a}}
\newcommand{\fb}{{\mathfrak b}}

\newcommand{\fq}{{\mathfrak q}}
\newcommand{\fm}{{\mathfrak m}}

\newcommand{\bv}{{\mathbf v}}

\newcommand{\cL}{\ensuremath{\mathcal L}}
\newcommand{\cS}{\ensuremath{\mathcal S}}

\newcommand{\cK}{\ensuremath{\mathcal K}}
\newcommand{\cD}{\ensuremath{\mathcal D}}

\newcommand{\cM}{\ensuremath{\mathcal M}}

\newcommand{\cH}{\ensuremath{\mathcal H}}

\newcommand{\cR}{\ensuremath{\mathcal R}}

\newcommand{\cT}{\ensuremath{\mathcal T}}

\newcommand{\R}{\ensuremath{\mathbb R}}
\newcommand{\C}{\ensuremath{\mathbb C}}

\newcommand{\Z}{\ensuremath{\mathbb Z}}

\newcommand{\half}{\ensuremath{\frac{1}{2}}}
\newcommand{\qtr}{\ensuremath{\frac{1}{4}}}
\newcommand{\N}{{\mathcal N}}

\newcommand{\V}{{\mathcal V}}

\newcommand{\hk}{hyperk\"ahler\xspace}

\newcommand{\al}{\alpha}

\newcommand{\I}{{\mathrm i}}

\newcommand{\de}{\mathrm{d}}

\newcommand{\IP}[1]{\left\langle#1\right\rangle}

\newcommand{\fM}{\underline{\overline{\cM}}}

\DeclareMathOperator{\Tr}{Tr}

\bibliography{comparevev}

\begin{document}

\title{On 't Hooft Defects, Monopole Bubbling and Supersymmetric Quantum Mechanics}
\date{}
\author[1]{T. Daniel Brennan}
\author[2]{Anindya Dey}
\author[3]{Gregory W. Moore}
\affil[1,2,3]{NHETC and
Department of Physics and Astronomy, Rutgers University \\
126 Frelinghuysen Rd., Piscataway NJ 08855, USA}
\maketitle

{\abstract{We revisit the localization computation of the expectation values of 't Hooft operators
in $\N=2^*$ $SU(N)$ theory on $\R^3 \times S^1$. We show that the part of the answer
arising from ``monopole bubbling'' on $\R^3$ can be understood as an equivariant integral over a Kronheimer-Nakajima
moduli space of instantons on an orbifold of $\C^2$. It can also be described as a Witten index
of a certain supersymmetric quiver quantum mechanics with $\N=(4,4)$ supersymmetry. The map between the defect data
and the quiver quantum mechanics is worked out for all values of $N$. For the $SU(2)$ theory,
we compute several examples of these line defect expectation values using the Witten index formula and
confirm that the expressions agree with the formula derived by Okuda, Ito and Taki \cite{Ito:2011ea}.
In addition, we present a Type IIB construction -- involving D1-D3-NS5-branes -- for monopole bubbling in
$\N=2^*$ $SU(N)$ SYM and demonstrate how the quiver quantum mechanics arises in this brane picture.}}

\tableofcontents
\section{Introduction and summary of the main results}
\subsection{Generalities}

 't Hooft-Wilson defects are the simplest class of non-local operators in gauge theories and have been studied
from various perspectives, starting with the pioneering work of 't Hooft \cite{tHooft:1977nqb, tHooft:1979rtg, tHooft:1981bkw}.
In this paper we study 't Hooft defects in four-dimensional $\N=2^*$ $SU(N)$ gauge theory on $\R^3 \times S^1$,
where the defect is inserted at the origin of $\R^3$.
In a $4d,~ \N=2$ Lagrangian theory on $\R^3 \times S^1$, the vev of an 't Hooft defect, characterized by an element of the
cocharacter lattice $B$ and denoted as $T_B$,  can be understood as a supersymmetric index:
\begin{align}\label{def-dyonic}
\IP{T_{B}}& = \Tr_{\mathcal H_{(B)}} (-1)^{F}e^{-2\pi R \cdot H}\, e^{2\pi \I \lambda \cdot (J_3+J_R)} \,e^{2\pi \I \mu_f \cdot F_f}\, e^{2\pi \I \theta\, \cdot Q},
\end{align}
where $\mathcal H_{(B)}$ denotes the Hilbert space of the theory with the line defect, $F$ is the fermion number, $R$ is the radius of the circle, and
$H$ is the Hamiltonian. Here $J_3$ generates rotation in the $1-2$ plane of $\R^3$, $J_R$ is the Cartan generator for the R-symmetry group $SU(2)_R$, $\{F_f\}$ generate the flavor symmetries in theories with matter. Additionally, $\lambda$ is the chemical potential for $(J_3+J_R)$, $\{\mu_f\}$ are chemical
potentials for $\{F_f\}$,
and $\theta=(\theta_e, \theta_m)$ are background electric and magnetic Wilson lines (which are chemical potentials for the electric and magnetic charges at spatial infinity $Q=(Q_e, Q_m)$).\\

The above index should be interpreted as a path integral with the appropriate boundary conditions at the origin of $\R^3$ and at spatial infinity.
The boundary conditions at the origin are
\begin{equation}
\begin{split}\label{'t Hooft b.c.}
& A_\mu \,\de x^\mu  \sim - g^2 \vartheta \frac{B}{16\pi^2 } \frac{1}{r} \,\de \tau + \frac{B}{2}\, \cos{\theta} \, \de \phi,\\
& Y \sim  - g^2 \vartheta \frac{B}{16\pi^2 } \frac{1}{r}, \quad X \sim \frac{B}{2r},
\end{split}
\end{equation}
where $r= |\vec{x}|$, and $g^2, \vartheta$ are the 4d gauge coupling and theta-angle respectively, and we view the cocharacter $B$ as an element of
a Cartan subalgebra of the Lie algebra of $SU(N)$. $X$ and $Y$ are real scalars of the
$\N=2$ vector multiplet.
For vanishing theta-angle, the above equations reduce to the simplified form:
\begin{align}
& F \sim -\frac{B}{2} \sin{\theta}\, \de \theta \wedge \de \phi =- \frac{B}{2}\epsilon_{ijk} \frac{x^i}{r^3}\, \de x^j \wedge \de x^k\quad,\qquad X \sim \frac{B}{2 r},
\end{align}
and $Y$ is regular at $r=0$. At spatial infinity, the field configurations approach a vacuum associated with a generic point on the Seiberg-Witten moduli space $\cM$ \cite{Seiberg:1996nz}, which is a fibration over the Coulomb branch of the 4d theory by a torus of electric and magnetic Wilson lines. The magnetic Wilson line $\theta_m$ is introduced  in the path integral by first working with a fixed magnetic charge $\gamma_m$ at infinity, and then defining $\theta_m$ as the Fourier dual of $\gamma_m$.
In other words, we first introduce a path integral $\IP{T_{B}}(\gamma_m)$ with boundary conditions \eref{'t Hooft b.c.} at the origin of $\R^3$ and the following boundary conditions at spatial infinity \footnote{Note that the superscript $(\infty)$ implies the vev of the respective field at the spatial infinity
$r \to \infty$.}:
\begin{equation} \label{bc-fixedm}
\begin{split}
& F \to \frac{\gamma_m}{2} \sin{\theta} \de \theta \wedge \de \phi~, \qquad X \to X^{(\infty)} - \frac{\gamma_m}{2r}+ O(r^{-1-\delta})~,\\
&\oint_{S^1_\tau|_{r \to \infty}} A_\tau \de \tau = \theta_e\quad~~,\qquad Y \to Y^{(\infty)}+ O(r^{-\delta}), \qquad \delta >0~,
\end{split}
\end{equation}
and then define the Fourier dual of the path integral:
\begin{align}
\IP{T_{B}} (\theta_m) =  \sum_{\gamma_m} \IP{T_{B}}(\gamma_m) e^{-2 \pi \I \gamma_m \theta_m}~.
\end{align}

The Seiberg-Witten moduli space $\cM$ is a \hk manifold with a $\C\mathbb{P}^1$ worth of complex structures parametrized by $\zeta \in \C^\times$
\footnote{$\zeta$ should not be confused with FI parameters of quiver gauge theories that appear later.}.
The 't Hooft operator vev $\IP{T_{B}} (\theta_m)$ is a holomorphic function on $\cM$ with respect to a chosen
complex structure $\zeta$ associated with the 't Hooft defect. In this paper, we will set $\zeta=1$, and indeed we have done so in writing
\eref{'t Hooft b.c.}.\\

Recently, extremely powerful techniques for computing vevs of 't Hooft-Wilson defects
were devised for theories in class $\mathcal{S}$ using the AGT correspondence \cite{Alday:2009aq}.
In this approach, vevs of 4d line operators are related to correlation functions of appropriate loop operators in
Liouville/Toda CFT which live on the Riemann surface associated with the class $\mathcal{S}$ construction of the 4d theory.
The latter can then be computed using the standard Verlinde operator approach \cite{Verlinde:1988sn, Moore:1989vd}, as discussed in \cite{Alday:2009fs, Drukker:2009id}, leading to explicit expressions for the 4d line operator vevs.\\

In a parallel set of developments, vevs of Wilson defects were computed for $4d, \N=2$ theories on compact space-time manifolds like ellipsoids and four-spheres \cite{Pestun:2007rz, Hama:2012bg} using localization techniques. Localization of 't Hooft defects in $4d, \N=2^*$ $SU(N)$ theory
on a round four-sphere was addressed by Gomis, Okuda, and Pestun (GOP) in \cite{Gomis:2011pf}. It is important to note that GOP did not compute the vev $\IP{T_{B}}$ directly. Instead they computed the vev of a product of 't Hooft operators in a minimal representation (the fundamental representation for the case of $\N=2^*$ $SU(N)$) in the coincident limit of collinear insertion points. 
Rather than computing the $SU(2)$ defect $T_B$ with $B={\rm diag}(\frac{p}{2}, -\frac{p}{2})$, GOP computes the following correlation function:
\be\label{defect-GOP}
\IP{\widetilde{T}_{B}} = \lim_{\{z_i\} \to 0} \IP{\prod^{p}_{i=1} T_{B_{\rm min}}(z_i)}~,
\ee
where $B_{\rm min}={\rm diag}(\frac{1}{2}, -\frac{1}{2})$. This 't Hooft defect is S-dual to a Wilson defect in the
representation $\cR = \Big(\cR_{\rm fund}\Big)^p$, where $\{\cR_{\rm fund}\}$ is the fundamental representation of $SU(2)$,
as opposed to the irreducible $j=\frac{p}{2}$ representation. Using the operator product algebra for line defects,
one can of course extract $\IP{{T}_{B}}$ from a knowledge of $\langle \tilde{T}_{B^\prime}\rangle$ for various $B^\prime$.\\

In \cite{Ito:2011ea},  Ito, Okuda, and Taki (IOT) extended the computation of GOP \cite{Gomis:2011pf} for an 't Hooft operator on
$\R^3 \times S^1_R$ inserted at the origin of $\R^3$ and wrapping $S^1_R$, where $\R^3 \times S^1_R$ has the coordinates
$\{x^\mu\}=(\vec x, \tau)$ and a metric $\de s^2 = \de \vec x^2 + \de \tau^2$, where $\tau$ is a periodic coordinate: $\tau \sim \tau + 2\pi R$.
They primarily considered $\N=2^*$ SYM, and $\N=2$ SYM with fundamental hypers, with a single $SU(N)$ gauge group, although their formula
can be generalized to include other gauge groups and matter representations.\\

These 't Hooft operators $\IP{\widetilde{T}_{B}}$ are holomorphic functions on the Seiberg-Witten moduli space $\cM$.
Therefore, it is convenient to write the localization answer in terms of a particular set of holomorphic coordinates
-- the complexified Fenchel-Nielsen coordinates $(\fa, \fb)$ \cite{Dimofte:2011jd, Nekrasov:2011bc} --  which have the following expressions in terms physical parameters defined in the weak coupling expansion: 
\begin{align}
\begin{split}
&\fa = \Big(\theta_e + \I \, R\, Y^{(\infty)}\Big) + \ldots\quad,\qquad
\fb= \Big(\frac{\theta_m}{2\pi}-\frac{4\pi \I R}{g^2} X^{(\infty)} + \frac{\I \vartheta R}{2\pi} Y^{(\infty)}\Big) + \ldots~,
\end{split}\end{align}
where we have written the classical contribution explicitly in the weak-coupling expansion of $(\fa,\fb)$, while the ellipsis indicate non-perturbative corrections. A systematic discussion of these non-perturbative contributions will be discussed in a future paper.\\

Given the boundary conditions in \eref{'t Hooft b.c.} and \eref{bc-fixedm}, the localization formula for the 't Hooft operator vev can be written as a
Fourier series w.r.t. a complexified Fenchel-Nielsen coordinate $\fb$:
\begin{align}\label{Loc-IOT}
\IP{\widetilde{T}_{B}}(\theta_m) & = \sum_{\{\mathbf{v} \in \Lambda_{\rm cr} +B|\, (\mathbf{v},\mathbf{v}) \leq (B,B)\}} e^{2\pi \I (\fb,\, \mathbf{v})} Z_\text{1-loop}(\fa,\mu_f,\lambda;\mathbf{v}) Z_\text{mono}(\fa,\mu_f,\lambda;B,\mathbf{v})~,
\end{align}
where $\lambda,\,\mu_f$ are chemical potentials defined in \eref{def-dyonic}, $\mathbf{v}$ is a cocharacter such that $\mathbf{v} - B$ is an element of
the coroot lattice $\Lambda_{\rm cr}$, and $(\cdot\, ,\, \cdot)$ denotes a Killing form on the Lie algebra of $SU(N)$. The factorization of the Fourier coefficient into $Z_{\rm 1-loop}  Z_{\rm mono} $  is discussed in the next paragraph.

The sum over $\mathbf{v}$ in \eref{Loc-IOT} can be physically interpreted as a sum over the monopole bubbling sectors where $\mathbf{v}$ is the effective 't Hooft charge
after bubbling in a given sector. 
As shown in GOP and IOT \cite{Gomis:2011pf, Ito:2011ea}, this sum arises from a sum over the isolated fixed points of the Q-fixed locus of the 4d path
 integral with 't Hooft defect. These can be described as the fixed points of a certain group action on the moduli space of
 $U(1)_K$--invariant\footnote{$U(1)_K$ acts on $\C^2 $ as $(z_1,z_2) \to (e^{2\pi \I \nu}z_1, e^{-2\pi \I \nu}z_2)$ and this induces an action on the moduli space of instantons
on $\C^2$. See section \ref{ADHM-KN} for a review of the ADHM construction of
the moduli space $\cM (B, \mathbf{v})$. } instantons on $\C^2$ where the $U(1)_K$-action on the instanton bundle is specified by the defect data $(B,\mathbf{v})$. We will denote this moduli space as $\cM (B, \mathbf{v})$.  
 The fixed points of $\cM (B, \mathbf{v})$ with respect to the $U(1)_K$ action are then labelled by tuples of Young diagrams consistent with the $U(1)_K$ invariance
(see appendix \ref{IOT} for a quick review of the results of IOT). 
Similarly, the one-loop determinant
from fluctuations of fields around these fixed points are obtained by restricting to the $U(1)_K$--invariant
weights of the group action at each fixed point. The universal part of this determinant is called
$Z_\text{1-loop}$ while the remaining part (dependent on the fixed points) is identified as $Z_\text{mono}$.


In reference [16] IOT have given a formula for $Z_{\rm mono}$ of the form
\begin{align}
Z^{\rm IOT}_\text{mono}(\fa,\mu_f,\lambda;B,\mathbf{v})=Z^{S^1 \times \R^4}_{\rm{inst}}|^{U(1)_K \, \rm{inv.}}_{k=k(B,\mathbf{v})} = \int_{\cM (B, \mathbf{v})} C(T\cM) C(E)~, \label{IOT-mono}
\end{align}
where $\cM (B, \mathbf{v})$ is the moduli space of $U(1)_K$--invariant instantons on $\C^2$, $k=k(B,\bv)$ is the instanton number, and the integrand of the equivariant integral for $Z^{\rm IOT}_\text{mono}$ is the appropriate characteristic class for the 5d instanton partition function on $S^1 \times \R^4$ for a given theory \footnote{We discuss these characteristic classes in detail in appendix \ref{equiv structure}.}.
This formula is not precise, in part because the integral is over a
singular space. In the case of $SU(N)$ $\N=2^*$,  a natural regularization of the integral (explained below) yields answers for the the ’t
Hooft line defect vev’s in agreement with those given by the AGT prescription. However, as noted in \cite{Ito:2011ea}
for other groups and hypermultiplet representations the prescription
for defining the integrals in \eref{IOT-mono} in general does not agree with the relevant AGT computations.
We will comment on this issue in  more detail after \eref{WI formula-main} in section \ref{summary}.\\

Before summarizing the results of this paper, we would like to mention briefly a couple of important issues that
we do not pursue in this paper but hope to address in a future work :
\begin{itemize}
\item The path integral expression for the vev  $\langle T_B \rangle(\gamma_m)$  can be reduced to an integral over the moduli space of singular monopoles on $\R^3$ with an 't Hooft defect of charge $B$ at the origin and asymptotic charge $\gamma_m$ at spatial infinity. We will denote this space $\fM(B, \gamma_m, X^\infty)$.
The expansion \eref{Loc-IOT} of the path integral is closely related to the recent analysis of singular monopole moduli spaces by Nakajima and Takayama \cite{Nakajima:2016guo}
in the context of bow construction \cite{Blair:2010vh,Cherkis:2009jm,Cherkis:2010bn,Cherkis:2016gmo} for moduli spaces of instantons on a Taub-NUT space. In
particular, the authors of \cite{Nakajima:2016guo} show that the space $\fM(B, \gamma_m, X^\infty)$ admits a stratification
\be
\fM(B, \gamma_m, X^\infty)=\coprod_{\substack{0\leq v\leq B\\v\in \Lambda_{\rm cr}+B}} \fM^{(s)}(\mathbf{v}, \gamma_m, X^\infty)~,
\ee
where $\fM^{(s)}(v,\gamma_m,X^\infty)$ is the smooth component (i.e. the complement of the singular locus) of $\fM(v,\gamma_m,X^\infty)$,
and that $\cM (B, \mathbf{v})$ is the transversal slice to $\fM^{\rm (s)}(\mathbf{v}, \gamma_m, X^\infty)$ in $\fM(B, \gamma_m, X^\infty)$.\\

\item 't Hooft defects in 4d $\N=2$ theories are closely related to Coulomb branch physics of $3d, \N=4$ theories.
Given the formula for $\IP{T_B}_{\R^3 \times S_R^1}$, one can compute expectation values of monopole operators
in the $3d, \N=4$ theory on $\R^3$ by taking the $S^1$ radius $R \to 0$ carefully. In particular, this allows one to compute precise
equivariant expressions for coefficients of the ``Abelianization Maps" introduced by Bullimore, Dimofte and Gaiotto \cite{Bullimore:2015lsa}.

\end{itemize}

\subsection{Summary} \label{summary}
In this work, we revisit the localization computation of the vev of 't Hooft defects of the form \eref{defect-GOP}
in a $4d$ $\N=2^*$ theory on $\R^3 \times S^1$. In particular, we show that the non-perturbative part of the
path integral is an equivariant integral over a Kronheimer-Nakajima moduli space of instantons on an orbifold of $\C^2$,
and is given by the Witten index of a $\N=(4,4)$ SQM living on $S^1$. The main results of our paper are summarized
as follows:

\subsection*{$U(1)_K$--invariant moduli space of instantons as a KN moduli space}

From the ADHM construction of $U(1)_K$-invariant $SU(N)$ instantons on $\mathbb{C}^2$,
we show that the moduli space $\cM (B, \mathbf{v})$ is isomorphic to a  Kronheimer-Nakajima (KN) space\footnote{This was also noted in \cite{Nakajima:2016guo}.}, which describes the moduli space of
$U(N)$ instantons on an orbifold of $\mathbb{C}^2$ .
The space $\cM (B, \mathbf{v})$ can therefore be described as a linear quiver
variety $\Gamma_{\vec k, \vec w}$, where the quiver data $(\vec k, \vec w)$ can be derived from the defect data $(B, \mathbf{v})$.
\begin{equation}\label{KN-main}
\boxed{\cM (B, \mathbf{v},\, SU(N)) \cong  \cM^{\C^2/\Z_n}_{\text{inst.}} (\vec k, \vec w, \, U(N)) }
\end{equation}
where $n$ is sufficiently large. 
This is a crucial observation which allows one to realize the moduli space of
$U(1)_K$--invariant instantons in terms of a very well-known moduli space.
We discuss the derivation in section \ref{ADHM-KN}.

\subsection*{Monopole bubbling Index as Witten index of an SQM}
Given the identification of $\cM (B, \mathbf{v})$ with a KN moduli space, the result \eref{IOT-mono}
implies that $Z_{\rm mono}$ for 't Hooft defects in an $\N=2^*$ $SU(N)$ theory is equal to a 5d $\N=1^*$ $SU(N)$
instanton partition function of instanton number $k$, on $S^1 \times \C^2/\Z_n$ for a sufficiently large $n$. The instanton number $k$ is
determined by the defect data $(B, \mathbf{v})$.\\

The linear quiver  $\Gamma_{\vec k, \vec w}$ therefore
encodes the data of a (4,4) supersymmetric quiver quantum mechanics, such that the moduli space $\cM (B, \mathbf{v})$
is realized as the Higgs branch of this quantum mechanics. In other words, $\Gamma_{\vec k, \vec w}$ arises as the ADHM
quiver for the KN instantons in \eref{KN-main}. The moduli space is singular, and can be resolved by
introducing real stability parameters in the ADHM construction. This corresponds to turning on FI
parameters for $U(1)$ factors in the linear quiver $\Gamma_{\vec k, \vec w}$.\\

The 5d instanton partition function is given by the Witten index of the SQM computed in the Higgs scaling limit,
where we take the SQM gauge coupling $e^2 \to 0$ and the FI parameter $\zeta \to \infty$ such that $\zeta' = e^2 \zeta$
is held fixed\footnote{For multiple unitary gauge groups, one sets $e_i=e$ and $\zeta_i=\zeta$ for all $i$, and then takes
the Higgs scaling limit.}.
Therefore, one can write a formula for $Z^{\R^3 \times S^1}_{\rm mono}$ in terms of the SQM Witten index
\footnote{We drop the dependence on some of the equivariant parameters in this equation for brevity.}:
\begin{empheq}[box=\widefbox]{align}\begin{split}\label{WI formula-main}
Z
_{\rm mono} \Big(B,\mathbf{v};G=SU(N)\Big) &= Z^{S^1 \times \C^2/\Z_n}_{\rm inst} \Big(\vec k, \vec w;G'=U(N)| \sum^N_{i=1} a_i=0\Big)\\
&= Z_{\rm SQM}\Big(\Gamma_{\vec k, \vec w}|\sum^N_{i=1} a_i=0\Big)
\end{split}\end{empheq}
Generically, the Witten index and the 5d partition function will depend on the sign of $\zeta'$.
However, the $\N=1^*$ $SU(N)$ instanton partition function and the associated Witten index are invariant under the
transformation $\zeta' \to - \zeta'$. Therefore, the above equation is well-defined.\\

In the general case, where the partition function is dependent on the sign of $\zeta'$, setting
$Z_{\rm mono}$ naively equal to the partition function in the $\zeta' > 0$ or $\zeta' < 0$ chamber gives a
wrong result. For example, in the $SU(2)$ theory with $N_f=4$ flavors, the naive answer for $Z_{\rm mono}$ computed in any chamber differs from the
AGT expression by certain extra terms. These extra terms are closely related to the non-trivial wall-crossing
of the Witten index as $\zeta' \to -\zeta'$. A further investigation into this discrepency is in progress.

\subsection*{Defect SQMs for $\N=2^*$ $SU(2)$}
As an illustrative example, we work out the linear quivers associated with 't Hooft defects in $\N=2^*$ $SU(2)$
explicitly. Consider a defect labelled by $B={\rm diag}(\frac{p}{2}, -\frac{p}{2})$, and a monopole bubbling sector labelled by
$\mathbf{v}={\rm diag}(\frac{v}{2}, -\frac{v}{2})$, with integer $(p,v)$ and $v=-p, -p+2,\ldots, p$. The quiver SQMs associated
with $Z^{\R^3 \times S^1}_{\rm mono}$ for the cases $v=0$ and $v \neq 0$ are given as:

\begin{center}
\begin{tikzpicture}[
cnode/.style={circle,draw,thick,minimum size=4mm},snode/.style={rectangle,draw,thick,minimum size=6mm}]
\node[cnode] (1) {1};
\node[cnode] (2) [right=.5cm  of 1]{2};
\node[cnode] (3) [right=.5cm of 2]{3};
\node[cnode] (4) [right=1cm of 3]{\tiny{$\frac{p}{2}-1$}};
\node[cnode] (5) [right=0.5cm of 4]{$\frac{p}{2}$};
\node[cnode] (6) [right=0.5cm of 5]{\tiny{$\frac{p}{2}-1$}};
\node[cnode] (7) [right=1cm of 6]{{$3$}};
\node[cnode] (8) [right=0.5cm of 7]{$2$};
\node[cnode] (9) [right=0.5cm of 8]{1};
\node[snode] (10) [below=0.5cm of 5]{2};
\draw[-] (1) -- (2);
\draw[-] (2)-- (3);
\draw[dashed] (3) -- (4);
\draw[-] (4) --(5);
\draw[-] (5) --(6);
\draw[dashed] (6) -- (7);
\draw[-] (7) -- (8);
\draw[-] (8) --(9);
\draw[-] (5) -- (10);
\end{tikzpicture}

\qquad \qquad

\begin{tikzpicture}[
cnode/.style={circle,draw,thick,minimum size=4mm},snode/.style={rectangle,draw,thick,minimum size=6mm}]
\node[cnode] (1) {1};
\node[cnode] (2) [right=.5cm  of 1]{2};
\node[cnode] (3) [right=.5cm of 2]{3};
\node[cnode] (5) [right=1cm of 3]{\tiny{$\frac{p-v}{2}-1$}};
\node[cnode] (6) [right=0.5cm of 5]{$\frac{p-v}{2}$};
\node[cnode] (9) [right=1cm of 6]{$\frac{p-v}{2}$};
\node[cnode] (10) [right=0.5cm of 9]{\tiny{$\frac{p-v}{2}-1$}};
\node[cnode] (13) [right=1cm of 10]{{$3$}};
\node[cnode] (14) [right=0.5cm of 13]{$2$};
\node[cnode] (17) [right=0.5cm of 14]{1};
\node[snode] (18) [below=0.5cm of 6]{1};
\node[snode] (19) [below=0.5cm of 9]{1};
\draw[-] (1) -- (2);
\draw[-] (2)-- (3);
\draw[dashed] (3) -- (5);
\draw[-] (5) --(6);
\draw[dashed] (6) -- (9);
\draw[-] (9) -- (10);
\draw[dashed] (10) -- (13);
\draw[-] (13) -- (14);
\draw[-] (14) -- (17);
\draw[-] (6) -- (18);
\draw[-] (9) -- (19);
\end{tikzpicture}
\end{center}

\noindent respectively. The quiver SQMs for 't Hooft defects in $\N=2^*$ $SU(2)$ are discussed in section \ref{Ex-WI}. Using the Witten index formula
\eref{WI formula-main}, we compute $Z^{\R^3 \times S^1}_{\rm mono}$ for a few examples with small $p$ and $v$
(in section \ref{Ex-WI} and Appendix \ref{Okuda-2*}) and check that they agree with the IOT expressions.

\subsection*{Hanany-Witten construction and $SU(N)$ quiver}
 We present a Type IIB Hanany-Witten type construction of singular monopoles which can be used to describe monopole bubbling in a 4d $\N=2$ $U(N)$ SYM
(or $\N=2^*$ $U(N)$ SYM). This construction is described by the worldvolume theory of a stack of D3-branes with decorating D1- and NS5-branes. We show that using this construction, we can derive the Higgs branch quiver (a quiver gauge theory
whose Higgs branch is isomorphic to the moduli space in question) for
$\cM (B, \mathbf{v})$ from the world volume theory on the D1-branes. For generic $N >2$, we write down a general form of the Higgs branch quiver, built out of a linear array of
$N-1$ superconformal sub-quivers $\mathcal{S}_i$ ($i=1,\ldots, N-1$). These superconformal subquivers are connected by exactly $N-2$ unbalanced
\footnote{A balanced $U(k_i)$ gauge node in a linear quiver gauge theory is one for which the one-loop $\beta$ function vanishes.
This happens when $2k_i = k_{i+1} + k_{i-1} + w_i$ in the notation of figure \ref{fig:linearquiver} below.} gauge nodes,
such that two adjacent sub-quivers are separated by a single unbalanced gauge node: \\

\begin{tikzpicture}[
cnode/.style={circle,draw,thick,minimum size=5mm},snode/.style={rectangle,draw,thick,minimum size=8mm}]
\tikzstyle{oct} = [regular polygon,regular polygon sides=8, draw,
    text width=2em, text centered]
\tikzstyle{line} = [draw, -latex']
\node [oct] (1) {$\mathcal{S}_1$};
\node [cnode](2) [right=0.5cm of  1] {$k_{n_1}$};
\node (3) [oct,right=0.5cm of 2] {$\mathcal{S}_2$};
\node [cnode](4) [right=0.5cm of 3] {$k_{n_1 + n_2}$};
\node (5) [oct,right=0.5cm of 4] {$\mathcal{S}_3$};
\node [cnode](6) [right=1.5cm of 5] {$k_{n_{tot}}$};
\node (7) [oct,right=0.5cm of  6] {$\mathcal{S}_{N-1}$};
\node[snode] (8) [below=0.5cm of 2]{$w_{n_1}$};
\node[snode] (9) [below=0.5cm of 4]{$w_{n_1 + n_2}$};
\node[snode] (10) [below=0.5cm of 6]{$w_{n_{tot}}$};
\draw[-] (1) --  (2);
\draw[-] (2) --  (3);
\draw[-] (3) --  (4);
\draw[-] (4) --  (5);
\draw[dashed] (5) --  (6);
\draw[-] (6) --  (7);
\draw[-] (2) --  (8);
\draw[-] (4) --  (9);
\draw[-] (6) --  (10);
\end{tikzpicture}

where the circular nodes denote the unbalanced gauge nodes. Details of this quiver are discussed in section \ref{HW-U(N)}.\\

The plan of the paper is as follows.
Section \ref{N=2*}, the core of the paper, discusses the contribution of monoopole bubbling to the expectation value of `t Hooft line defects. There we show how this contribution can be given by an equivariant integral over a certain Kronheimer-Nakajima quiver variety describing the moduli space of $U(1)_K$-invariant instantons on $\C^2$ $\big(\cM(B,\bv)\big)$  which can equivalently be written as a Witten index for the associated quiver SQM.
 Then in Section \ref{HW}, we introduce a
D-brane description of singular monopoles and monopole bubbling.  Using this description, we give a derivation and physical explanation of the quiver SQM associated to $\cM(B,\bv)$. In the appendices we provide additional background material on computing the Witten index of ADHM SQM's and previous work on computing the $Z_{\rm mono}$ contributions to `t Hooft defects. We also explicitly compute several examples and discuss equivariant integrals associated to these Witten indices.

\section{Defect SQM for 't Hooft operators in $SU(N)$ $\N=2^*$ theories}\label{N=2*}

In \cite{Ito:2011ea, Gomis:2011pf}, the authors showed that the monopole bubbling contribution $Z_\text{mono}$
to the 't Hooft operator vev is given by an equivariant integral of certain trigonometric characteristic classes
over $\cM(B, \mathbf{v})$: the moduli space of $U(1)_K$--invariant instantons on $\C^2$. In addition, these
characteristic classes were shown to be precisely those which that appear in the equivariant integral formula for
a 5d instanton partition function on $S^1 \times \C^2$ \footnote{By 5d instanton partition function, we will mean the non-perturbative part of the 5d index
only and therefore not including the one-loop part.}. In other words, $Z_\text{mono}$ is given by the $U(1)_K$--invariant
part of a 5d instanton partition function on $S^1 \times \C^2$.\\

In this section, we derive that for a given $SU(N)$ defect labelled by a cocharacter
$B$, 
the space
$\cM(B, \mathbf{v})$ can be thought of as a Kronheimer-Nakajima (KN)  space describing $U(N)$ instantons
on an orbifold $\C^2/\Z_n$ for a sufficiently large positive integer $n$. We will show that the fact that $\cM(B, \mathbf{v})$ can be described as a KN space implies that $Z_\text{mono}$ is an equivariant integral of a characteristic class over the KN space, and therefore can
be identified with a 5d instanton partition function on $S^1 \times \C^2/\Z_n$ specified by the defect data.

In order to write $Z_{\rm mono}$ as an equivariant integral, we must address the singularities of $\cM(B,\bv)$. The resolution of singularities in KN moduli spaces is a
well-studied problem and one can unambiguously define equivariant integrals on such spaces. This consists of taking the closure (adding point instantons) and then resolving the singularities by introducing stability parameters (FI parameters). For an $\N=2^*$ theory, this leads to a well-defined
equivariant integral formula for $Z_\text{mono}$, which we discuss in appendix \ref{equiv structure}.
However, in addition to resolving $\cM(B,\mathbf{v})$, in a generic Lagrangian $\N=2$ theory one must address the chamber-dependence
of $Z_{\rm mono}$ with respect to the stability/FI parameters. However, for the case of the $\N=2^*$ theory, this dependence is trivial as we discuss in section \ref{WI}.

From a string theory perspective, instantons
on  $\mathbb{C}^2/\mathbb{Z}_n$ can be realized in Type IIA, by considering the world volume theory of D4-branes wrapped on $\mathbb{C}^2/\mathbb{Z}_n$ with $k$ dissolved, fractional D0-branes  \cite{Douglas:1996sw}. 
 The moduli space of these configurations can also be realized as the Higgs branch of the world volume theory
on the D0 branes (which we will refer to as the KN quiver). 
From this construction, it is clear that the 5d instanton partition function
is given by the Witten index\footnote{For review, see appendix \ref{WittenIndex} and \ref{App-WI}.} of the D0-brane world
volume theory \cite{Nekrasov:1996cz, Nekrasov:2004vw}.
Therefore, by exploiting the relation between a 5d instanton partition function and the Witten index of an SQM, one can write
$Z_\text{mono}$ as the Witten index of the SQM 
corresponding to the fractional instanton. This allows us to write $Z_{\rm mono}$ as an equivariant integral over a characteristic class which can be reduced to a contour integral whose solution is a sum over poles enumerated by Young diagrams.\\

In summary, for a 4d $\N=2^*$ theory with gauge group $SU(N)$ and an associated 5d $\N=1^*$ theory with gauge group $U(N)$, $Z_{\rm mono}$ satisfies
\begin{empheq}[box=\widefbox]{align}
Z
_{\rm mono} \Big(B,\mathbf{v};SU(N)\Big) = & Z^{S^1 \times \C^2/\Z_n}_{\rm inst} \Big(\vec k, \vec w;U(N)| \epsilon_-=0,\,\sum^N_{i=1} a_i=0\Big)\\
=& Z_{SQM}\Big(\Gamma_{\vec k, \vec w}|\epsilon_-=0,\,\sum^N_{i=1} a_i=0\Big) \nonumber
\end{empheq}
where the equivariant parameters in $Z^{\R^3 \times S^1}_{\rm mono}$ ($\fa, \fm, \lambda$) and $Z^{S^1 \times \C^2/\Z_n}_{\rm inst}$ ($a, m, \epsilon_+$) are related in a simple fashion:
\begin{align}
a= 2\I \pi \fa\quad, \qquad m=2 \I \pi \fm\quad, \qquad \epsilon_+ =\I \pi {\lambda}~.
\end{align}

Here the $SU(N)$ defect data $B$ and bubbling data $\mathbf{v}$ is mapped to $U(N)$ instanton data on $\C^2/\Z_n$ specified by vectors $(\vec k, \vec w)$.
Also, the lower bound of $n$ is determined by the defect data. We discuss the defect data/instanton data map as well as the bound on $n$ in detail
in section \ref{ADHM-KN}.


\subsection{Brief Review of the KN quiver variety} \label{KN-review}
We begin with a brief review of instanton moduli spaces on $\C^2/\Z_n$ and KN quiver variety relevant
for the subsequent discussion. We will restrict our discussion to $U(N)$ instantons on $\C^2/\Z_n$.
Consider the standard ADHM complex
\begin{equation}\label{ADHM seq}
 0 \longrightarrow V \stackrel{\sigma(z)}\longrightarrow \mathbb C^2\otimes V\oplus W
\stackrel{\tau(z)}\longrightarrow V \longrightarrow 0~,
\end{equation}
where $V= \C^k$ and $W=\C^N$. Recall that $V$ is the space of Dirac zero modes on $\R^4$ in an instanton background,
while $W$ is the fiber of the associated bundle in the fundamental representation at a base-point at infinity.\\

The maps $\sigma(z)$ and $\tau(z)$, explicitly given as
\begin{align}
  \sigma(z)=
  \begin{pmatrix}
B_1-z_1\\
B_2-z_2\\
J
  \end{pmatrix}
\quad,
\qquad
\tau(z)=
\begin{pmatrix}
 z_2-B_2& B_1 - z_1 & I
\end{pmatrix}
\,,
\end{align}
obey the condition $\tau(z) \sigma(z) =0$, so that the sequence \eref{ADHM seq} is a complex. The ADHM data consists of the following matrices:

 \begin{empheq}{align}\begin{split}\label{ADHM data}
 B_1 & \in  \text{Hom}(V, V)\quad,\qquad  I  \in  \text{Hom}(W,V),~\\
 B_2 & \in  \text{Hom}(V, V)\quad, \qquad  J \in   \text{Hom}(V,W)~.
\end{split} \end{empheq}

The moduli space of instantons on $\C^2/\Z_n$ is a hyperk\"ahler quotient
of the ADHM data invariant under the $\Z_n$ orbifold action, induced from the action on $\C^2$ given by
 $(z_1,z_2) \mapsto (\omega z_1, \omega^{-1} z_2)$, where $\omega=e^{2\pi \I/n}$. The invariance condition on the ADHM
 variables under $\Z_n$ action is given by the following equations:
\begin{align} \begin{split}\label{ADHM-orb action}
& B_a = R_a^{\; b} (g) \gamma_V(g) B_b \gamma^{-1}_V(g) \quad,\quad (a,b=1,2)   ~,\\
& I= \gamma_V(g) I \gamma^{-1}_W(g)~,\\
& J=\gamma_W(g) J \gamma^{-1}_V(g)~, 
\end{split}\end{align}
where $g$ is a generator of  $\Z_n$,  the matrix $R_a^{\; b}$ implements an $SU(2)$ rotation on $\C^2$ while $\gamma_V(g), \gamma_W(g)$ implement the $\Z_n$ orbifold action on the vector spaces $V$ (${\rm dim}\,V =k$) and $W$ (${\rm dim}\,W =N$) respectively. In terms of the one-dimensional irreps of $\Z_n$, defined as
\begin{equation}
R_j\,:\, \omega= e^{2\pi \I/n} \mapsto \omega^j = e^{2\pi \I j /n}\quad, \qquad j \sim j \mod n~,
\end{equation}
the spaces $V$ and $W$ admit the following isotypical decomposition:
\begin{align}
& V= \oplus^{n-1}_{j=0}�� V_j \otimes R_j \quad,\qquad
 W = \oplus^{n-1}_{j=0}� W_j�� \otimes R_j~,
\end{align}

Let the integers $k_j= {\rm dim} V_j$ and $w_j= {\rm dim} W_j$ count dimensions of the degeneracy spaces,
i.e. the number of times the $j$-th one-dimensional irrep appears in the isotypical decomposition,
such that $\sum^{n-1}_{j=0} k_j =\text{dim}\, V=k$ and $\sum^{n-1}_{j=0} w_j =\text{dim}\, W=N$. This data
is summarized in terms of the KN vector $\vec k =\{k_0,k_1, \ldots, k_{n-1}\}$ and the monodromy vector
$\vec w=\{w_0,w_1, \ldots, w_{n-1}\}$.

Explicitly, the matrices $R_a^{\; b},  \gamma_V$ and $\gamma_W$ can be written, in some suitable basis, as follows:
\begin{align}\begin{split}
&R_a^{\; b} (g)= \mbox{diag} (\omega^{-1}, \omega)~,\\
&\gamma_V(g)=\mbox{diag} (\omega^{n_1}, \omega^{n_2}, \ldots, \omega^{n_k})~,\\
& \gamma_W(g)=\mbox{diag} (\omega^{r_1}, \omega^{r_2}, \ldots, \omega^{r_N})~, \,\,
\end{split}\end{align}
where $n_i $ ($i=1,\ldots, k$) and $ r_\al$ ($\al=1,\ldots,N$) are integers defined modulo $n$, and can be repeated.
The multiplicities of the integers $\{n_i\}$ and $\{r_\al\}$ are given by the entry $k_{n_i}$ in the KN vector $\vec k $
and $w_{r_\al}$ in the monodromy vector $\vec w$ respectively.
For $SU(N)$ instantons, one must also impose $\sum^N_{\al=1} r_\al=0 \, {\rm mod}\, n$.\\

A generic solution of the equation \eref{ADHM-orb action} 
 is given as follows:
 \begin{align}\begin{split}\label{Zn-inv data}
 B_1 & \in \oplus^{n-1}_{i=0} \text{Hom}(V_{{i+1}}, V_{i})\quad,\qquad  I  \in  \oplus^{n-1}_{i=0} \text{Hom}(V_{i}, W_{i} )~,\\
 B_2 & \in  \oplus^{n-1}_{i=0} \text{Hom}(V_{{i-1}}, V_{i})\quad,\qquad
 J  \in  \oplus^{n-1}_{i=0} \text{Hom}(W_i, V_{i})~.
\end{split}\end{align}
In the final step, we take the hyperk\"ahler quotient of the $\Z_n$-invariant ADHM data w.r.t. the group $\prod^{n-1}_{i=0} U(k_i)$, i.e.
 \begin{align}
\cM_{KN} (U(N)) := \left \{(B_1,B_2,I,J)_{\Z_n}  \right\} {\Big /}{\Big /}{\Big /} \prod_i U(k_i)~,
\end{align}
where the quotient is implemented via the ADHM equations:
\begin{align}\begin{split}\label{eq:complex-ADHM}
 \mu_{\mathbb C}&\equiv [B_1,B_2]+IJ=0\quad,\qquad
\mu_{\mathbb R}\equiv [B_1^\dagger, B_1]+[B_2^\dagger, B_2]+II^\dagger-J^\dagger J=0~.
\end{split}\end{align}

The resultant space is a quiver variety labelled by the vectors $\vec k$ and $\vec w$.
For our study of line defects, we will be interested in KN instantons  where one or more
integers $k_i$ may be zero,
such that the KN vector $\vec k$ and monodromy vector $\vec w$
are given as:
\begin{align}\begin{split}
&\vec k = \mbox{diag} (0, \ldots, 0, k_{i_{min}}, k_{i_{min} +1}, k_{i_{min} +2},\ldots,k_{i_{max}}, 0, \ldots,0)~, \\
&\vec w= \mbox{diag} (0, \ldots, 0, w_{i_{min}}, w_{i_{min} +1}, w_{i_{min} +2},\ldots, w_{i_{max}}, 0, \ldots,0)~.
\end{split}\end{align}

The KN data is related to topological data of the instanton bundle on the orbifold/ALE space. We mention a
few useful results here and refer the reader to \cite{Fucito:2004ry, Nak:1990kr, Witten:2009xu} for details.
Given an ALE space of $A_{n-1}$ type, one can introduce a tautological bundle $\cT$ over the ALE base
with a regular representation of $\Z_n$ being the fiber. $\cT$ admits a decomposition $\cT = \oplus^{n-1}_{j=0} \cT_j \otimes R_j$,
where $R_j$ is the $j$-th irrep of $\Z_n$, and $\cT_j$ are certain vector bundles on the ALE space such that their first Chern classes -- $c_1(\cT_j)$
-- form a basis for $H^2({\rm ALE}, \Z)$ for $j \neq 0$ (we set $c_1(\cT_0)=0$). The first and the second Chern classes of the instanton bundle
can be written in terms of the first and second Chern classes of the bundles $\cT_j$:
\begin{align}\begin{split}\label{cc}
& c_1=\sum^{n-1}_{j=0} (w_j -2k_j +k_{j+1} +k_{j-1})\, c_1(\cT_j)~,\\
& c_2= \sum^{n-1}_{j=0} (w_j -2k_j +k_{j+1} +k_{j-1})\, c_2(\cT_j) +\frac{1}{n}\sum^{n-1}_{i=0} k_i~,
\end{split}\end{align}

The number $\frac{1}{n}\sum^{n-1}_{i=0} k_i$ is often referred to as the
instanton number, which coincides with the second Chern class of the instanton bundle only for a balanced quiver. In addition,
we do not require $\text{dim}\, V$ to be an integer multiple of $n$ which implies that the KN instantons are generically fractional.\\

\subsection{Moduli spaces for $U(1)_K$-invariant instantons on $\C^2$ as Kronheimer-Nakajima quiver varieties} \label{ADHM-KN}

Kronheimer's correspondence states that smooth monopoles in the presence of a single `t Hooft defect can be described by $U(1)_K$-invariant instantons on $\mathbb{C}^2$ \cite{KronCorr}.
For this purpose, the ADHM construction for $U(1)_K$-invariant $SU(N)$ instantons on $\C^2$ was presented in \cite{Kapustin:2006pk, Gomis:2011pf, Ito:2011ea}. We now demonstrate how this ADHM moduli space can be thought of as a special case of a KN moduli space of $U(N)$ instantons  on an orbifold of $\C^2$. The basic result of this subsection may be summarized as follows.

 The $SU(N)$ defect data on $\R^3$ consists of a cocharacter $B={\rm diag}(p_1,p_2,\ldots,p_N)$ and $\mathbf{v} ={\rm diag}(v_1,v_2,\ldots, v_N) \in \Lambda_{\rm cr} + B$, such that $(\mathbf{v},\mathbf{v}) \leq (B,B)$, with $p_i, v_i \in \Z $, and $\sum^N_{i=1} p_i=0$, $\sum^N_{i=1} v_i=0$. Given a pair of cocharacters $(B, \mathbf{v})$,  let $\cM (B, \mathbf{v},\, SU(N))$ be the moduli space of $U(1)_K$-invariant $SU(N)$ instantons on $\C^2$, where $B$ and $\mathbf{v}$ determine the $U(1)_K$ action on the fibers of the instanton bundle at the origin and at infinity respectively. For sufficiently large $n$, we claim

\begin{equation}
\boxed{\cM (B, \mathbf{v},\, SU(N)) \cong  \cM^{\C^2/\Z_n}_{\text{frac inst.}} (\vec k, \vec w, \, U(N)) }
\end{equation}

\noindent where $\cM^{\C^2/\Z_n}_{\text{frac inst.}} (\vec k, \vec w, \, U(N))$ is the moduli space of a $U(N)$ instanton on the orbifold $\C^2/\Z_n$ with a monodromy vector $\vec w$ and a Kronheimer-Nakajima vector $\vec k$, as discussed above. The relation between the defect data $(B, \mathbf{v})$ and the KN data $(\vec k, \vec w)$ is explained later in this subsection. The isomorphism implies that $\cM (B, \mathbf{v},\, SU(N))$ can be understood as a linear quiver variety.\\

Let us review the ADHM construction of the $U(1)_K$-invariant instanton moduli space of instanton number $k$.
Consider the following $U(1)$ action on $\C^2$: $z=(z_1, z_2) \mapsto (e^{2\pi \I \nu} z_1, e^{-2\pi \I \nu} z_2)$, where
$e^{2\pi \I \nu} \in U(1)$. This is the action of $U(1)_K$.
To discuss the equivariant version of the ADHM construction under the $U(1)_K$ action, it is convenient
to write the standard ADHM complex in a slightly different (but equivalent) form :

\begin{equation}\label{ADHM seq-2}
 0 \longrightarrow V\otimes L_{-} \stackrel{\sigma(z)}\longrightarrow  S^{+} \otimes V\oplus W
\stackrel{\tau(z)}\longrightarrow V \otimes L_+ \longrightarrow 0\,~,
\end{equation}
where $V= \C^k$, $W=\C^N$, and $\cS^{\pm}$ are the chiral spinor bundles on $\C^2$ (with fibers $S^\pm$ at a point
$z \in \C^2$). Under the $U(1)$ action, $\cS^-$ decomposes into line bundles: $\cS^-= \cL_- \oplus \cL_+$, and $L_\pm$
denote the corresponding fibers of the line bundles.
The maps $\sigma(z)$ and $\tau(z)$, explicitly given as
\begin{align}
  \sigma(z)=
  \begin{pmatrix}
B_1-z_1\\
B_2-z_2\\
J
  \end{pmatrix}
\,\quad,\qquad
\quad
\tau(z)=
\begin{pmatrix}
 z_2-B_2& B_1 - z_1 & I
\end{pmatrix}
\,~,
\end{align}
obey the condition $\tau(z) \sigma(z) =0$, so that the sequence \eref{ADHM seq-2} is a complex. The ADHM data is
given by \eref{ADHM data}.\\

Next, we promote the vector spaces $V,W$ to $U(1)_K$ representations so that the maps $\sigma(z), \tau(z)$ are themselves equivariant.
The representations are of the following form:
\be
\rho_V\Big(e^{2\pi \I \nu}\Big)=e^{2\pi \I K \nu}\quad, \qquad \rho_W \Big(e^{2\pi \I \nu}\Big)=e^{2\pi \I \mathbf{v} \nu}~,
\ee
where $K$ and $\mathbf{v}$ are cocharacters. Explicitly the complex is $U(1)_K$ equivariant if the ADHM variables obey the following relations:
\begin{align}\begin{split}\label{U(1)K action}
&e^{2\pi \I K \nu} B_1 e^{-2\pi \I K \nu} = e^{2\pi \I \nu} B_1 \qquad,\qquad e^{2\pi \I K \nu} I e^{-2\pi \I \mathbf{v} \nu} = I~,\\
&e^{2\pi \I K \nu} B_2 e^{-2\pi \I K \nu} = e^{-2\pi \I \nu} B_2\quad~,\qquad   e^{2\pi \I \mathbf{v} \nu} J e^{-2\pi \I K \nu}=J~.
\end{split}\end{align}
Given the equivariant complex, one can define the fibers of the gauge bundle using cohomology groups of
the complex:
\begin{align}
&H^0_z=\text{Ker} [\sigma(z)]\,\quad,\quad H^1_z=\text{Ker}[\tau(z)]/\text{Im}[\sigma(z)]\,\quad,\quad H^2_z=V/\text{Im} [\tau(z)]~.
\end{align}
If $H^0_z=H^2_z=0$, then $E_z=H^1_z$ describes the fiber of a smooth irreducible
instanton bundle over $\mathbb C^2$. In particular, the fiber $E_\infty$ is identified with W, (${\rm dim}\, W =N$) and
$E_0$ is the fiber at the origin $z_1=0,z_2=0$ (${\rm dim}\, E_0 =N$). 
Therefore, the $U(1)_K$ representation associated with the fiber $E_0$ is of the form:
\be
\rho_{E_0} \Big(e^{2\pi \I \nu}\Big)=e^{2\pi \I B \nu}~.
\ee
The cocharacters $(B, \mathbf{v}, K)$ are related. From the Euler-Poincare principle, the $U(1)_K$ characters must
obey the following equation:

\be
{\rm ch}_{S^+ \otimes V} +  {\rm ch}_W - {\rm ch}_{S^- \otimes V} = {\rm ch}_{E_0}~.
\ee

Noting that ${\rm ch}_{S^+ \otimes V}  - {\rm ch}_{S^- \otimes V} = (e^{2\pi i\nu}+e^{-2\pi i\nu}-2)\,\Tr_V e^{2\pi i K\nu}$,
and that $E_0 \cong W$ as vector spaces, we arrive at the equation \cite{Kapustin:2006pk}:

\begin{equation}\label{char-eqn-main}
\boxed{  \Tr_W e^{2\pi i B \nu}
=\Tr_W e^{2\pi i \mathbf{v} \nu}+(e^{2\pi i\nu}+e^{-2\pi i\nu}-2) \Tr_V e^{2\pi i K\nu}}
\end{equation}

Given $(B,\mathbf{v})$, the above equation determines the cocharacter $K$ up to conjugacy.
Note that the equation doesn't always have a solution. Taking the limit $\nu \to 0$, we have the following relations
in the leading and sub-leading order:
\begin{align}
& \Tr_W B = \Tr_W {\bf v}\quad,\qquad  k = {\rm dim} V = \half (\Tr_W B^2 - \Tr_W {\bf v}^2)~,
\end{align}
where the second equation implies that $(B,B) \geq (\bf v, \bf v)$ for \eref{char-eqn-main} to have a solution.

This $U(1)_K$ action descends to an action on the ADHM \hk quotient.
The resultant fixed point subspace is the moduli space of $U(1)_K$-invariant instantons, which we have denoted as
$\cM(B, \mathbf{v}, SU(N))$.\\

We will now show that $\cM (B, \mathbf{v},\, SU(N))$ is a linear quiver variety.
Let us perform the following transformation of the triplet of matrices $(B, \mathbf{v},K)$:
\begin{equation}\label{shift-defectdata}
\begin{split}
& B \to B' = B - p_{\rm min} \, \mathbf{I}~,\\
& \mathbf{v} \to \mathbf{v}' = \mathbf{v} - p_{\rm min} \, \mathbf{I}~,\\
& K \to K'= K - p_{\rm min} \, \mathbf{I}~.
\end{split}
\end{equation}
The resultant triple $(B', \mathbf{v}', K')$ is a solution of the Euler-Poincare character formula \eref{char-eqn-main}, where the eigenvalues of the
matrices $(B', \mathbf{v}',K')$ can be taken to be non-negative integers \footnote{Note that this is an arbitrary choice. However, the quiver variety is stable under any such overall shift transformation of $(B', \mathbf{v}',K')$}. Note that for $B'={\rm diag}(p'_1,p'_2,\ldots,p'_N)$, for $p'_i \in \Z$, we have $\sum^N_{i=1} p'_i \neq 0$, which implies that $B'$ is an element of the cocharacter lattice of $U(N)$ as opposed to $SU(N)$. In addition, the conditions of $U(1)_K$ equivariance of the ADHM data $(B_1,B_2, I, J)$ are invariant under the shift \eqref{shift-defectdata}. This leads to the following isomorphism of moduli spaces:
\begin{align}\label{SU(N) vs U(N)}
\cM(B, \mathbf{v}, SU(N)) \cong \cM(B', \mathbf{v}', U(N))~,
\end{align}
where $(B', \mathbf{v}')$ for a given pair $(B,\mathbf{v})$ are given by the transformation in \eqref{shift-defectdata}.

The vector spaces $V$ and $W$ are now associated with $U(1)$ representations labelled by the cocharacters $K'$ and
$\mathbf{v}'$ respectively. Let $\rho_q$ denote an irreducible representation of $U(1)_K$ with charge $q$:
\begin{equation}
\rho_q\,:\, x=e^{2\pi \I \nu} \mapsto x^q= e^{2\pi \I q \nu}, \quad q \in \Z.
\end{equation}

One can now write the isotypical decompositions of $V$ and $W$ under this $U(1)$ action. Since all eigenvalues of the operators $K'$ and $\mathbf{v}'$ are non-negative integers, isotypical decompositions of $V$ and $W$ will only involve irreps with non-negative $U(1)_K$ charges, i.e.
\begin{align}\begin{split}\label{isotyp}
& V= \bigoplus_{q \in \Z_{\geq 0}} V^{(q)} \otimes \rho_q\quad,\qquad
W= \bigoplus_{q \in \Z_{\geq 0}}W^{(q)} \otimes \rho_q~.
\end{split}\end{align}
where $V^{(q)}$ and $ W^{(q)}$ are degeneracy spaces.

 Invariance of the ADHM data under $U(1)_K$ action implies invariance under any subgroup of $U(1)_K$ and in particular, the subgroup of $n$-th roots of unity, $\Z_n$. Under the inclusion
\begin{align}
\iota:� \Z_n \hookrightarrow U(1)~,
\end{align}
one can write the $\Z/n\Z$ irrep $R_j$, defined as
\begin{equation}
R_j\,:\, \omega= e^{2\pi \I/n} \mapsto \omega^j = e^{2\pi \I j /n}\quad, \quad j \sim j \mod n~,
\end{equation}
as a pull back of the $U(1)$ irrep $\rho_q$:
\begin{equation}
\iota^*(\rho_q) = R_j\quad , \quad�j = q\mod n~.
\end{equation}

The isotypical decompositions can therefore be rewritten in terms of the $\Z/n\Z$ irreps as follows:
\begin{align}
& V= \oplus_{j}�� V_j \otimes R_j \quad,\qquad W = \oplus_{j}� W_j�� \otimes R_j~,
\end{align}
where $V_j$, $W_j$ are the corresponding degeneracy spaces, and $j = q\mod n$ with $q \in \Z_{\geq 0}$. We can now choose $n$ such that the
labels of the  $\Z/n\Z$ irreps $R_j$ can be taken in the fundamental domain, i.e. $j = 0, \ldots, n-1$, and one can unambiguously
set $j=q$. This can be done if $n$ is greater than the maximal $U(1)_K$ charge $q_{max}$ which appears in the isotypical decomposition
\eref{isotyp}.
\begin{equation}\label{n-lowerbound}
\boxed{n > q_{\rm max}}
\end{equation}
which is what we mean by a sufficiently large $n$. Given $j$ in the fundamental domain, the isotypical decompositions assume the form
\begin{align}
& V= \bigoplus^{n-1}_{j=0}�� V_j \otimes R_j \quad,\qquad
 W = \bigoplus^{n-1}_{j=0}� W_j�� \otimes R_j~.
\end{align}
Analogous to the Kronheimer-Nakajima construction, one can now define the vectors $\vec k$ and $\vec w$
which count the dimensions of the degeneracy spaces:
\begin{align}\begin{split}
&\vec k= \Big(k_0, k_1, \ldots, k_{n-1} \Big)\qquad, \quad k_j= {\rm dim} V_j~, \\
&\vec w= \Big(w_0, w_1, \ldots, w_{n-1}\Big)\quad, \quad w_j= {\rm dim} W_j~.
\end{split}\end{align}
In addition, some of the integers $k_j$ may be zero. For example, writing the character equation \eref{char-eqn-main} for the triple
$(B',\mathbf{v}',K')$ as
\begin{align}
\Tr_k x^{K'} = \frac{(\Tr_N x^{B'} - \Tr_N x^{\mathbf{v}'})}{(x^{\half} - x^{-\half})^2} = \sum^{n-1}_{j=0} k_j x^{j}\quad, \quad q \geq 0~,
\end{align}
and taking $x \to 0$ limit, one can see that $k_0=0$, if the eigenvalues $p'_i, v'_i \geq 0$, $\forall i$.
Also, $k_j$ for all $j > q_{max}$ will vanish. \\

Therefore, a more precise way
of writing the isotypical decompositions is:
\begin{align}
& V= \bigoplus^{q_{max}}_{j=q_{min}}�� V_j \otimes R_j \quad,\qquad  W = \bigoplus^{q_{max}}_{j=q_{min}}� W_j�� \otimes R_j~.
\end{align}
where $q_{min} > 0$ and $q_{max} < n$. The vectors $\vec k, \vec w$ are given as
\begin{align}\begin{split}\label{KNvec-def}
&\vec k =\Big(k_0, k_1, \ldots, k_{n-1} \Big) =\Big(0, \ldots, 0, k_{q_{min}}, k_{q_{min} +1}, k_{q_{min} +2},\ldots,k_{q_{max}}, 0, \ldots,0 \Big)~,  \\
&\vec w=\Big(w_0, w_1, \ldots, w_{n-1}\Big) =\Big(0, \ldots, 0, w_{q_{min}}, w_{q_{min} +1}, w_{q_{min} +2},\ldots, w_{q_{max}}, 0, \ldots,0 \Big)~.
\end{split}\end{align}

One can write down the explicit solution for the $U(1)_K$ invariant ADHM variables $\{B_1,B_2, I,J\}_K$ from equation \eref{U(1)K action}:
 \begin{align}\begin{split}\label{ADHM-inv-data}
 B_1 & \in \oplus^{q_{max}-1}_{j=q_{min}} \text{Hom}(V_{j+1}, V_{j})\quad~~,\qquad  I  \in  \oplus^{q_{max}}_{j=q_{min}} \text{Hom}(V_{j}, W_{j} )~,\\
 B_2 & \in  \oplus^{q_{max}}_{j=q_{min}+1} \text{Hom}(V_{j-1}, V_{j})\quad, \qquad
 J \in  \oplus^{q_{max}}_{j=q_{min}} \text{Hom}(W_{j}, V_{j})~.
\end{split}\end{align}
In particular, note that $B_1,B_2$ does not have a component of the form $\text{Hom}(V_{q_{max}}, V_{q_{min}})$
or $\text{Hom}(V_{q_{min}}, V_{q_{max}})$, since $n > \, q_{max}$. It is obvious from the discussion above that the
$U(1)_K$ invariant ADHM data \eref{ADHM-inv-data} 
 can be thought of as solutions of the $\Z_n$ invariance equation
\eref{ADHM-orb action} provided we make the following identification:
\begin{align}\label{KN-IOT}
\boxed{ n_i := K'_i,\quad \quad r_\al := v'_\al}
\end{align}
where the integers $n_i$ and $r_\al$ are in the fundamental domain, i.e. $0 \leq n_i \leq n-1$ for all $i$,
and $0 \leq r_\al \leq n-1$ for all $\al$. Note that the integers $K'_i$ and $v'_\al$ are non-negative.\\

Finally, the moduli space $\cM(B', \mathbf{v}', U(N))$ is given by the hyperK\"ahler quotient
 \begin{eqnarray} \label{moduli space - U(1)_K}
\boxed{ \cM(B', \mathbf{v}', U(N))=
\left   \{
 (B_1,B_2,I,J)_K
 \right\}
{\Big /}{\Big /}{\Big /}
\prod_j U(k_j)\, \cong  \cM(B, \mathbf{v}, SU(N))}
 \end{eqnarray}
where the last equality follows from \eqref{SU(N) vs U(N)}. The hyperk\"ahler quotient is implemented via the ADHM equations
(the first of which follows from the condition $\tau(z) \sigma(z)=0$):
\begin{align}\begin{split}\label{eq:complex-ADHM}
 \mu_{\mathbb C}&\equiv [B_1,B_2]+IJ=0~,\\
\mu_{\mathbb R}&\equiv [B_1^\dagger, B_1]+
[B_2^\dagger, B_2]+
II^\dagger-J^\dagger J=0~.
\end{split}\end{align}
$\cM(B', \mathbf{v}', U(N))$ is therefore a linear KN quiver variety,
with generic form of vectors $\vec k$ and $\vec w$, given in \eref{KNvec-def}.
Note that, the quiver variety stabilizes as a function of $n$ for sufficiently large $n$.
Consider shifting the triple $(B, \mathbf{v}, K)$ to $(B'', \mathbf{v}'', K'')$ such that
\begin{equation}\label{shift-defectdata}
\begin{split}
& B \to B'' = B - (p_{\rm min} - u )\, \mathbf{I},\\
& \mathbf{v} \to \mathbf{v}'' = \mathbf{v} - (p_{\rm min} -u)\, \mathbf{I},\\
& K \to K''= K - (p_{\rm min}-u) \, \mathbf{I}.
\end{split}
\end{equation}
where $u \in \Z_{>0}$, such that the eigenvalues of  $K''$ are positive integers, different from the eigenvalues of $K'$ defined earlier.
Using the same line of argument as above, one can show that $\cM(B'', \mathbf{v}'', U(N))$ is isomorphic to the same linear quiver variety for a sufficiently large $n$.

\subsection{Defect SQM and Witten Index}\label{WI}
In the previous subsection, we established that the moduli space of $U(1)_K$-invariant instantons  on $\C^2$ can be understood in terms of certain KN instantons on $\C^2/\Z_n$. Given this description of $U(1)_K$--invariant instanton moduli space, one can now express the bubbling index of an 't Hooft defect vev in an $\N=2^*$ $SU(N)$ SYM as a 5d instanton partition function of an $\N=1^*$ $U(N)$ theory on $S^1 \times \C^2/\Z_n$ following the discussion in the beginning of section \ref{N=2*}.

\begin{align}\label{4d-5d}
\boxed{Z^{\R^3 \times S^1}_{\rm mono} (B,\mathbf{v}; \fa,\fm, \lambda| G=SU(N)) =Z^{S^1 \times \C^2/\Z_n}_{\text{ inst.}}(\vec k, \vec w; a, m,\epsilon_{+}, \epsilon_{-}| G'=U(N), \sum_i a_i=0)}
\end{align}
where the equivariant parameters on both sides of the equation are related as
\be \label{eqpar-map}
a=2\I \pi \fa\quad,\quad\, m=2\I \pi \fm\quad,\quad\,\epsilon_{+}=\I \pi {\lambda}\quad,\quad\, \epsilon_{-}=0~.
\ee
Unfortunately, the RHS of equation \eref{4d-5d} is not well-defined since the instanton moduli space on the RHS suffers from UV singularities arising from zero-size instantons. As discussed in section \ref{KN-equiv}, the singularities can be resolved which
introduces suitable stability/FI parameters $\{\zeta^i_\R \}$ (with $i=q_{min}, \ldots, q_{max}$) that deform the real moment map.
There exists two natural chambers defined by: $\zeta^i_{\R} <0$ (or $\zeta^i_{\R} > 0$) for all $i$, where the
partition function $Z^{S^1 \times \C^2/\Z_n}$ is given by a $\Z_n$-projection of the partition function $Z^{S^1 \times \C^2}$.

This is the partition
 function that appears in the RHS of \eref{4d-5d} and will be studied in this section. For a generic 5d $\N=1$ theory, the answer would still depend on the sign of the stability parameters. However, for the specific case of $\N=1^*$ theory, the instanton partition function is invariant under an overall sign flip of the FI parameters, which allows one to write down the RHS of \eref{4d-5d} unambiguously.\\

The 5d instanton partition function $Z^{S^1 \times \C^2/\Z_n}_{\text{ inst.}}$ is given by an equivariant integral over a KN moduli space, which can also
be realized as the Higgs branch of a (4,4) quiver SQM (ADHM SQM). Following \cite{Nekrasov:2004vw}, the instanton partition function is given by the Witten index of this SQM (reviewed in appendix \ref{Loc-app}). An effective way to read off the quiver SQM is to realize the 5d instanton particles in a  Type IIA brane construction, i.e. as a stack of fractional D0-branes probing $N$ D4-branes wrapping the orbifold $\mathbb{C}^2/\mathbb{Z}_n$ \cite{Douglas:1996sw}. The (4,4) quiver SQM then arises as the D0-brane world volume theory.\\

\begin{figure}[ht!]
\begin{center}
\includegraphics[scale=0.4]{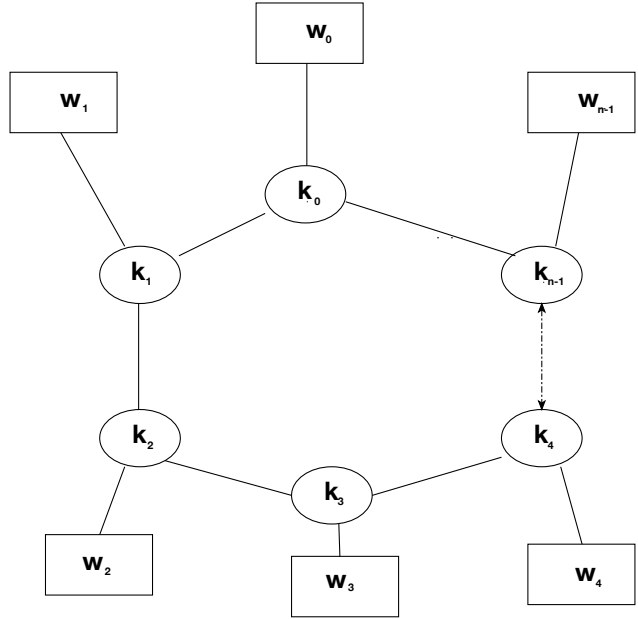}
\caption{The Kronheimer-Nakajima quiver for a regular $U(N)$ instantons on $\C^2/\Z_n$. Each node denotes the unitary group with the labelled rank. The circular nodes denote gauge groups and the square nodes denote the flavour symmetries.}
\label{fig:necklace}
\end{center}
\end{figure}

\begin{figure}[ht!]
\begin{center}
\includegraphics[scale=0.6]{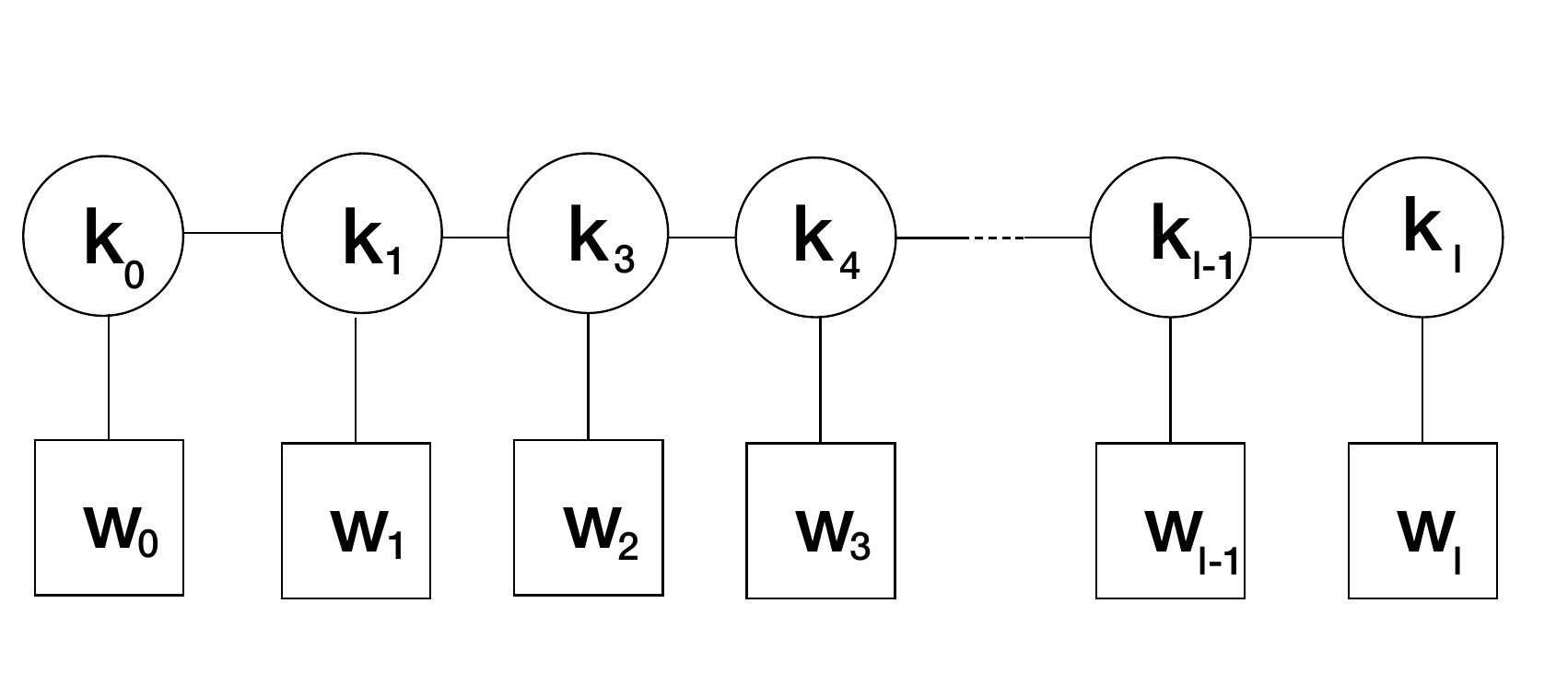}
\caption{The Kronheimer-Nakajima quiver for a fractional $U(N)$ instanton on $\C^2/\Z_n$, with KN vector $\vec k= (k_0,k_1,k_2,\ldots,k_l,0,\ldots,0)$ and monodromy vector $\vec w= (w_0,w_1, w_2, \ldots, w_l,0,\ldots,0)$.}
\label{fig:linearquiver}
\end{center}
\end{figure}

We now discuss some general features of these quiver SQMs and write down a formula for their Witten index.
A generic circular quiver associated with the instanton moduli space $\cM^{\C^2/\Z_n}_{\rm inst} (\vec k, \vec w)$ is given in figure \ref{fig:necklace},
while figure \ref{fig:linearquiver} shows a generic linear quiver -- these are known as the Kronheimer-Nakajima (KN) quivers \cite{Nak:1990kr}.
In each case, the quiver is specified by the following data:
\begin{enumerate}
\item  Kronheimer-Nakajima vector $\vec k =(k_0, k_1, \ldots, k_{n-1})$ with $k_i \in \Z_{\geq 0}$ for $i =0,1, \ldots, n-1$.
Figure \ref{fig:necklace} corresponds to the case where $k_i \neq 0$ $\forall i$ --
the gauge group is $ {G}=\prod^{n-1}_{i=0} U(k_i)$ with bifundamental hypers
forming an affine $A_{n-1}$-type quiver. For linear quivers, where one or more entries of
the vector $\vec k$ are zero, one simply deletes the corresponding nodes in the quiver along with the bifundamentals, leading to a linear quiver.

\item The monodromy vector $\vec w =(w_0, \ldots, w_{n-1})$ associated with holonomy vector $\vec r$ of the gauge fields
such that
\begin{equation}
w_i=\sum^N_{\al=1} \delta_{i,r_\al}\quad, \qquad \quad N =  w_0+\ldots + w_{n-1}~,
\end{equation}
with $w_i \in \Z_{\geq 0}$ for all $i =0, \ldots, n-1$, denoting the number of fundamental hyper associated with each gauge node $U(k_i)$.
\end{enumerate}

As mentioned earlier, the ADHM construction of the instanton moduli space is equivalent to the description of the Higgs branch of the above quiver SQMs as a hyperk\"ahler quotient. From the $\Z_n$-invariant ADHM data in equation \eref{ADHM-inv-data}, one can clearly see that the variables $B_1, B_2$ assemble themselves as scalar vevs of hypers in the bifundamental of $U(k_{i+1}) \times U(k_{i})$ 
 while $I,J$ give the scalar vevs of hypers in the fundamental of the $U(k_i)$s. The moment map equations arise as F-term and D-term
equations. In addition, the stability parameters $\{\zeta^i_\R\}$ arise as FI parameters for the gauge groups $U(k_i)$.\\

We can now write down the Witten index of the quiver SQM following the general approach in \cite{Hwang:2014uwa, Hori:2014tda, Cordova:2014oxa}. For computing the index using localization, various flat directions in the space of supersymmetric vacua should be lifted. The global symmetry twists in the definition of the Witten index ensure that the flat directions coming from various hypermultiplet scalars are lifted. Flat directions associated with one of the adjoint scalars\footnote{This is the scalar component of the $(0,2)$ vector multiplet inside the $(0,4)$ vector multiplet which, in turn, lives inside the $(4,4)$ vector multiplet.} which is neutral under these global symmetries, is lifted by turning on the FI parameters $\{\zeta^i_\R \}$.
We will be interested in studying the partition function in a chamber where all the FI parameters have the same sign. Furthermore it will be convenient to set the SQM gauge couplings $e_i=e$ and FI parameters $\zeta^i_\R =\zeta$ for all $i$.

 Further, since we are interested in computing the instanton partition function, which is given by an equivariant integral on the Higgs branch of the SQM, it is natural to compute the Witten index in the Higgs scaling limit \cite{Hori:2014tda} which introduces large masses for all the vector multiplet scalars. This limit is defined by taking $e^2\to 0$ and $\zeta\to \infty$ while holding $\zeta'=e^2 \zeta$ fixed. The Witten index computed in this fashion generically depends only on the sign of $\zeta'$. Therefore, we have

\be
\boxed{Z^{S^1 \times \C^2/\Z_n}_{\text{inst.}}(\vec k, \vec w; a, m,\epsilon_{\pm}
; \pm \zeta^i_\R \to \infty |U(N))= Z_{SQM}(\Gamma_{\vec k, \vec w}| a, m,\epsilon_{\pm}
; \pm \zeta' > 0)}
\ee
where the signs on the two sides of the equation are correlated.\\

Following the basic recipe given in appendix \ref{Loc-app}, the Witten index $Z_{SQM}$ can be written as a contour integral over a
real and compact $k$-dimensional cycle in  $(\mathfrak{t}_G \otimes \C) / \Lambda_{\rm cr} \cong (\C^*)^k$, where $k=\sum^{q_{max}}_{i=q_{min}} k_i$ and $\Lambda_{\rm cr}$ is the coroot lattice. For a linear quiver quantum mechanics with (4,4) supersymmetry, as shown in figure \ref{fig:linearquiver}, the Witten index is
\footnote{Note that the formula can be easily extended to the affine quiver, where $q_{min} =0$ and $q_{max}=n-1$, and one bifundamental hyper connecting the nodes labelled by $q_{min}$ and $q_{max}$.}
\begin{align} \label{D0quiver-orbifold}
	&Z_{SQM} (\Gamma_{\vec k,\vec w}| a, m,\epsilon_{+}, \epsilon_{-}; \zeta') = \nonumber \\
	=& \frac{1}{\prod^{q_{max}}_{i=q_{min}}k_i!}\oint_{JK(\zeta')} \prod^{q_{max}}_{i=q_{min}} \prod^{k_i}_{I=1}\left[\frac{d\phi^i_I}{2\pi i}\right] \,  {Z}_{\vec k}^{\rm vector}(\phi,m;\epsilon_{1,2}) \cdot  & {Z}_{\vec k}^{\rm bifund}(\phi,m;\epsilon_{1,2})  \cdot {Z}_{\vec k,\vec w}^{\rm fund}(\phi,a,m;\epsilon_{1,2})~,
\end{align}
The integrand is written as contributions of various $(4,4)$ supermutiplets (gauge and matter) of the SQM. Explicitly, these functions are
\footnote{We use the following notation in all subsequent Witten index expressions
$$2\sinh(x\pm y) = 2\sinh(x+y)\, 2\sinh(x-y)~.$$}:

\begin{align}\label{explicitfunctions}
	 {Z}^{\rm vec}_{\vec k}(\phi,a,m;\epsilon_{1,2}) &= \prod^{q_{max}}_{i=q_{min}}\Big(\prod_{I,J=1}^{k_i} \frac{2\sinh\frac{(\phi^i_{IJ}+2\epsilon_+)}{2}}{2\sinh{\frac{(\phi^i_{IJ} + m \pm \epsilon_+)}{2}}} \times \prod_{I \neq J}^{k_i} 2 \sinh \frac{\phi^i_{IJ}}{2} \Big)\ ~, \cr
	 {Z}^{\rm bifund}_{\vec k}(\phi,a,m;\epsilon_{1,2}) &=\prod^{q_{max} -1}_{j=q_{min}}\prod_{I=1}^{k_{j+1}} \prod_{J=1}^{k_j} \frac{2\sinh{\frac{(\phi^{j+1}_{I}-\phi^j_{J} + m + \epsilon_-)}{2}} \, 2\sinh{\frac{(\phi^{j}_{J}-\phi^{j+1}_{I} +m -\epsilon_-)}{2}}}{2\sinh{\frac{(\phi^{j+1}_{I}-\phi^j_{J}+\epsilon_+ + \epsilon_-)}{2}}\, 2\sinh{\frac{(\phi^j_{J}-\phi^{j+1}_{I}+\epsilon_+ - \epsilon_-)}{2}}}  ~ \ , \cr
	 {Z}_{\vec k, \vec w}^{\rm fund}(\phi,a,m;\epsilon_{1,2})&=\prod^{q_{max}}_{i=q_{min}}\prod_{I=1}^{k_i}\prod_{l=1}^{w_i}\frac{2\sinh{\frac{(\phi^i_I-a_l^i + m)}{2}} \, 2\sinh{\frac{(-\phi^i_I + a_l^i + m)}{2}}}{2\sinh{\frac{(\phi^i_I-a_l ^i+ \epsilon_+)}{2}}\, 2\sinh{\frac{(-\phi^i_I+a_l^i + \epsilon_+)}{2}}}~.
\end{align}
Here the parameters $\{a_l^i\}$ are related to the $U(1)^N$ equivariant parameters $a_\ell$ (with $\ell=1,\ldots, N$) as follows.
\be
a_{\ell(i,l)} = a^i_l, \quad \ell(i,l)= N+1 - \sum^{i}_{j=q_{min}}w_{j-1} -l,
\ee
where $ i=q_{min},\ldots, q_{max},\, l=1,\ldots, w_i$, and $w_{q_{min} -1} =0$. Note that this ordering of the $a_{\ell(i,j)}$ is a convenient choice which does not affect the final result because $Z_{\rm mono}$ is is invariant under the action of the Weyl group.


As discussed in \cite{Hwang:2014uwa, Hori:2014tda, Cordova:2014oxa, Benini:2013xpa}, these contour integrals should be evaluated using the JK residue prescription  (reviewed in appendix \ref{Loc-app})
with the covector $\eta$ of dimension $k$ being set to $\eta = \zeta' (1, 1, \cdots,1)$, where $\pm \zeta' >0$ depending on the chamber.

Equivalently, one can evaluate the contour integral by a  colored version of the Young diagram prescription \cite{Fucito:2004ry}. For $\zeta' >0$, for example,
this proceeds as follows\footnote{The prescription below is essentially a $\Z_n$-projection of Nekrasov's original prescription for instantons on $\C^2$.}:
\begin{enumerate}
\item Consider all $N$--tuples of Young diagrams consisting of a total number of $k=\sum^{n-1}_{i=0}k_i$ boxes. Label each box by the $\Z_n$ charge: the $(i,j)$ box in the $\ell$-th Young diagram\footnote{Our convention for Young diagrams is to draw them in the first
quadrant with $i$ and $j$ labelling the horizontal and vertical axes respectively, increasing away from the origin. Also, note that $r_\ell=\bv_\ell$ from \eref{KN-IOT}.} is assigned the integer $s={r}_\ell +i-j=\mathbf{v}_\ell +i-j$ 
\footnote{Note that the subset of Young diagrams $\cR(\vec k, \vec w)$ depend on $r_\ell$, and therefore on the monodromy vector $\vec w$.}.
\item Each N-tuple of Young diagrams in $\cR(\vec k, \vec w)$ labels a pole in the contour integral \eref{D0quiver-orbifold}. Given an N-tuple of Young diagrams $D \in \cR(\vec k, \vec w)$, let $\cT^D_s(\vec k, \vec w)$ denote the collection of $k_s$ boxes labelled by the $\Z_n$ charge $s$. Then the poles in the variables $\phi^s_I$, corresponding to $D$, will be given as
\begin{align} \label{poles-orbifold}
\phi^s_\ell= a^s_\ell + \epsilon_+ - i \epsilon_1 - j \epsilon_2~, \quad  \forall  (i,j) \in \cT^D_s(\vec k, \vec w)~,\quad\, \ell=1,\ldots,k_s~.
\end{align}
\item Compute the sum of all residues coming from such poles.
\end{enumerate}

As explained in appendix \ref{equiv structure}, flipping the sign of $\zeta'$ corresponds to the transformation $\epsilon_+ \to - \epsilon_+$
in the Witten index, with all other equivariant parameters held fixed. The expression for the Witten index in the $\zeta' < 0$ chamber
can therefore be readily obtained from the expression for the $\zeta' > 0$ chamber by
the following equation:
\be
Z_{SQM}(\Gamma_{\vec k, \vec w}| a, m,\epsilon_{+}, \epsilon_{-}=0;\zeta' < 0) = Z_{SQM}(\Gamma_{\vec k, \vec w}| a, m, -\epsilon_{+}, \epsilon_{-}=0;\zeta' > 0)~.
\ee
It turns out that $Z_{SQM}$ is an even function of $\epsilon_+$ for SQMs associated with $\N=1^*$ instanton partition functions, so that the former is invariant under a sign change of $\zeta'$. Therefore, we can unambiguously define a 5d instanton function for this theory.\\

Given the relation between 5d instanton partition function on $S^1 \times \C^2/\Z_n$ and $Z
_{\rm mono}$ stated in \eref{4d-5d},
we therefore have a concrete formula for the monopole bubbling contribution to line defects in $\N=2^*$ $SU(N)$ SYM, where the RHS is explicitly
given by the equations \eref{D0quiver-orbifold}-\eref{explicitfunctions}, i.e.
\begin{align}\label{quiver-defect1}
\boxed{Z
_{\rm mono} (B,\mathbf{v}; \fa,\fm, \lambda| 
SU(N))
= Z_{SQM}(\Gamma_{\vec k, \vec w}| a, m,\epsilon_{+}, \epsilon_{-}=0; \pm \zeta' > 0)}
\end{align}
where the equality holds for both signs of $\zeta'$ . The map between equivariant parameters on the two sides of the equation is given in equation \eref{eqpar-map}, and the map between the defect data $(B,\mathbf{v})$ on one side and the instanton data $(\vec k, \vec w)$ on the other is discussed in section \ref{ADHM-KN}.

\subsection{Examples of Defect SQMs}\label{Ex-WI}
\subsubsection{$SU(2)$ SYM}
We now proceed to write down explicitly the contour integral formula for $Z_{\rm mono}$ in $4d, \N=2^*$ $SU(2)$ SYM using \eref{quiver-defect1}.
The Dirac quantization condition for an $\N=2^*$ $SU(2)$ theory allows for the charges $B$ and $\mathbf{v}$
to be labelled by half integers, i.e.
\begin{align}
B:=(p/2,-p/2) \quad,\qquad   \mathbf{v}:=(v/2,-v/2)~,
\end{align}
where $p,v$ are integers, and $p=v\,{\rm mod}\,2$. As discussed above, $Z_{\rm mono}(B,\mathbf{v})$ in this case is given by the instanton partition function $Z^{S^1 \times \C^2/\Z_n}_{\vec k, \vec w}$ where $\vec w$ is determined by $\mathbf{v}$ and $\vec k$ is determined by the matrix $K$.

From the character equation \eref{char-eqn-main} one can write down an explicit solution for the matrix $K$ in this case:
\begin{align}
\Tr e^{2\pi \I K \nu}= e^{2\pi \I (\frac{p}{2}-1)\nu} + 2e^{2\pi \I (\frac{p}{2}-2)\nu}+ \ldots + &\frac{p-v}{2} e^{2\pi \I (\frac{v}{2})\nu}+ \ldots + \frac{p-v}{2} e^{2\pi \I (\frac{-v}{2})\nu}+ \ldots \nonumber\\ & +2e^{-2\pi \I (\frac{p}{2}-2)\nu} +e^{-2\pi \I (\frac{p}{2}-1)\nu}~,
\end{align}
such that one has exactly $p-1$ distinct entries $K_i = \frac{p}{2}-i$, where $i=1,\ldots, p-1$, with the multiplicities shown above. Using the redefinition
$$(B, \mathbf{v}, K) \mapsto (B+ \frac{p}{2} \mathbf{I}, \mathbf{v}+ \frac{p}{2} \mathbf{I}, K+ \frac{p}{2} \mathbf{I})~, $$ as discussed in \eref{shift-defectdata}, we have:
\begin{align}
\Tr e^{2\pi \I K \nu}=& e^{2\pi \I (p-1)\nu} + 2e^{2\pi \I (p-2)\nu}+ \ldots + \frac{p-v}{2} e^{2\pi \I (\frac{p+v}{2})\nu}+ \ldots + \frac{p-v}{2} e^{2\pi \I (\frac{p-v}{2})\nu}+ \ldots \nonumber\\ & +2e^{2\pi \I (2)\nu} +e^{2\pi \I (1)\nu}~, \nonumber \\
& \implies K = \text{diag} (1,2,2,3,3,3,\ldots,p-2,p-2,p-1)~,
\end{align}
and a redefined $\mathbf{v}$:
\begin{align}
\mathbf{v}=&(\frac{p+v}{2},\frac{p-v}{2})~.
\end{align}

The redefined $K$ and $\mathbf{v}$ can be packaged into KN data for a fractional $U(2)$ instanton (not $SU(2)$) on $\C^2/\Z_n$ as follows:
\begin{align}
\vec k =&(k_0,k_1, k_2,\ldots, k_{\frac{p-v}{2}}, \ldots, k_{\frac{p+v}{2}}, \ldots, k_{p-2}, k_{p-1}, k_p, \ldots, k_{n-1})\nonumber \\
=& (0,1,2,\ldots,\frac{p-v}{2}, \ldots, \frac{p-v}{2}, \ldots,2,1,0,\ldots,0)~,\\
\vec w=& (w_0,w_1,\ldots,w_{\frac{p-v}{2}}, \ldots,w_{\frac{p+v}{2}}, \ldots, w_p,\ldots,w_{n-1})\nonumber \\
=& (0,0,\ldots,w_{p/2}=2, 0,\ldots,0) \quad \text{if} \; v=0,\\ =& (0,0,\ldots,w_{\frac{p-v}{2}}=1,0,\ldots,0,w_{\frac{p+v}{2}}=1,0,\ldots,0 ) \quad \text{if} \; v \neq 0 ~,
\end{align}
where $\frac{p-v}{2}$ is repeated $v+1$ times in $\vec k$. Note that $k_i =0$, $\forall i \geq p$, since these integers do not appear as entries in the matrix $K$.\\

\begin{figure}[ht!]
\begin{center}
\begin{tikzpicture}[
cnode/.style={circle,draw,thick,minimum size=4mm},snode/.style={rectangle,draw,thick,minimum size=6mm}]
\node[cnode] (1) {1};
\node[cnode] (2) [right=.5cm  of 1]{2};
\node[cnode] (3) [right=.5cm of 2]{3};
\node[cnode] (4) [right=1cm of 3]{\tiny{$\frac{p}{2}-1$}};
\node[cnode] (5) [right=0.5cm of 4]{$\frac{p}{2}$};
\node[cnode] (6) [right=0.5cm of 5]{\tiny{$\frac{p}{2}-1$}};
\node[cnode] (7) [right=1cm of 6]{{$3$}};
\node[cnode] (8) [right=0.5cm of 7]{$2$};
\node[cnode] (9) [right=0.5cm of 8]{1};
\node[snode] (10) [below=0.5cm of 5]{2};
\draw[-] (1) -- (2);
\draw[-] (2)-- (3);
\draw[dashed] (3) -- (4);
\draw[-] (4) --(5);
\draw[-] (5) --(6);
\draw[dashed] (6) -- (7);
\draw[-] (7) -- (8);
\draw[-] (8) --(9);
\draw[-] (5) -- (10);
\end{tikzpicture}
\caption{The Kronheimer-Nakajima quiver associated to a 't Hooft loop labelled by $B=(p/2,\,-p/2)$ ( with $p$ even) in the sector $\mathbf{v}=(0,0)$ in an $\N=2^*$ $SU(2)$ theory.}
\label{fig:v0quiver}
\end{center}
\end{figure}
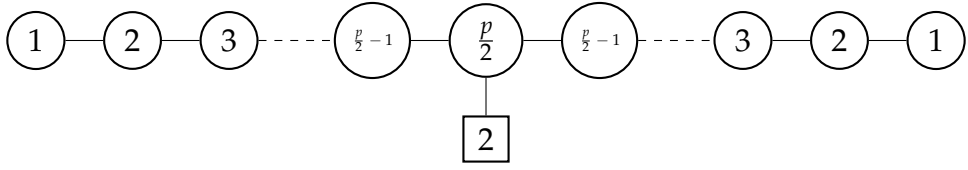

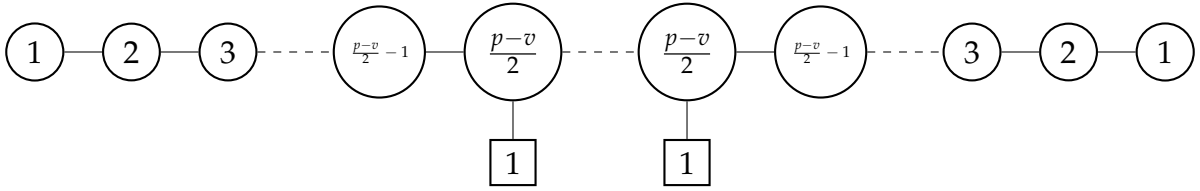
\begin{figure}[ht!]
\begin{center}
\begin{tikzpicture}[
cnode/.style={circle,draw,thick,minimum size=4mm},snode/.style={rectangle,draw,thick,minimum size=6mm}]
\node[cnode] (1) {1};
\node[cnode] (2) [right=.5cm  of 1]{2};
\node[cnode] (3) [right=.5cm of 2]{3};
\node[cnode] (5) [right=1cm of 3]{\tiny{$\frac{p-v}{2}-1$}};
\node[cnode] (6) [right=0.5cm of 5]{$\frac{p-v}{2}$};
\node[cnode] (9) [right=1cm of 6]{$\frac{p-v}{2}$};
\node[cnode] (10) [right=0.5cm of 9]{\tiny{$\frac{p-v}{2}-1$}};
\node[cnode] (13) [right=1cm of 10]{{$3$}};
\node[cnode] (14) [right=0.5cm of 13]{$2$};
\node[cnode] (17) [right=0.5cm of 14]{1};
\node[snode] (18) [below=0.5cm of 6]{1};
\node[snode] (19) [below=0.5cm of 9]{1};
\draw[-] (1) -- (2);
\draw[-] (2)-- (3);
\draw[dashed] (3) -- (5);
\draw[-] (5) --(6);
\draw[dashed] (6) -- (9);
\draw[-] (9) -- (10);
\draw[dashed] (10) -- (13);
\draw[-] (13) -- (14);
\draw[-] (14) -- (17);
\draw[-] (6) -- (18);
\draw[-] (9) -- (19);
\end{tikzpicture}
\caption{The Kronheimer-Nakajima quiver associated to a 't Hooft loop labelled by $B=(p/2,\,-p/2)$ in the sector $\mathbf{v}=(v/2,-v/2)$ with
$v \neq 0$ in an $\N=2^*$ $SU(2)$ theory.
The gauge node $U(\frac{p-v}{2})$ is repeated $v+1$ times.}
\label{fig:vn0quiver}
\end{center}
\end{figure}

The above data completely fixes the D0 world volume theory -- a linear quiver (not a necklace quiver since $k_i=0$, $\forall i \geq p$) with a gauge group $ {G}= \prod^{p-1}_{i=1} U(k_i)$ with bifundamentals and two fundamental hypers distributed among the gauge nodes (as dictated by $\vec w$), as shown in the figures \ref{fig:v0quiver} and \ref{fig:vn0quiver}. The monopole bubbling contribution to the line operator can then be computed using \eref{quiver-defect1}.\\

The complex dimension of the vector space $V$ is given by the quaternionic dimension of the Coulomb branch quiver which can be computed as a function of $p$ and $v$:
\begin{align}
k = \text{dim}_{\C}V={\rm dim}_{\mathbb{H}}\,\cM_C(\Gamma_{\vec k, \vec w})=\sum^{n-1}_{i=0}k_i =\frac{p-v}{2}\times \frac{p+v}{2}~,
\end{align}
while the quaternionic dimension of the quiver variety $\cM(B, \mathbf{v})$ is given by the Higgs branch dimension of the quiver
\begin{align}
{\rm dim}_{\mathbb{H}}\,\cM(B, \mathbf{v}) = {\rm dim}_{\mathbb{H}}\,\cM_H(\Gamma_{\vec k, \vec w}) = \frac{p-v}{2}~.
\end{align}

\begin{figure}[ht!]
\begin{center}
\includegraphics[scale=0.5]{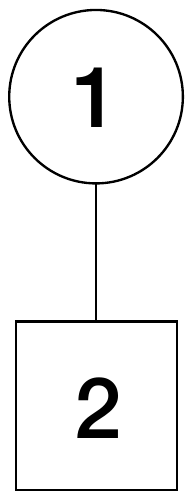}
\caption{The Kronheimer-Nakajima quiver associated to a 't Hooft loop labelled by $B=\half(2,\,-2)$  in the sector $\mathbf{v}=(0,0)$ in an $\N=2^*$ $SU(2)$ theory. This corresponds to the integer $p=2$ in the figure \ref{fig:v0quiver}.}
\label{fig:v0quiverex1}
\end{center}
\end{figure}

One can now proceed to compute some simple examples and check that the above contour integral indeed reproduces the IOT result.
Consider the simplest example of $\mathbf{Z_{\rm mono}}(p=2,v=0)$ :\quad the character equation \eref{char-eqn-main} for
$p=2,v=0$ gives a one-dimensional matrix $K=0$.
After the aforementioned shift in $K$ and $v$, we get $K=1$ and $\mathbf{v}=(1,1)$. The KN quiver is therefore characterized by the instanton data
$\vec k=(0,1,0,\ldots,0)$ and $\vec w=(0,2,0,\ldots,0)$ for a $U(2)$ theory on a $\C^2/\Z_n$ orbifold. This gives a $(4,4)$ theory with gauge group $U(1)$ and two fundamental hypers, as shown in figure \ref{fig:v0quiverex1}. The Witten index in the $\zeta' > 0$ chamber can be read off from \eref{D0quiver-orbifold} :
\begin{align}
	&Z_{SQM}(\Gamma_{\vec k, \vec w}| a, m,\epsilon_{+}, \epsilon_{-};\zeta' > 0) =  \oint_{JK(\zeta')} \left[\frac{d\phi}{2\pi i}\right] \,  {Z}^{\rm vector}(\phi,m;\epsilon_{1,2}) \cdot   {Z}^{\rm fund}(\phi,a,m;\epsilon_{1,2}) \ ~,\cr
	&  {Z}^{\rm vec}(\phi,m;\epsilon_{1,2}) =  \Big(\frac{2\sinh{(\epsilon_+)}}{2\sinh{\frac{(m \pm \epsilon_+)}{2}}}\Big) \ ~, \cr
	&	 {Z}^{\rm fund}(\phi,a,m;\epsilon_{1,2})=\prod^2_{\ell=1}\frac{2\sinh{\frac{(\phi-a_\ell + m)}{2}}\, 2\sinh{\frac{(-\phi+a_\ell + m)}{2}}}{2\sinh{\frac{(\phi-a_\ell + \epsilon_+)}{2}}\,2\sinh{\frac{(-\phi+a_\ell + \epsilon_+)}{2}}} ~.
\end{align}

The poles of the above contour integral correspond to doublets (since we have a $U(2)$ theory) of colored Young diagrams
with total number of boxes equal to $\sum_i k_i=1$, where every box is assigned the integer $s=\mathbf{v}_\ell + i-j$ ($l=1,2$ indexes the doublet of Young diagrams and $(i,j)$ in the first quadrant) such that the number of boxes labelled by integer $s$ is $k_s$.  From equation \eref{poles-orbifold}, the poles are then explicitly given as
\begin{align}\begin{split}
& (1) \vec Y =(Y_1,Y_2): Y_1=\Yvcentermath1 \begin{Young} 1  \cr \end{Young}\;, Y_2 = \emptyset, \implies \phi= a_1 - \epsilon_+, \\
& (2)\vec Y =(Y_1,Y_2): Y_1 = \emptyset\;, Y_2=\Yvcentermath1 \begin{Young} 1 \cr \end{Young}\,, \implies \phi= a_2 - \epsilon_+.
\end{split}\end{align}
Computing the residues at these two poles, one obtains
\begin{align}
Z_{\rm mono}&(p=2,v=0; a,m,\epsilon_{+})= Z_{SQM}(\Gamma_{\vec k, \vec w}| a, m,\epsilon_{+}, \epsilon_{-};\zeta' > 0)|_{\epsilon_{-}=0} \nonumber \\ &= \frac{\sinh{\frac{(2a+m +\epsilon_+)}{2}} \sinh{\frac{(2a-m +\epsilon_+)}{2}}}{\sinh{a}\sinh{(a+\epsilon_+)}} + \frac{\sinh{\frac{(2a+m-\epsilon_+)}{2}} \sinh{\frac{(2a-m - \epsilon_+)}{2}}}{\sinh{a}\sinh{(a-\epsilon_+)}}~.
\end{align}
The above formulae matches IOT's expressions with the redefinition of equivariant parameters as given in \eref{eqpar-map}.

We compute more examples of 't Hooft operators and check their agreement with the results of \cite{Ito:2011ea} in appendix \ref{Okuda-2*}.
 We discuss quivers arising in $\N=2^*$ $SU(N)$ theory for $N > 2$ in section \ref{HW}, after discussing the Type IIB construction of singular monopole moduli spaces and its relation to the SQMs associated to 't Hooft defects.

\subsubsection{$U(2)$ SYM}
We now proceed to write down explicitly the contour integral for $Z_{\rm mono}$ for line defects in $\N=2^*$ $U(2)$ SYM. Consider a line defect $T_B$ and the screening charge $\mathbf{v}$ labelled by
\begin{align}
B:=(p,0)\quad, \qquad \quad \mathbf{v}:=(v,p-v)~,
\end{align}
where $p,v$ are non-negative integers with $v \leq p$. Similar to the $SU(2)$ case, $Z_{\rm mono}(B,\mathbf{v})$ in this case is given by the instanton partition function $Z^{S^1 \times \C^2/\Z_n}_{\vec k, \vec w}$ where $\vec r$ is determined by $\mathbf{v}$ and $\vec k$ is determined by the matrix $K$. We determine the instanton data and the associated quiver description of the answer in the usual fashion.

\begin{figure}[ht!]
\begin{center}
\includegraphics[scale=0.5]{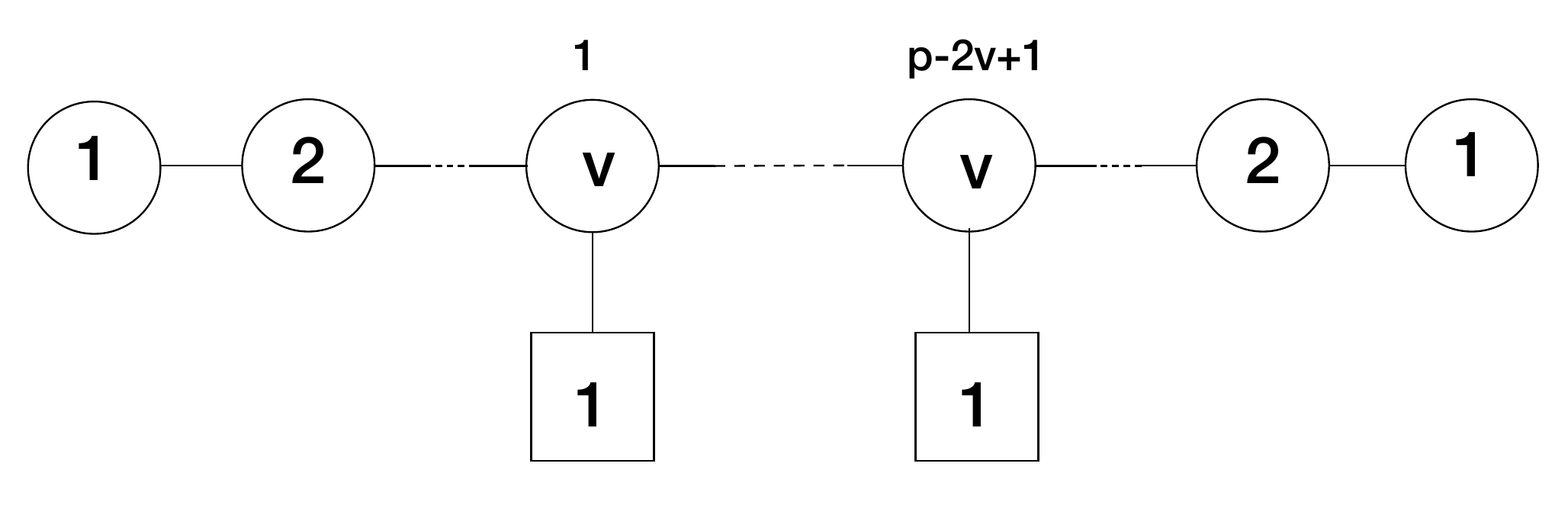}
\caption{The Kronheimer-Nakajima quiver associated to a 't Hooft loop labelled by $B=(p,\,0)$ in the sector $\mathbf{v}=(v,p-v)$ (where $v \neq p-v$) in an $\N=2^*$ $U(2)$ theory.}
\label{fig:vn0Uquiver}
\end{center}
\end{figure}

\begin{figure}[ht!]
\begin{center}
\includegraphics[scale=0.5]{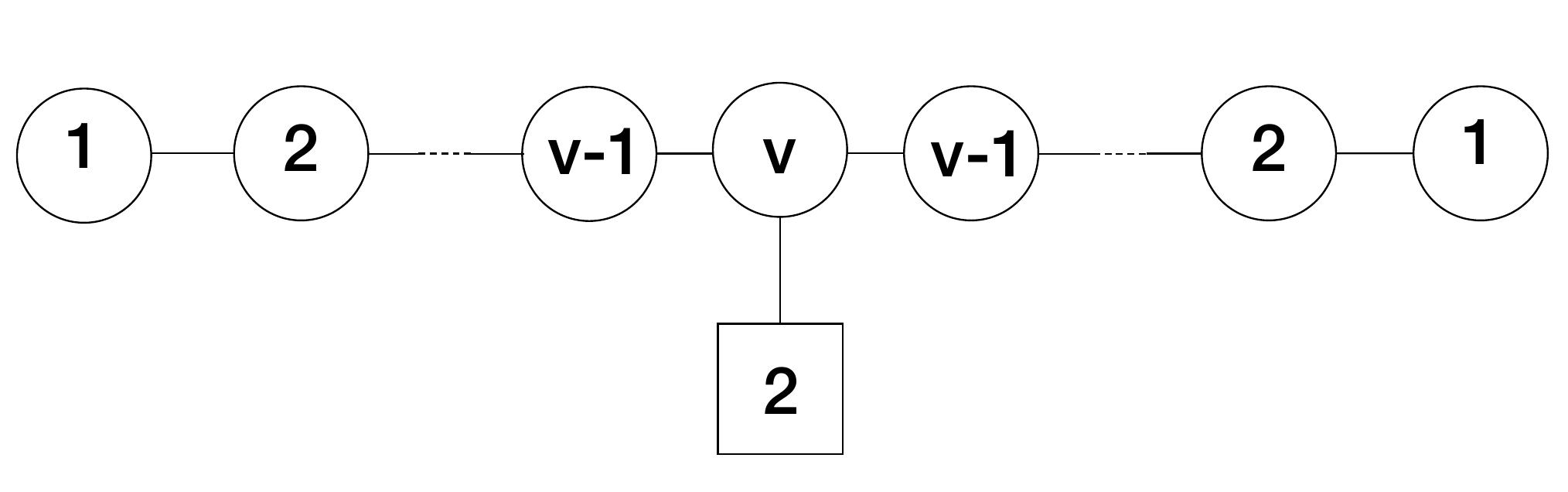}
\caption{The Kronheimer-Nakajima quiver associated to a 't Hooft loop labelled by $B=(p,\,0)$ ( with $p$ even) in the sector $\mathbf{v}=(p/2,p/2)$ in an $\N=2^*$ $U(2)$ theory.}
\label{fig:v0Uquiver}
\end{center}
\end{figure}

From the character equation \eref{char-eqn-main} one can write down an explicit solution for the matrix $K$ in this case:
\begin{align}
\Tr e^{2\pi \I K \nu}= e^{2\pi \I (1)\nu} + 2e^{2\pi \I (2)\nu}&+ \ldots +  v e^{2\pi \I v \nu}+ v e^{2\pi \I (v+1) \nu}+ \ldots + v e^{2\pi \I (p-v-1)\nu}
+ v e^{2\pi \I (p-v)\nu} \nonumber\\ & + (v-1) e^{2\pi \I (p-v+1)\nu}+ \ldots +2e^{2\pi \I (p-2)\nu} +e^{2\pi \I (p-1)\nu}~,
\end{align}
which translates to the following KN instanton data of a $U(2)$ theory on $\C^2/\Z_n$:
\begin{align}\begin{split}
& \vec k=(0,1,2,\ldots,v-1,v,\ldots,v, v-1,\ldots,2,1,0,\ldots,0)~,\\
& \vec w = (0,1,2,\ldots,0,1,\ldots,1,0,\ldots,0,0,0,\ldots,0)~.
\end{split}\end{align}
where $v$ is repeated $p-2v+1$ times. The associated quiver quantum mechanics are given in figure \ref{fig:vn0Uquiver} and \ref{fig:v0Uquiver} (for $v \neq p/2$ and $v =p/2$) and its Witten index can be computed as before. Line defects labelled by $B=(p,-p)$ work out in ways similar to the $SU(2)$ SYM with a defect B labelled by an even spin.

\section{String Theory description of singular monopole moduli spaces associated to line defects}\label{HW}
In this section, we present a Type IIB string theory description of monopole bubbling on $\R^3$, and demonstrate
how one can derive the quiver variety $\cM(B, \mathbf{v})$ from a
configuration involving D1-D3-NS5-branes.
Without monopole bubbling, the Type IIB description presented in this section is U-dual to the brane configuration of Cherkis and Kapustin
\cite{Cherkis:1997aa} -- the new element is the incorporation of monopole bubbling in the picture. The brane picture gives an alternative derivation of
the quiver variety $\cM(B, \mathbf{v})$ for an $SU(N)$ line defect with $N > 2$, the general form of which is rather difficult to derive directly from the character
equation \eref{char-eqn-main}.

\subsection{Review of D1-D3 system for smooth monopoles}
Let us first review the standard Type IIB description of smooth monopoles in terms of finite segments of D1 branes ending on D3 branes,
using the Nahm construction \cite{Diaconescu:1996rk}. Consider the D1-D3-brane configuration:

\begin{center}
\text{Type IIB}\\
\begin{tabular}[c]{c|cccccccccc}
&0&1&2&3&4&5&6&7&8&9\\
\hline
D3&\textbf{--}&\textbf{--}&\textbf{--}&\textbf{--}&&&&&&\\
D1&\textbf{--}&&&&\textbf{--}&&&&&\\
\end{tabular}
\end{center}

where $\textbf{--}$ indicates that the D-brane extends along that direction and blanks mean a Dirichlet boundary condition is imposed for that coordiinate. Here $x^4$ is a coordinate on a compact direction transverse to the
D3-brane. We will often denote it by $s$.
 A Yang-Mills-Higgs system is naturally realized in the low energy string theory on the world volume of D3 branes. These extend along
the directions $x^\mu,~\mu=0,1,2,3,$ in the 10d spacetime of Type IIB string theory and sit at definite values of $x^{\alpha}, ~\alpha=4,5,6,7,8,9$.
The low energy world volume gauge theory on a stack of $N$ coincident D3-branes is 4d $\N=4$ $U(N)$ SYM, which consists of a gauge
field, six real adjoint scalars and four adjoint Weyl fermions. The adjoint scalars encode the profile of the D3-branes in the
six directions $x^\al,~\al=4,\ldots,9$ \cite{Witten:1995im}. For the rest of this section, we will consider a classically truncated version of the D3-brane
world volume theory where we set all fermions and five of the six scalar fields to zero, choosing only the scalar field $X$ associated
with the $x^4$ direction to be non-zero\footnote{This is a consistent truncation because the
equations of motion for these fields have no source terms built only out of $(A_{\mu=0,1,2,3}, X)$.}.\\

The world volume theory on a stack of D1-branes is a 2d $(8,8)$ SYM theory, while the D3 branes act as half-BPS boundary
conditions that reduce the supersymmetry to $(4,4)$. The 2d $(8,8)$ vector multiplet consists of a 2d gauge field and eight
real scalars which encode the position of the D1-brane along the eight transverse directions in the 10-dimensional space-time.
Let $(X_i)_{i=1,2,3}$ denote the three real scalar fields which are associated with the positions of D1-branes in the spatial $\R^3_{1,2,3}$
of the D3-brane world volume. In the effective $0+1$ dimensional theory obtained by KK-reducing the D1-brane world volume theory along the compact
direction, the scalars $(X_i)$ combine with the scalar $A_4$ to give the bosonic part of a $(4,4)$ hypermultiplet.\\

For the sake of brevity, we will specialize to the case of smooth $SU(2)$ monopoles in this subsection. The Type IIB picture in this case consists of
two D3-branes located at $s= \pm s_0$, and $m$ D1-brane segments ending on them.
It was shown \cite{Diaconescu:1996rk} that the moduli space of supersymmetric ground states (preserving $(4,4)$ supersymmetry)
of this brane configuration is isomorphic to the moduli space of smooth $SU(2)$ monopoles with asymptotic magnetic charge
$\gamma_m = {\rm diag}(-m,m)$ and Higgs vev $X_\infty = {\rm diag}(-s_0, s_0)$. The moduli space of supersymmetric ground states
of the brane configuration is given by the moduli space of solutions of the following BPS equations in the D1 world volume gauge theory:
\be \label{Nahm-smoothD1}
\frac{\de X_i}{\de s} - \I [A_4, X_i]+ \frac{\I}{2} \epsilon_{ijk} [X_j, X_k] =0~,
\ee
where $A_4, X_i$ are $m \times m$ Hermitian matrices, transforming under a $SU(m)$ gauge transformation $g(s)$ as follows:
\be
(A_4, X_i) \to (g^{-1} A_4 g + \I g^{-1} \de g, g^{-1} X_i g)~.
\ee
This $SU(m)$ gauge transformation can be used to gauge-fix $A_4$ to zero.
In addition, the fields $X_i$ encounter Nahm poles in the vicinity of D3-branes, i.e. around $s \to \pm s_0$,
\be \label{Nahm-smooth-bc}
X_i = \frac{L^{\pm}_i}{s \mp s_0} + O(1)\quad, \qquad [L^{\pm}_i, L^{\pm}_j]=i\epsilon_{ijk} L^{\pm}_k,
\ee
where the $L^{\pm}_i$s form a spin-$(m-1)/2$ representation of the $SU(2)$ Lie algebra.
Equation \eref{Nahm-smoothD1} is equivalent to Nahm's equation \cite{Nahm:1979yw, Nahm:1981nb, Nahm:1982jb} -- the moduli space of
solutions of this equation subject to the boundary condition in equation \eref{Nahm-smooth-bc} gives the
moduli space of smooth $SU(2)$ monopoles on $\R^3$ with asymptotic charge $\gamma_m$. 
 The scalar
fields $X_i(s)$ together with the boundary condition constitute the Nahm data.

In addition to the moduli space, the explicit monopole solution $(A_{i}, X)$ in the $SU(2)$ Yang-Mills-Higgs system can be
constructed from the Nahm data using the reconstruction procedure \cite{Weinberg:2006rq} in the following fashion.
Let us define a linear differential operator
\be
\Delta (\vec x, s) := \frac{\de}{\de s} - X_i(s) \otimes \sigma_i + x_i \textbf{I}_m \otimes \sigma_i~,
\ee
and compute solutions to the equation:
\be
\Delta^\dagger (\vec x, s) w(\vec x, s) = \Big[-\frac{\de}{\de s} - X_i(s) \otimes \sigma_i + x_i \textbf{I}_m \otimes \sigma_i\Big] w(\vec x, s) =0~,
\ee
where $w(\vec x, s)$ is a $2m$-dimensional vector. Let $\{w_a(\vec x, s)\}$ denote a basis of normalizable linearly independent solutions
of the above equation with $a=1,2$ in the present case \footnote{It was shown in \cite{Weinberg:2006rq} that there are precisely $N$ basis vectors labelled by $a=1,\ldots,N$ for $SU(N)$. Normalizability of the solution requires that $w(\vec x, s)$ be regular as $s \to \pm s_0$. See section 4.4.3 of \cite{Weinberg:2006rq}
for more details.}.
Given these solutions, the Yang-Mills-Higgs fields $(A_{i}, X)$ are given as
\begin{align}
& X^{ab}(\vec x)= \langle w_a|x^4|w_{b}\rangle=\int^{s_0}_{-s_0} \de s\, s\, w^\dagger_a(\vec x, s) \, w_{b}(\vec x, s)~,\\
& A^{ab}_i(\vec x)=\langle w_a|p_i|w_{b}\rangle=  \int^{s_0}_{-s_0} \de s\, w^\dagger_a(\vec x, s) \, (- \I\partial_i) \, w_{b}(\vec x, s)~.
\end{align}
It can be explicitly shown that the classical field configurations constructed by the above procedure satisfies the Bogomolnyi equation for an $SU(2)$ Yang-Mills-Higgs system on $\R^3$ and gives the correct asymptotic behavior at infinity. We refer the reader to section 4 of \cite{Weinberg:2006rq} for details.

\subsection{D1-D3-NS5 system for $SU(2)$ singular monopoles and monopole bubbling}\label{HW-U(2)}
We now discuss how singular monopoles on $\R^3$ can be realized in Type IIB string theory by introducing
NS5-branes in the D1-D3 configuration described above. As mentioned earlier, the relevant brane set-up is
closely related to a U-dual version of the brane configuration studied in \cite{Cherkis:1997aa}.
We discuss in detail the case of a product of minimal singular $SU(2)$  monopoles with total 't Hooft charge $B={\rm diag}(-p, p)$ and asymptotic
charge $\gamma_m={\rm diag}(-m,m)$, where $p, m$ are positive integers.
The bubbling sectors are labelled by $\mathbf v={\rm diag}(-v, v)$, where $v \leq p$ is a positive integer.
Also, let $\tilde{\gamma}_m = \gamma_m + B = {\rm diag}(-p-m,\, p+m)$.
Generalization to the $SU(N)$ case is straightforward, and will be discussed in the next subsection.

\begin{figure}[h]
\centering
\includegraphics[scale=1.7]{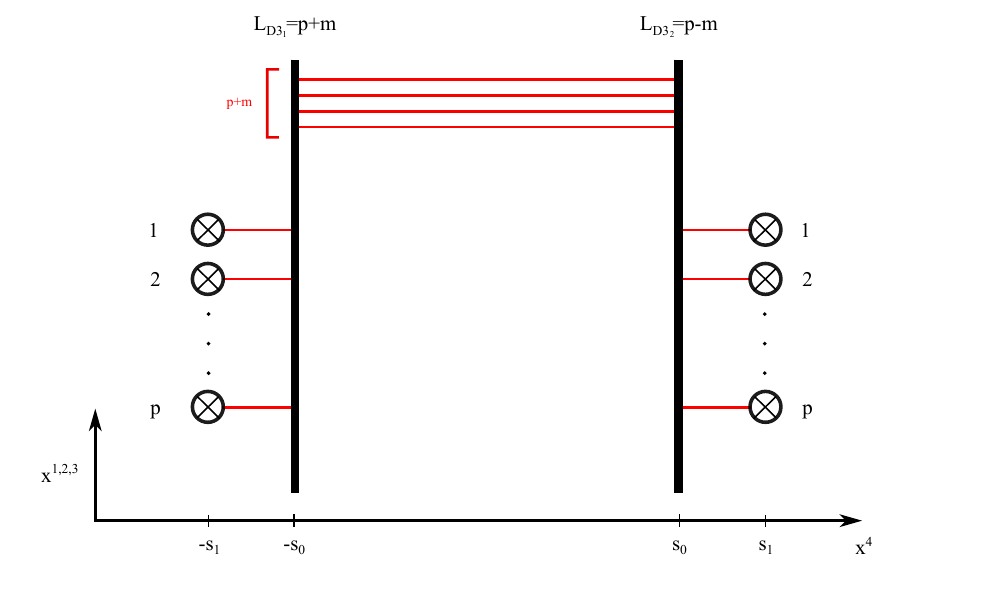}
\caption{D1-D3-NS5 brane configuration for singular monopoles in an $SU(2)$ theory.
Circles with crosses, horizontal lines in red, and vertical black lines, denote NS5-branes,
D1-branes and D3-branes respectively.
Linking numbers of the two D3 branes are $m+p$ and $-m + p$ respectively, while the NS5 linking numbers are all equal to 1, as described below. }
\label{SingMono}
\end{figure}

Consider the Type IIB configuration
consisting of 2 D3-branes, $n=2p$ NS5-branes and $(m+p)$ D1-branes, summarized in the table (and in figure \ref{SingMono}):
\begin{center}
\text{Type IIB}\\
\begin{tabular}[c]{c|cccccccccc}
&0&1&2&3&4&5&6&7&8&9\\
\hline
D3&\textbf{--}&\textbf{--}&\textbf{--}&\textbf{--}&&&&&&\\
D1&\textbf{--}&&&&\textbf{--}&&&&&\\
$n$-NS5&\textbf{--}&&&&&\textbf{--}&\textbf{--}&\textbf{--}&\textbf{--}&\textbf{--}\\
\end{tabular}\end{center}

\noindent As before, -- indicates that the corresponding brane extends in that particular direction, while other directions have Dirichlet boundary
conditions.

Specifically, the D3-branes are located at $s=\pm s_0$ along the compact direction $x^4$ and the $2p$ NS5-branes are located at points $(x_1,x_2,x_3)$ in the $\R^3$ of the D3-brane world volume. For an $SU(2)$ monopole, we will take the $\R^3$ positions of the NS5-branes to pairwise coincide such that there are exactly $p$ independent positions $\vec x_\al$ ($\al=1,\ldots,p$) and each pair has an NS5-brane located at $s=\pm s_1$ in the $x^4$ direction \footnote{For generic positions, we will end up with 2p insertions of minimal $SU(2)$ `t Hooft defects.}. Additionally, we take a single D1-brane connecting every NS5-brane to the nearest D3-brane and $(p+m)$ other D1-branes connecting the two D3-branes at points on $\R^3$ (generically distinct from $\vec x_\al$) as shown in figure \ref{SingMono}.

The moduli space of supersymmetric ground states of this Type IIB brane configuration gives the moduli space
of multiple singular $SU(2)$ monopoles\footnote{Note that each singular monopole is the coincident limit of a pair of singular monopoles which are S-dual to a Wilson defect in the fundamental representation.} on $\R^3$ with total 't Hooft charge $B$ and asymptotic charge $\gamma_m$
. In the limit where all the $\vec{x}_\al$ coincide\footnote{Note that in order to take this coincident limit in the brane construction, we require displacing the NS5-branes in the $x^4$-direction so that they are all at distinct points.}, this describes a configuration with a single magnetic defect of magnetic charge $B$.

This Type IIB picture admits a nice physical description for monopole bubbling.
Given the D1/D3/NS5-brane configuration, one can check that it corresponds to singular monopoles by directly constructing the classical solutions for the Yang-Mills-Higgs system $(A_i, X)$ on the D3-brane world volume theory. This can be accomplished by solving Nahm's equations along the compact direction $x^4$ and then using the reconstruction procedure as outlined earlier (a procedure that requires the bow diagram technology \cite{Blair:2010vh}). However, solving Nahm's equations for arbitrary $p,m$ of course is a technically difficult problem of computing non-Abelian solutions of the Nahm equation. Our goal in this section is to give an intuitive D-brane picture of the bubbling locus of singular monopoles. We will see that this will give a clear, physical interpretation of the space $\mathcal{M}(B,\bv)$.


To begin, consider the D1/D3/NS5-brane configuration shown in figure \ref{SingMono}
for $p=1$ and $m=0$. Here there are two NS5-branes and a
single D1-brane stretched between the D3-branes. This has the interpretation of a single smooth monopole in the presence of an
't Hooft singularity. In order to construct the field configuration of these branes, we want to solve Nahm's equations on the interval between the NS5-branes. For this configuration, the D3- and NS5-branes introduce boundary conditions to the Nahm equations.

 It was shown in \cite{Gaiotto:2008sa, Hanany:1996ie} that the NS5-branes (located at $s= \pm s_1$) impose Dirichlet boundary conditions while the D3-branes introduce Nahm pole boundary conditions (located at $s=\pm s_0$).
Since $m+p=1$, the $X_i$'s are $1\times1$ matrices and the Nahm equation away from the boundaries reduces to its Abelian version, i.e.
\begin{align}
\frac{\de X_i}{\de s} + \frac{\I}{2} \epsilon_{ijk} [X_j, X_k] =0~, \quad \implies \quad X_i = {\rm constant}~.
\end{align}
This implies that the fields $X_i(s)$ are piece-wise constant and can jump discontinuously across a D3-brane. Explicitly, one
can write solution corresponding to an $SU(2)$ monopole as:
\begin{equation}
\vec X=
\begin{cases}
\vec x_1 & \text{for}\, -s_1 < s < -s_0 \\
\vec x'_1    & \text{for}\, -s_0 < s < s_0 \\
\vec x_1 & \text{for}\,\quad  s_0 < s < s_1
\end{cases}
\end{equation}
Physically, the solutions simply correspond to the position of the respective D1-brane segment in the spatial $\R^3$
of the D3-brane world volume -- in particular, $\vec x'_1$ is the position of the smooth monopole on $\R^3$.
Given the above solution, the Yang-Mills-Higgs system $(A_i, X)$ can be obtained
by the standard reconstruction procedure of Nahm data. Such problems
have been analyzed in \cite{Cherkis:2007qa, Cherkis:2007jm}, and therefore we can use their results instead
of going through the details of the reconstruction procedure. In the limit $|s_1| \to \infty$, the Higgs field $X$ and
the gauge connection $A$ are given by \cite{Cherkis:2007qa}
\begin{align}\begin{split}
& X = \vec \sigma \cdot \vec \phi\quad,\qquad A= \vec \sigma \cdot \vec A~,\\
& \vec \phi = \Big(\Big(s_0 + \frac{1}{2 r }\Big)\frac{\cK}{\cL} -\frac{1}{2l} \Big)\frac{\vec l}{l} - \frac{l}{r \cL} (\vec d - \frac{\vec l \cdot \vec d}{l^2} \vec l)~,\\
& \vec A = \Big(\Big(s_0 + \frac{r+d}{\cD }\Big)\frac{\cD}{\cL} -\frac{1}{2l} \Big)\frac{\vec l \times \de \vec x}{l}
- \frac{l}{\cL} \Big(\frac{\vec r \times \de \vec x}{r} + (\frac{\cK}{\cD}-1)\frac{\vec l \cdot (\vec r \times \de \vec x)}{l\, r} \frac{\vec l}{l} \Big)~,
\end{split}\end{align}
where $\vec \sigma$ are the Pauli matrices, the various relative position vectors and the functions $\cK,\cL$ are given as
\begin{align}\begin{split}
& \vec r = \vec x - \vec x_1\quad,\qquad \, \vec l = \vec x - \vec x'_1\quad,\qquad \, \vec d = \vec x'_1 - \vec x_1~,\\
& \cK = ((r+d)^2 + l^2)\cosh{(2s_0 l)} + 2 l (r+d)\sinh{(2 s_0 l)}~, \,\\
& \cL= ((r+d)^2 + l^2)\sinh{(2s_0 l)} + 2 l (r+d)\cosh{(2 s_0 l)}~,\\
& \cD= ((r+d)^2 - l^2)~.
\end{split}\end{align}
To begin with, consider the situation where the D1-brane segment between the pair of D3-branes is far away from the location of the NS5-branes,
i.e. $|\vec x'_1| \to \infty$, and $r=|\vec r|$ is finite. From the perspective of the D3 world volume theory, this corresponds to the smooth monopole being far away
from the location of the 't Hooft defect. In this limit, $d =|\vec d | \to \infty$, $l =|\vec l |\to \infty$, $\frac{\cK}{\cL} \sim 1$, $\cL \sim l^2 e^{2\,s_0\, l}$,
which leads to the Dirac monopole solution at $\vec x = \vec x_1$ with 't Hooft charge $B_1= {\rm diag}(-1,1)$
\footnote{We can apply a constant $SU(2)$ gauge transformation to diagonalize $X$ and $A$ in the neighborhood of $x=x_1$.
Here we are using the convention of \cite{Cherkis:2007qa, Cherkis:2007jm} to write down the solutions of $X,A$.}:
\begin{align}
X \sim \Big(s_0 + \frac{1}{2r} \Big) \frac{\vec \sigma \cdot \vec l}{l}\quad,\qquad \quad |X| = \sqrt{\vec \phi \cdot \vec \phi } \sim \Big(s_0 + \frac{1}{2r} \Big)~.
\end{align}
Now, let us use this to study monopole bubbling. In this description, monopole bubbling corresponds to when the position of the D1-brane on $\R^3_{1,2,3}$ coincides with that of the NS5 branes.
In the D3 world volume theory, this corresponds to a smooth monopole dissolving in the 't Hooft defect, thereby screening the 't Hooft charge. In the present example, this happens when $\vec x'_1 \to \vec x_1$,
which implies $\vec r \to \vec l$, $\vec d \to 0$, and therefore leads to complete screening of the 't Hooft charge, i.e.
\begin{align}
X \sim \Big(s_0 + \frac{1}{2r}  -  \frac{1}{2r} \Big) \frac{\vec \sigma \cdot \vec r}{r}\quad \implies \quad |X| \sim s_0~.
\end{align}
This monopole bubbling configuration is labelled by the effective 't Hooft charge $\mathbf{v}=(0,0)$.\\

Now consider the case of arbitrary $p,m$. In the limit where the D1-branes are far away,
 the $p$-pairs of NS5-branes introduce $p$ Dirac monopoles of 't Hooft charge
$B_\al = {\rm diag}(-1, 1)$ at positions $\vec x_\al$, $\al=1,\ldots,p$, on $\R^3$ of the D3 world volume.
A single Dirac monopole of 't Hooft charge $B={\rm diag}(-p,p)$ can be obtained by making the positions $\vec x_\al$
of the $p$ pairs NS5-branes coincide, while keeping their positions in the $x^4$-direction unchanged.

\begin{figure}[t!]
\centering
\includegraphics[scale=1.7,trim=1cm 20.5cm 11cm 0.5cm,clip]{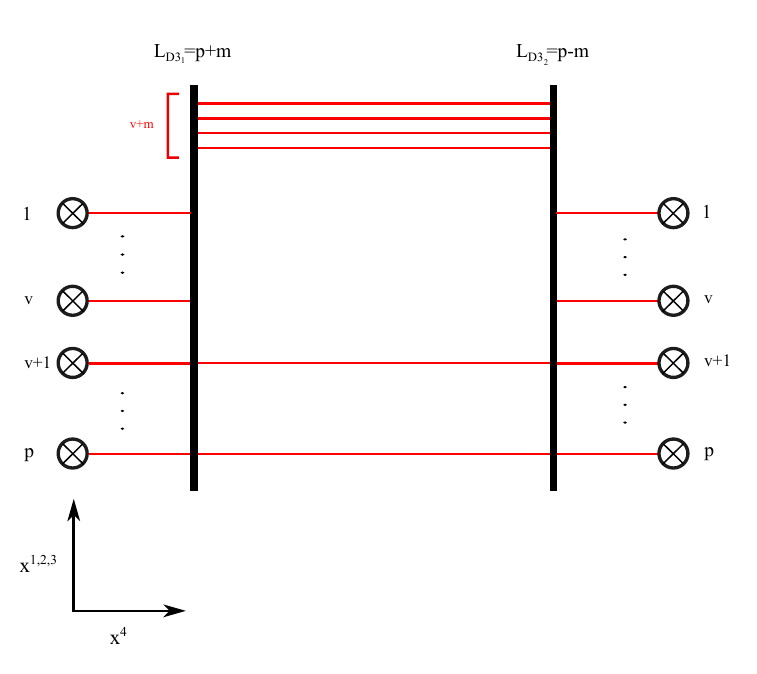}
\caption{D1-D3-NS5-brane configuration for bubbling monopole in a $SU(2)$ theory in the sector $\mathbf{v}={\rm diag}(-v,\,v)$.
The $\R^3$ positions of the $p$ pairs of NS5-branes are distinct. The $\R^3$ positions of $(p-v)$ D1-branes coincide
with the $\R^3$ positions of $(p-v)$ pairs of NS5-branes, thereby completely screening their 't Hooft charge, as described above.
In the figure, the pairs labelled $v+1$ through $p$ are screened, while the pairs labelled $ 1$ through $v$ are not.
On taking the limit where the $\R^3$ positions coincide, one obtains a single
't Hooft defect with charge $\mathbf{v}={\rm diag}(-v,\,v)$. }
\label{fig:monobubb}
\end{figure}

Monopole bubbling can be observed in this set-up in the following fashion.
Consider the configuration in which the pairs of NS5-branes are well-separated.
Now let us move a total of $(p-v)$ D1-branes such that their $\R^3$ positions coincide with that of $(p-v)$
pairs of NS5-branes, thereby completely screening their 't Hooft charge, as described above.
The 't Hooft charges of the remaining $p-(p-v)=v$ Dirac monopoles are not screened.
Therefore, in taking the limit where $\R^3$ positions of the $p$ Dirac monopoles coincide,
we obtain a product of  't Hooft defects with effective charge $\mathbf{v}=(-v,\,v)$. This corresponds to the
bubbling configuration labelled by the effective 't Hooft charge $\mathbf{v}=(-v,\,v)$.
The Type IIB description is shown  in figure \ref{fig:monobubb}.

Now, one can use the Type IIB brane configuration to derive the quiver variety $\cM(B, \mathbf{v})$. Recall that $\cM(B,\mathbf{v})$ is the transversal slice to the smooth space $$\cM^{(s)}(v,\gamma_m;X_\infty)\subset \cM(B,\gamma_m;X_\infty)~.$$ Since the smooth space $\cM^{(s)}$ describes the moduli of unbubbled monopoles in the bulk away from the singular monopole, this means that $\cM(B,\mathbf{v})$ describes the moduli of the bubbled monopoles.\\

Now recall from \cite{Hanany:1996ie} that, given a Type IIB configuration of D1/D3/NS5-branes, one can associate a {\it linking number}
to every D3 and NS5-brane\footnote{Notice that this construction is T-dual to a D3/D5/NS5-brane configuration as studied in \cite{Hanany:1996ie}.}. The linking numbers of these three and five-branes can be read off from the brane configurations in figure \ref{SingMono} or figure \ref{fig:monobubb}. This quantity measures the
effective D1-brane number at infinity on the respective D3 or NS5-brane \cite{Hanany:1996ie}. We will define a {\it Hanany-Witten frame}
as a brane configuration obtained by moving NS5 and D3-branes in the original configuration past each other --
creating or destroying D1-branes in the process -- such that the linking numbers of the D3 and NS5-branes are preserved. Explicitly, using the convention of \cite{Gaiotto:2008ak}, we have
\begin{align}\begin{split}
& L_{D3_1} :=n_{\rm left}(NS5) + n_{\rm right}(D1) - n_{\rm left}(D1)=  m+p~, \\
& L_{D3_2} :=n_{\rm left}(NS5) + n_{\rm right}(D1) - n_{\rm left}(D1)= -m + p~,\\
& L_{NS5_\al} :=n_{\rm left}(D3) + n_{\rm right}(D1) - n_{\rm left}(D1)=1~,\quad\forall \al~,
\end{split}\end{align}
where $n_{\rm left}(NS5)$ denotes the number of NS5 to the left of a given D3-brane, $n_{\rm left}(D3)$ denotes the number of D3-branes to the
left of a given NS5-brane, and $n_{\rm right}(D1)$, $n_{\rm left}(D1)$ denote the number of D1-branes ending on a D3 or an NS5-brane from the
right and the left respectively.

Consider only D1-branes corresponding to bubbled monopoles. To read off the quiver gauge theory whose Higgs branch corresponds to $\cM(B,\mathbf{v})$, we need to go to a specific Hanany-Witten frame,
where these D1-branes begin and end only on NS5 branes \footnote{This is related to the fact that an NS5-brane
imposes Neumann boundary conditions on the (4,4) vector multiplet and Dirichlet boundary conditions on the adjoint (4,4) hypermultiplet, in the
D1-brane world volume theory. We refer the reader to \cite{Gaiotto:2008sa, Gaiotto:2008ak, Hanany:1996ie} for details.}.
The brane configuration resulting from these transitions is shown in figure \ref{fig:magdual}. The associated quiver, which arises as the low energy effective theory
on the D1 world volume, can be easily read off  from the massless open string spectrum (see figure \ref{quiverU(2)}), as summarized in
\cite{Hanany:1996ie} :

\begin{figure}[t]
\centering
\includegraphics[scale=1.8,clip,trim=0.5cm 0cm 0cm 0cm]{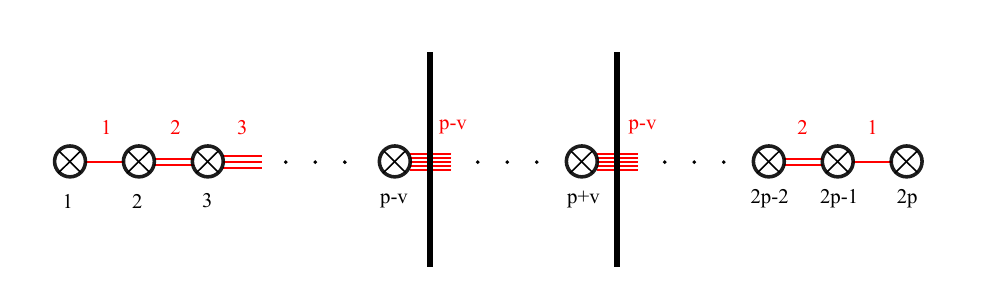}
\caption{This figure shows the brane configuration of figure \ref{fig:monobubb} in a specific Hanany-Witten frame where the D1-branes localized
at the origin begin and end only on NS5 branes. The number in red is the number of D1-branes in an interval between two NS5-branes.}
\label{fig:magdual}
\end{figure}

\ben
\item D1-D1 open strings beginning/ending on $k_i$ D1-branes between the $i$-th and the $(i+1)$-th NS5 branes give a $U(k_i)$ vector multiplet.
\item D1-D1 open strings connecting D1-branes in adjacent intervals give bifundamental hypers.
\item D1-D3 open strings in the interval between the $i$-th and the $(i+1)$-th NS5 branes give $w_i$ hypers in the fundamental
representation of $U(k_i)$, where $w_i$ is the number of D3 branes in the interval.
\een

\noindent As a consistency check, one can see that the quiver agrees with figure \ref{fig:vn0quiver} in section \ref{N=2*}, with $p \to 2p, v \to 2v$.

Note that this construction of line defects in the brane description is different from that studied in \cite{Moore:2014gua}. There the authors introduced singular monopoles to the world volume theory of a stack of D3-branes by taking the limit of a D3-brane with finite D1-branes (smooth monopoles) attached to infinity, thus creating semi-infinite D1-branes (singular monopoles). It is not obvious to us if we can derive the description of singular monopoles and monopole bubbling in \cite{Moore:2014gua} from the picture here by a chain of U-dualities. This will be discussed in more detail in a future paper.


\begin{center}
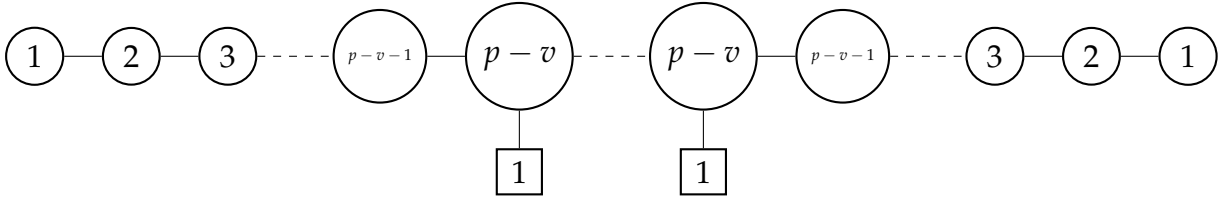
\begin{figure}
\begin{tikzpicture}[
cnode/.style={circle,draw,thick,minimum size=4mm},snode/.style={rectangle,draw,thick,minimum size=6mm}]
\node[cnode] (1) {1};
\node[cnode] (2) [right=.5cm  of 1]{2};
\node[cnode] (3) [right=.5cm of 2]{3};
\node[cnode] (5) [right=1cm of 3]{\tiny{$p-v-1$}};
\node[cnode] (6) [right=0.5cm of 5]{$p-v$};
\node[cnode] (9) [right=1cm of 6]{$p-v$};
\node[cnode] (10) [right=0.5cm of 9]{\tiny{$p-v-1$}};
\node[cnode] (13) [right=1cm of 10]{{$3$}};
\node[cnode] (14) [right=0.5cm of 13]{$2$};
\node[cnode] (17) [right=0.5cm of 14]{1};
\node[snode] (18) [below=0.5cm of 6]{1};
\node[snode] (19) [below=0.5cm of 9]{1};
\draw[-] (1) -- (2);
\draw[-] (2)-- (3);
\draw[dashed] (3) -- (5);
\draw[-] (5) --(6);
\draw[dashed] (6) -- (9);
\draw[-] (9) -- (10);
\draw[dashed] (10) -- (13);
\draw[-] (13) -- (14);
\draw[-] (14) -- (17);
\draw[-] (6) -- (18);
\draw[-] (9) -- (19);
\end{tikzpicture}
\caption{Higgs branch quiver for $\cM(B, \mathbf{v})$ in a $SU(2)$ theory for $B=(-p, p)$ and $\mathbf{v}=(-v, v)$ as deduced from the
D3-D1-NS5-brane system. The quiver is the same as the one
given in figure \ref{fig:vn0quiver} with $p \to 2p, v \to 2v$.}
\label{quiverU(2)}
\end{figure}
\end{center}

\subsection{$SU(N)$ defect SQM for $N>2$}\label{HW-U(N)}
In this subsection, we extend the construction above to $SU(N)$ singular monopoles for $N >2$
and discuss a prescription to determine 
from the defect data in a given bubbling sector.
The defect data associated with a given bubbling sector in the path integral is specified by the $SU(N)$ cocharacters:
\begin{equation}
B={\rm diag}(p_1, p_2,....,p_N)\quad, \qquad \mathbf{v}={\rm diag} (v_1,v_2,..., v_N)~,
\end{equation}
where the diagonal entries are integers arranged in a non-decreasing order.

\begin{figure}[t]
\begin{center}
\includegraphics[scale=2.2,trim=0.5cm 1cm 0cm 0cm,clip]{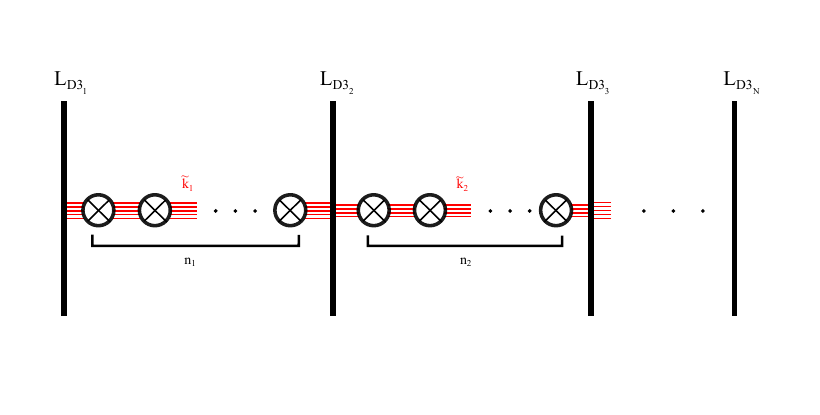}
\caption{D1-D3-NS5 configuration for monopole bubbling in a $SU(N)$ theory. The number of NS5-branes in the $i$-th chamber is $n_i$.
The number of D1-branes in the $i$-th chamber (beginning/ending on D3-branes) is the same and denoted by $\widetilde{k}_i$,
where $\widetilde{k}_i$s are defined in \eref{Brane2defect-map}.}
\label{U(N)brane}
\end{center}
\end{figure}

The Type IIB description for this configuration consists of D1-D3-NS5-branes
such that D1-branes end on $N$ parallel D3-branes. We can then introduce a singular monopole  by adding a certain number of NS5-branes
in each chamber defined by consecutive D3-branes whose positions in $\R^3$ coincide at the origin: $x^{1,2,3}=0$.
The generic Type IIB configuration is shown in figure \ref{U(N)brane}, where we only show the D1-branes localized
at the origin\footnote{There can also be D1-branes away from $x^{1,2,3}=0$ in each interval. They are related to smooth
monopoles in the presence of the 't Hooft defect.}.

Let $n_i$ be the number of NS5 branes in the $i$-th chamber (i.e. the chamber between the $i$-th and the $i+1$-th D3-brane with $i=1,\ldots, N-1$)
and let $L_{D3_i}$ be the linking number of the $i$-th D3 brane ($i$ increasing left to right).
Also, let $\widetilde{k}_i$ be the number of D1-branes localized
at $x^{1,2,3}=0$ in the $i$-th chamber\footnote{Only D1-branes localized at $x^{1,2,3}=0$ are relevant for the quiver data.
There could be other freely moving D1-branes, as in figure \ref{fig:monobubb}, but their presence (or absence) will not affect our discussion.}.

The data of the integers $(n_i, L_{D3_i})$ for all $i$ suffices to determine the entire Type IIB brane configuration. In order to see this note that for the $i$-th D3 brane we have
\begin{align}
L_{D3_i} = \sum^{i-1}_{j=0} n_j +  \widetilde{k}_i  -\widetilde{k}_{i-1}\quad, \quad i=1,...,N~,
\end{align}
where $\widetilde{k}_i$ ($\widetilde{k}_{i-1}$) is the number of D1-branes ending on the right (left) of the $i$-th D3-brane and $\widetilde{k}_0=0$, $\widetilde{k}_N=0$, and $n_0=0$.
 Therefore, one can readily compute $\{\widetilde{k}_i\}$ from the data $(n_i, L_{D3_i})$, thereby completely specifying the Type IIB configuration.

 The above data also fixes the
NS5-brane linking numbers:
\begin{align}
L_{NS5_\al}=n_{\rm left}(D3) + n_{\rm right}(D1) - n_{\rm left}(D1)= i ~, \quad {\rm where} \,\,\sum^{i-1}_{j=0}n_{j} + 1 \leq \al \leq \sum^{i}_{j=0} n_j~,
\end{align}
where $i=1,\ldots, N-1$, and $\al=1,\ldots, \sum^{N-1}_{j=1} n_j$ labels the NS5-branes. This condition on $\al$ implies that it is located in the
$i$-th chamber.\\

In analogy to the case of an $SU(2)$ defect, the map between the Type IIB data and the defect data $(B,\mathbf{v})$ is given as
\be \label{Brane2defect-map}
B=\sum^{N-1}_{i=1} n_i h^i\quad,\qquad B-\mathbf{v} =\sum^{N-1}_{i=1} \widetilde{k}_i H_i ~,
\ee
where the $H_i$ are simple coroots\footnote{In our convention, $H_i=-e_{i,i} + e_{i+1,i+1}$, where $e_{i,j}$ is an $N \times N$ matrix
with the $(ij)$-th entry equal to 1, and all other entries zero. } and
the $h^i$ are magnetic weights satisfying $(h^i,H_j)=\delta^i_{~j}$.
These translate to the following relations between the Type IIB data ${(n_i, L_{D3_i})}$ and the defect data $(B,\mathbf{v})$:
\begin{align}
& n_i = p_{i+1} -p_i, (i=1,..,N-1)\quad,\qquad
 L_{D3_i} = v_i - p_1, (i=1,..,N)~.
\end{align}

Note that the above map is invariant under an overall shift of $\vec{p}$ and $\vec{v}$, which implies that
the Type IIB description is invariant under transformations of the defect data of the form \eref{shift-defectdata}. Thus, from the defect data $(B,\mathbf{v})$, we can construct the brane configuration described above (figure \ref{U(N)brane}).

As before, the quiver can be read off from this configuration after a series of standard Hanany-Witten moves, such that the
D1-branes, associated with monopole bubbling, end only on NS5-branes.
In this Hanany-Witten frame, let $\widetilde{n}_\al$ be the number of D3-branes between
the $\al$-th and the $(\al +1)$-th NS5-brane, and $k_\al$ and $k_{\al -1}$ be the number of D1-branes ending on the right and
left of the $\al$-th NS5-brane respectively\footnote{The integers $\{k_\al\}$ should be identified with the non-zero entries of the KN vector $\vec k$ in section \ref{N=2*}.}. Then, using the definition of linking number of an NS5-brane, we have
\begin{align}
& k_{\al +1} + k_{\al -1} - 2k_{\al} + \widetilde{n}_\al = L_{NS5_{\al +1}} - L_{NS5_\al}~, \nonumber \\
\implies & k_{\al +1} + k_{\al -1} - 2k_{\al} + \sum^{N}_{i=1} \delta_{\al, \, L_{D3_i}} = L_{NS5_{\al +1}} - L_{NS5_\al}~,
\end{align}
where $k_0=0$, and $k_{\sum^{N-1}_{j=1} n_j}=0$, and we have used the fact that $\widetilde{n}_\al = \sum^{N}_{i=1} \delta_{\al, \, L_{D3_i}}$.
This equation allows one to compute the ranks of the gauge and flavor symmetry groups of the Higgs branch quiver from the linking numbers of NS5 and D3-branes.

Note that the condition for the $\al$-th gauge node in the quiver to be balanced (i.e. to have zero $\beta$-function) is that the LHS of the above equation
has to vanish. This always happens if the $\al$-th and the $(\al +1)$-th NS5-brane are in the same D3 chamber in the original Hanany-Witten frame (see
figure \ref{U(N)brane}),
i.e. $L_{NS5_{\al +1}} =L_{NS5_\al}$. However, if there is a D3-brane between the  $\al$-th and the $(\al +1)$-th NS5-brane, the NS5 linking number
has an aditional contribution so that there is a single unbalanced node.

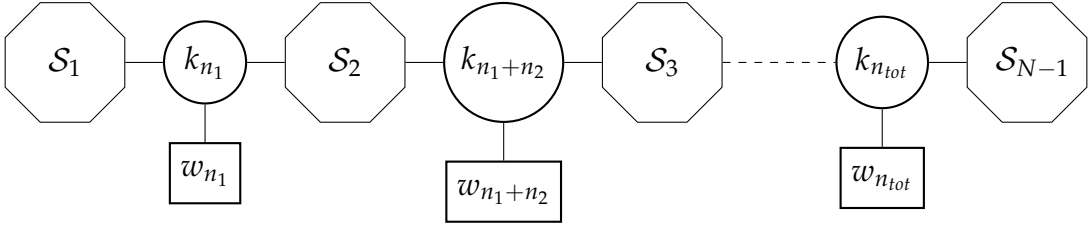
\begin{figure}
\begin{center}
\begin{tikzpicture}[
cnode/.style={circle,draw,thick,minimum size=5mm},snode/.style={rectangle,draw,thick,minimum size=8mm}]
\tikzstyle{oct} = [regular polygon,regular polygon sides=8, draw,
    text width=2em, text centered]
\tikzstyle{line} = [draw, -latex']
\node [oct] (1) {$\mathcal{S}_1$};
\node [cnode](2) [right=0.5cm of  1] {$k_{n_1}$};
\node (3) [oct,right=0.5cm of 2] {$\mathcal{S}_2$};
\node [cnode](4) [right=0.5cm of 3] {$k_{n_1 + n_2}$};
\node (5) [oct,right=0.5cm of 4] {$\mathcal{S}_3$};
\node [cnode](6) [right=1.5cm of 5] {$k_{n_{tot}}$};
\node (7) [oct,right=0.5cm of  6] {$\mathcal{S}_{N-1}$};
\node[snode] (8) [below=0.5cm of 2]{$w_{n_1}$};
\node[snode] (9) [below=0.5cm of 4]{$w_{n_1 + n_2}$};
\node[snode] (10) [below=0.5cm of 6]{$w_{n_{tot}}$};
\draw[-] (1) --  (2);
\draw[-] (2) --  (3);
\draw[-] (3) --  (4);
\draw[-] (4) --  (5);
\draw[dashed] (5) --  (6);
\draw[-] (6) --  (7);
\draw[-] (2) --  (8);
\draw[-] (4) --  (9);
\draw[-] (6) --  (10);
\end{tikzpicture}
\caption{General form for the Higgs branch quiver associated with $\cM(B, \mathbf{v})$ in a $SU(N)$ theory. Each octagon $\mathcal{S}_i$ denotes
a superconformal sub-quiver with precisely $n_i -1$ balanced nodes. The circular nodes denote the unbalanced gauge nodes, the total number of
such nodes being $N-2$. Note that $n_{tot} = \sum^{N-1}_{i=1} n_i =p_N - p_1$. The precise form of $\cS_i$ and expression for $w_{n_i}$ are given below.}
\label{U(N)quiver}
\end{center}
\end{figure}

This makes the general structure of the quiver manifest. It consists of $N-1$ superconformal sub-quivers $\mathcal{S}_i$ ($i=1,\ldots, N-1$) of length $n_i$ where all gauge nodes are balanced which
are connected by a single unbalanced gauge node, as shown in figure \ref{U(N)quiver}.
For $SU(2)$ monopoles, the quiver just consists of a single superconformal sub-quiver, as we found earlier, while for $SU(N)$ monopoles,
one generically ends up with a quiver containing exactly $N-2$ unbalanced nodes.\\

We now derive the detailed form of the superconformal sub-quivers from the Type IIB data, by performing a sequence of Hanany-Witten moves
on the configuration of figure \ref{U(N)brane}, to obtain a brane configuration from which the ADHM quiver can be read off. We will refer to the brane
configuration in figure \ref{U(N)brane}, where D1-branes end on D3-branes, as the ``electric" Hanany-Witten frame $(e)$.
In an intermediate brane configuration $(c)$, let $\ell^{(c)}_i$ denote the number of NS5-branes in the $i^{th}$ chamber\footnote{Here we introduce $\ell^{(c)}_i$ to account for the fact that in performing Hanany-Witten moves, the number of NS5-branes in a given chamber will change.}. The linking number of the $i^{th}$ D3-brane in this configuration is given as by
\be
L_{D3_i} =L^{NS5,(c)}_{\ell,i}+L^{D1,(c)}_{k,i}~,
\ee
where $L^{NS5,(c)}_{\ell,i}= \sum_{j=1}^{i-1}\ell^{(c)}_i$ and $L^{D1,(c)}_{k,i}$ denote the contributions from the NS5- and D1-branes respectively.
Note that $\ell^{(e)}_i=n_i$ in the electric frame.\\

We now want to perform a sequence of Hanany-Witten moves -- that is move NS5-branes across adjacent D3-branes -- to go to the ``magnetic'' Hanany-Witten frame $(m)$,where all the D1-branes end only on the NS5-branes: $L^{D1,(m)}_{k,i}=0$ $\forall i$. The quiver SQM can then be read off as the D1 world volume theory in this configuration.

Since we have the condition\footnote{This condition comes from the fact that each D1-brane screens 2 coincident NS5-branes as we saw in the last section. Therefore the completely screened condition is when $2\tilde{k}_i=n_i$.}
\be
|v|\leq |B|\quad \Longrightarrow \quad n_i \geq 2 \tilde{k}_i~,
\ee
this can be achieved by a sequence of HW-moves in which NS5-branes cross at most, a single D3-brane.

Let us denote the change of a generic linking number $L$ by HW-moves across the $i^{th}$ D3-brane as $\Delta_i L$. Then in going from the electric to the magnetic frame (where $L_{k,i}^{D1,(m)}=0$), we have the relations\footnote{$\Delta_i L_{D3_i}$ denotes the change in the linking number of the $i$-th D3-brane --there is no sum over $i$.}
\be
\Delta_i L_{D3_i}=\Delta_i L^{NS5}_{\ell,i}+\Delta_i L^{D1}_{k,i}=0\quad,\qquad \Delta_i L^{D1}_{k,i}=-L_{k,i}^{D1,(e)}\quad,\qquad \Delta_i \ell_i=-\Delta_i \ell_{i-1}~.
\ee
By combining these equations, we can solve   for the change in $\ell_i$:
\be
\Delta_i \ell_{i-1}= L_{k,i}^{D1, (e)}=\tilde{k}_{i}-\tilde{k}_{i-1}\quad\Longrightarrow\qquad \Delta_i \ell_i=\tilde{k}_{i-1}-\tilde{k}_i~.
\ee
The sign of $\Delta_i \ell_i$ tells us whether NS5-branes cross the D3-brane to the left or right.\\

Adding contributions from the HW-moves  involving the $i^{th}$ and $(i+1)^{th}$ D3-branes, gives the total
\be
\Delta \ell_i=\Delta_i \ell_i+\Delta_{i-1} \ell_i= \tilde{k}_{i+1}+\tilde{k}_{i-1}-2 \tilde{k}_i~.
\ee
Since $\ell_i^{(e)} \geq 2 \tilde{k}_i$, there always exists a solution to this set of equations so that $\ell^{(m)}_i\geq 0$, $\forall i$.

Now since moving an NS5-brane through a D3-brane changes the D3-brane contribution to the linking number by $\pm 1$, the number of D1-branes ending on the left and right of such an NS5-brane must differ by 1 as well. This means that generically the quiver describing the SQM on the D1-branes is of the form:

\tikzstyle{hex} = [regular polygon,regular polygon sides=6, draw,
    text width=2em, text centered]
\tikzstyle{hexs}= [regular polygon,regular polygon sides=5, draw,
    text width=1.7em, text centered]
\tikzstyle{line} = [draw, -latex']

\begin{center}
\begin{tikzpicture}[node distance = 2.2cm, auto]
\node [hex] (1) {$\Gamma_{0,1}$};
\node (2) [hexs,right of = 1] {$\Sigma_1$};
\node (3) [hex,right of =2] {$\Gamma_{1,2}$};
\node (4) [hexs,right of= 3] {$\Sigma_2$};
\node (5) [hex,right of=4] {$\Gamma_{2,3}$};
\node (6) [right of=5,xshift=-0.5cm] {};
\node (7) [right of=6,xshift=-1cm] {};
\node (8) [hexs,right of=7,xshift=-0.8cm] {$\Sigma_{N-1}$};
\node (9) [hex,right of=8] {$\Gamma_{N-1,N}$};
\draw[-] (1) --  (2);
\draw[-] (2) --  (3);
\draw[-] (3) --  (4);
\draw[-] (4) --  (5);
\draw[-] (5) --  (6);
\draw[dashed] (6) --  (7);
\draw[-] (7) --  (8);
\draw[-] (8) --  (9);
\end{tikzpicture}
\end{center}

The sub-quiver $\Sigma_i$ is given by

\begin{center}
\begin{tikzpicture}[node distance=1.3cm,
cnode/.style={circle,draw,thick,text centered},snode/.style={rectangle,draw,thick,minimum size=10mm,text centered}]
\node [cnode] (1) {$\tilde{k}_i$};
\node (2) [cnode, right of=1] {$\tilde{k}_i$};
\node (3) [right of=2,xshift=-0.1cm] {};
\node (4) [right of=3,xshift=-0.8] {};
\node (5) [cnode, right of=4,xshift=-0.1cm] {$\tilde{k}_i$};
\node (6) [cnode, right of=5] {$\tilde{k}_i$};
\node (7) [snode, below of=1] {$\omega_{i,i-1}$};
\node (8) [snode, below of =6] {$\omega_{i,i+1}$};
\node (9) [left of=1,xshift=.1cm] {=};
\node (10) [hexs,left of=9,xshift=.1cm] {$\Sigma_i$};
\draw[-] (1) -- (2);
\draw[-] (3) -- (2);
\draw[dashed] (3) -- (4);
\draw[-] (4) -- (5);
\draw[-] (5) -- (6);
\draw[-] (6) -- (8);
\draw[-] (1) -- (7);
\end{tikzpicture}
\end{center}
where $\Sigma_i$ is of length
\be
 |\Sigma_i|=n_i + 1 -|\tilde{k}_{i+1}-\tilde{k}_i|\omega_{i,i+1}-|\tilde{k}_{i-1}-\tilde{k}_i|\omega_{i,i-1}\quad,\qquad \omega_{i,j}=\begin{cases}
0&\tilde{k}_i \leq \tilde{k}_j\\
1&\tilde{k}_i >  \tilde{k}_j
\end{cases}
\ee
while the sub-quiver $\Gamma_{i,i+1}$ is given by (with $\tilde{k}_0=0$ and $\tilde{k}_N=0$)

\begin{center}
\begin{tikzpicture}[node distance=2.8cm,
cnode/.style={circle,draw,thick,text width=3.5em,text centered},snode/.style={rectangle,draw,thick,minimum size=10mm,text centered}]
\node [cnode] (1) {$\tilde{k}_i+1$};
\node (2) [cnode,right of=1] {$\tilde{k}_i+2$};
\node (3) [right of=2,xshift=-1cm] {};
\node (4) [right of=3,xshift=-2cm] {};
\node (5) [cnode, right of=4,xshift=-1cm] {$\tilde{k}_{i+1}-2$};
\node (6) [cnode, right of=5] {$\tilde{k}_{i+1}-1$};
\node (7) [left of=1,xshift=1cm] {=};
\node (8) [hex,left of=7,xshift=1cm] {$\Gamma_{i,i+1}$};
\draw[-] (1) -- (2);
\draw[-] (2) -- (3);
\draw[dashed] (3) -- (4);
\draw[-] (4) -- (5);
\draw[-] (5) -- (6);
\end{tikzpicture}
\end{center}
when $\tilde{k}_{i}<\tilde{k}_{i+1}$ and
\begin{center}
\begin{tikzpicture}[node distance=2.8cm,
cnode/.style={circle,draw,thick,text width=3.5em,text centered},snode/.style={rectangle,draw,thick,minimum size=10mm,text centered}]
\node [cnode] (1) {$\tilde{k}_i-1$};
\node (2) [cnode,right of=1] {$\tilde{k}_i-2$};
\node (3) [right of=2,xshift=-1cm] {};
\node (4) [right of=3,xshift=-2cm] {};
\node (5) [cnode, right of=4,xshift=-1cm] {$\tilde{k}_{i+1}+2$};
\node (6) [cnode, right of=5] {$\tilde{k}_{i+1}+1$};
\node (7) [left of=1,xshift=1cm] {=};
\node (8) [hex,left of=7,xshift=1cm] {$\Gamma_{i,i+1}$};
\draw[-] (1) -- (2);
\draw[-] (2) -- (3);
\draw[dashed] (3) -- (4);
\draw[-] (4) -- (5);
\draw[-] (5) -- (6);
\end{tikzpicture}
\end{center}
when $\tilde{k}_i>\tilde{k}_{i+1}$.

In the expressions above we have a few special cases:
\begin{itemize}
\item $\tilde{k}_i=\tilde{k}_{i+1}$: there is no $\Gamma_{i,i+1}$ quiver connecting $\Sigma_i$ and $\Sigma_{i+1}$, but rather the last node of $\Sigma_i$ is identified with the first node of $\Sigma_{i+1}$. Note that in this case $|\Sigma_i + \Sigma_{i+1}|=|\Sigma_i|+|\Sigma_{i+1}|-1.$
\item $\tilde{k}_i = \tilde{k}_{i+1} \pm 1$: $\Gamma_{i,i+1}$ is omitted and $\Sigma_i$ is directly connected to $\Sigma_{i+1}$.
\item $|\Sigma_i|=1$: there is a single gauge node of magnitude $\tilde{k}_i$ with two fundamental hypermultiplets.
\end{itemize}

Here the subquivers $\Gamma_{i,i+1}$ come from NS5-branes that change chambers in going to the magnetic Hanany-Witten frame and the
subquivers $\Sigma_i$ correspond to the NS5-branes which do not. Moving NS5-branes to the left or right across the D3$_{i+1}$-brane (determined by the ordering of $\tilde{k}_i,\tilde{k}_{i+1}$)  will give rise to an increasing or decreasing $\Gamma_{i,i+1}$ respectively and additionally endows the $\Sigma_{i+1}$ or $\Sigma_i$ subquiver respectively with a fundamental hypermultiplet on the gauge node of the adjacent end. This combination of the ordering of $\tilde{k}_i,\tilde{k}_{i+1}$ and $\tilde{k}_i,\tilde{k}_{i-1}$ and their corresponding hypermultiplet nodes give rise to 4 different types of $\Sigma_i$ subquivers.


One can now write down the superconformal sub-quivers $\mathcal{S}_i$ ($i=1,\ldots, N-1$) which appear in figure \ref{U(N)quiver}:

\begin{center}
\begin{tikzpicture}[
cnode/.style={circle,draw,thick,minimum size=5mm},snode/.style={rectangle,draw,thick,minimum size=8mm}]
\tikzstyle{oct} = [regular polygon,regular polygon sides=8, draw,
    text width=2em, text centered]
\tikzstyle{line} = [draw, -latex']
\node [oct] (1) {$\mathcal{S}_i$};
\node (2) [right=0.5cm of 1] {=};
\node (3) [hex,right=0.5cm of 2] {$\Gamma_{i-1,i}$};
\node [cnode](4) [right=0.5cm of 3] {$\tilde{k}_i$};
\node (5) [cnode,right=0.5cm of 4] {$\tilde{k}_i$};
\node (6) [right=0.5cm of 5] {};
\node (7) [right=0.5cm of 6] {};
\node (8) [cnode,right=0.5cm of  7] {$\tilde{k}_i$};
\node[hex] (9) [right=0.5cm of 8]{$\Gamma_{i,i+1}$};
\node [snode] (10) [below=0.5cm of 4]{1};
\node [snode] (11) [below=0.5cm of 8]{1};
\draw[-] (3) --  (4);
\draw[-] (4) --  (5);
\draw[-] (5) --  (6);
\draw[dashed] (6) --  (7);
\draw[-] (7) --  (8);
\draw[-] (8) --  (9);
\draw[-] (4) --  (10);
\draw[-] (11) -- (8);
\end{tikzpicture}
\end{center}
for $\tilde k_i> \tilde k_{i+1}, \tilde k_{i-1}$,
\begin{center}
\begin{tikzpicture}[
cnode/.style={circle,draw,thick,minimum size=5mm},snode/.style={rectangle,draw,thick,minimum size=8mm}]
\tikzstyle{oct} = [regular polygon,regular polygon sides=8, draw,
    text width=2em, text centered]
\tikzstyle{line} = [draw, -latex']
\node [oct] (1) {$\mathcal{S}_i$};
\node (2) [right=0.5cm of 1] {=};
\node [cnode](4) [right=0.5cm of 2] {$\tilde{k}_i$};
\node (5) [cnode,right=0.5cm of 4] {$\tilde{k}_i$};
\node (6) [right=0.5cm of 5] {};
\node (7) [right=0.5cm of 6] {};
\node (8) [cnode,right=0.5cm of  7] {$\tilde{k}_i$};
\node[hex] (9) [right=0.5cm of 8]{$\Gamma_{i,i+1}$};
\node[snode] (10) [below=0.5cm of 8]{1};
\draw[-] (4) --  (5);
\draw[-] (5) --  (6);
\draw[dashed] (6) --  (7);
\draw[-] (7) --  (8);
\draw[-] (8) --  (9);
\draw[-] (8) --  (10);
\end{tikzpicture}
\end{center}
for $\tilde{k}_{i-1}\leq\tilde{k}_i<\tilde{k}_{i+1}$,

\begin{center}
\begin{tikzpicture}[
cnode/.style={circle,draw,thick,minimum size=5mm},snode/.style={rectangle,draw,thick,minimum size=8mm}]
\tikzstyle{oct} = [regular polygon,regular polygon sides=8, draw,
    text width=2em, text centered]
\tikzstyle{line} = [draw, -latex']
\node [oct] (1) {$\mathcal{S}_i$};
\node (2) [right=0.5cm of 1] {=};
\node (3) [hex,right=0.5cm of 2] {$\Gamma_{i-1,i}$};
\node [cnode](4) [right=0.5cm of 3] {$\tilde{k}_i$};
\node (5) [cnode,right=0.5cm of 4] {$\tilde{k}_i$};
\node (6) [right=0.5cm of 5] {};
\node (7) [right=0.5cm of 6] {};
\node (8) [cnode,right=0.5cm of  7] {$\tilde{k}_i$};
\node[snode] (10) [below=0.5cm of 4]{1};
\draw[-] (3) --  (4);
\draw[-] (4) --  (5);
\draw[-] (5) --  (6);
\draw[dashed] (6) --  (7);
\draw[-] (7) --  (8);
\draw[-] (4) --  (10);
\end{tikzpicture}
\end{center}
for $\tilde{k}_{i+1}\leq\tilde{k}_i<\tilde{k}_{i-1}$,
\begin{center}
\begin{tikzpicture}[
cnode/.style={circle,draw,thick,minimum size=5mm},snode/.style={rectangle,draw,thick,minimum size=8mm}]
\tikzstyle{oct} = [regular polygon,regular polygon sides=8, draw,
    text width=2em, text centered]
\tikzstyle{line} = [draw, -latex']
\node [oct] (1) {$\mathcal{S}_i$};
\node (2) [right=0.5cm of 1] {=};
\node [cnode](4) [right=0.5cm of 2] {$\tilde{k}_i$};
\node (5) [cnode,right=0.5cm of 4] {$\tilde{k}_i$};
\node (6) [right=0.5cm of 5] {};
\node (7) [right=0.5cm of 6] {};
\node (8) [cnode,right=0.5cm of  7] {$\tilde{k}_i$};
\draw[-] (4) --  (5);
\draw[-] (5) --  (6);
\draw[dashed] (6) --  (7);
\draw[-] (7) --  (8);
\end{tikzpicture}
\end{center}
for when $\tilde{k}_i\geq\tilde{k}_{i+1},\tilde{k}_{i-1}$. Here the number of repeated $\tilde{k}_i$ nodes (without any fundamental hyper) are given by $\ell_i^{(m)}-2$, $\ell_i^{(m)}-1-\tilde{k}_i+\tilde{k}_{i-1}$, $\ell_i^{(m)}-1-\tilde{k}_i+\tilde{k}_{i+1}$, and $\ell_i^{(m)}-2\tilde{k}_i+\tilde{k}_{i-1}+\tilde{k}_{i+1}$ respectively and the $k_{n_i}$ and $w_{n_i}$ are given by
\be
k_{n_i}=\begin{cases}
\tilde{k}_i&\tilde{k}_i<\tilde{k}_{i+1}\\
\tilde{k}_{i+1}&\tilde{k}_i\geq\tilde{k}_{i+1}
\end{cases}\quad,\qquad
w_{n_1+n_2+\ldots+n_i}=\begin{cases}
0 & \tilde{k}_i\neq \tilde{k}_{i+1}\\
1 & \tilde{k}_i= \tilde{k}_{i+1}
\end{cases}
\ee
and
\begin{center}
\begin{tikzpicture}[
cnode/.style={circle,draw,thick,minimum size=5mm},snode/.style={rectangle,draw,thick,minimum size=8mm}]
\tikzstyle{oct} = [regular polygon,regular polygon sides=8, draw,
    text width=2em, text centered]
\tikzstyle{line} = [draw, -latex']
\node [oct] (1) {$\mathcal{S}_i$};
\node (2) [right=0.5cm of 1] {=};
\node (3) [hex,right=0.5cm of 2] {$\Gamma_{i-1,i}$};
\node [cnode](4) [right=0.5cm of 3] {$\tilde{k}_i$};
\node[hex] (9) [right=0.5cm of 4]{$\Gamma_{i,i+1}$};
\node [snode] (10) [below=0.5cm of 4]{2};
\draw[-] (3) --  (4);
\draw[-] (4) --  (9);
\draw[-] (4) --  (10);
\end{tikzpicture}
\end{center}
in the special case of $n_i=2k_i-k_{i+1}-k_{i+1}$. \\

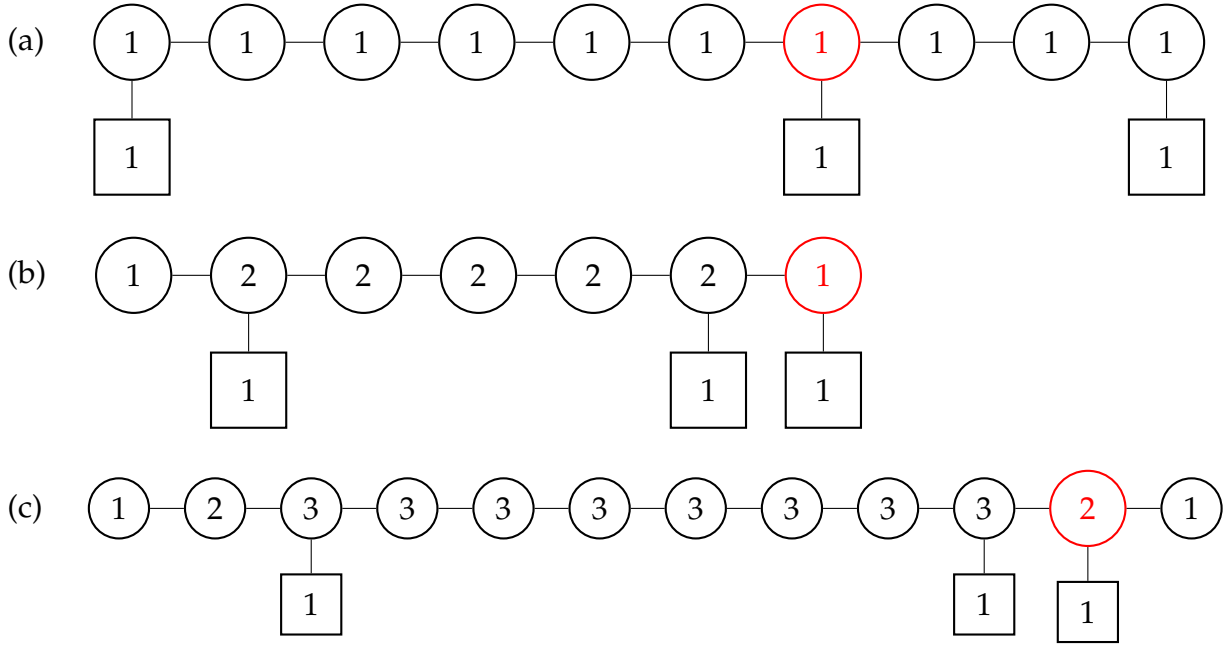
\begin{figure}[htbp]
\begin{tikzpicture}[
cnode/.style={circle,draw,thick,minimum size=10mm},snode/.style={rectangle,draw,thick,minimum size=10mm}, pnode/.style={circle,red, draw,thick,minimum size=10mm}]
\node[cnode] (1) {1};
\node (0) [left=.5cm of 1]{(a)};
\node[cnode] (2) [right=.5cm  of 1]{1};
\node[cnode] (3) [right=.5cm of 2]{1};
\node[cnode] (4) [right=.5cm of 3]{1};
\node[cnode] (5) [right=.5cm of 4]{1};
\node[cnode] (6) [right=.5cm of 5]{1};
\node[pnode] (7) [right=.5cm of 6]{1};
\node[cnode] (8) [right=.5cm of 7]{1};
\node[cnode] (9) [right=.5cm of 8]{1};
\node[cnode] (10) [right=.5cm of 9]{1};
\node[snode] (11) [below=0.5cm of 1]{1};
\node[snode] (12) [below=0.5cm of 7]{1};
\node[snode] (13) [below=0.5cm of 10]{1};
\draw[-] (1) -- (2);
\draw[-] (2)-- (3);
\draw[-] (3) -- (4);
\draw[-] (4) --(5);
\draw[-] (5) -- (6);
\draw[-] (6) -- (7);
\draw[-] (7) -- (8);
\draw[-] (8) -- (9);
\draw[-] (9) -- (10);
\draw[-] (1) -- (11);
\draw[-] (7) -- (12);
\draw[-] (10) -- (13);
\end{tikzpicture}\\
~~\\
\begin{tikzpicture}[
cnode/.style={circle,draw,thick,minimum size=10mm},snode/.style={rectangle,draw,thick,minimum size=10mm}, pnode/.style={circle,red, draw,thick,minimum size=10mm}]
\node[cnode] (1) {1};
\node (0) [left=.5cm of 1]{(b)};
\node[cnode] (2) [right=.5cm  of 1]{2};
\node[cnode] (3) [right=.5cm of 2]{2};
\node[cnode] (4) [right=.5cm of 3]{2};
\node[cnode] (5) [right=.5cm of 4]{2};
\node[cnode] (6) [right=.5cm of 5]{2};
\node[pnode] (7) [right=.5cm of 6]{1};
\node[snode] (11) [below=0.5cm of 2]{1};
\node[snode] (12) [below=0.5cm of 6]{1};
\node[snode] (13) [below=0.5cm of 7]{1};
\draw[-] (1) -- (2);
\draw[-] (2)-- (3);
\draw[-] (3) -- (4);
\draw[-] (4) --(5);
\draw[-] (5) -- (6);
\draw[-] (6) -- (7);
\draw[-] (2) -- (11);
\draw[-] (6) -- (12);
\draw[-] (7) -- (13);
\end{tikzpicture}\\
~~\\
\begin{tikzpicture}[
cnode/.style={circle,draw,thick,minimum size=8mm},snode/.style={rectangle,draw,thick,minimum size=8mm}, pnode/.style={circle,red, draw,thick,minimum size=10mm}]
\node[cnode] (1) {1};
\node (0) [left=.45 of 1]{(c)};
\node[cnode] (2) [right=.45  of 1]{2};
\node[cnode] (3) [right=.45 of 2]{3};
\node[cnode] (4) [right=.45 of 3]{3};
\node[cnode] (5) [right=.45 of 4]{3};
\node[cnode] (6) [right=.45 of 5]{3};
\node[cnode] (7) [right=.45 of 6]{3};
\node[cnode] (8) [right=.45 of 7]{3};
\node[cnode] (9) [right=.45 of 8]{3};
\node[cnode] (10) [right=.45 of 9]{3};
\node[pnode] (11) [right=0.45 of 10]{2};
\node[cnode] (12) [right=0.45 of 11]{1};
\node[snode] (13) [below=0.45 of 3]{1};
\node[snode] (14) [below=0.45 of 10]{1};
\node[snode] (15) [below=0.45 of 11]{1};
\draw[-] (1) -- (2);
\draw[-] (2)-- (3);
\draw[-] (3) -- (4);
\draw[-] (4) --(5);
\draw[-] (5) -- (6);
\draw[-] (6) -- (7);
\draw[-] (7) -- (8);
\draw[-] (8) -- (9);
\draw[-] (9) -- (10);
\draw[-] (10) -- (11);
\draw[-] (11) -- (12);
\draw[-] (3) -- (13);
\draw[-] (10) -- (14);
\draw[-] (11) -- (15);
\end{tikzpicture}
\caption{Example of quiver SQMs in an $SU(3)$ theory for (a) $B={\rm diag}(-6,1,5)$ and $\mathbf{v}={\rm diag}(-5,1,4)$, (b) $B={\rm diag}(-5, 2,3)$ and $\mathbf{v}={\rm diag}(-3,1,2)$,
and (c) $B={\rm diag}(-8,3,5)$ and $\mathbf{v}={\rm diag}(-5,2,3)$. Each quiver consists of two superconformal sub-quivers separated by a single
unbalanced gauge node, which is drawn in red.}
\label{Ex:U(N)quiver}
\end{figure}
Now we will consider a few examples of a $SU(3)$ defects. We will consider the examples: (a) $B={\rm diag}(p_1,p_2,p_3)={\rm diag}(-6,1,5)$ in the bubbling sector
labelled by $\mathbf{v}={\rm diag}(-5,1,4)$, (b)  $B={\rm diag}(-5,2,3)$ in the bubbling sector $\mathbf{v}={\rm diag}(-3,1,2)$, and (c) $B={\rm diag}(-8,3,5)$ in the bubbling sector $\mathbf{v}={\rm diag}(-5,2,3)$. The quivers associated with the corresponding $\cM(B, \mathbf{v})$ are shown
in figure \ref{Ex:U(N)quiver}.

\subsection{Relation to the character equation}
We now show that the quiver obtained from the brane description of monopole bubbling discussed in this section is indeed
the quiver that arises from the character equation \eref{char-eqn-main}. Recall that each $k_j$ in the KN vector $\vec k$, associated to the quiver SQM, contributes a term $k_j x^j$ to the trace Tr$_V~x^{K}$ (where $x=e^{2\pi \I \nu}$), up to some overall monomial which can be absorbed by a shifting $(B,\mathbf{v},K)$ (see equation \eref{shift-defectdata} and subsequent discussion). Since generically, the $k_j$ vary by at most one, multiplying Tr$_V~x^{K}$ by $(x+x^{-1}-2)$ cancels all contributions except for the terms of degree $s_i$, where $s_i$ is the eigenvalue of $K$ associated with the first or the last node of the $\Sigma_i$ subquiver (additionally one must include a term from the first and last node of the full quiver) which will lead to a contribution of terms
 \be
(x+x^{-1}-2){\rm Tr}_V~x^{K}= \sum_{i=1}^{2N} (-1)^{\sigma_i } x^{n_i}~,
 \ee
 where $\sigma_i=0,1$ $mod_2$ determines the sign of each contribution.

Note first that in the case where $\tilde{k}_I=\tilde{k}_{I+1}$, the prefactor $(x+x^{-1}-2)$ will cancel all contributions from the last and first nodes of the $\Sigma_I$ and $\Sigma_{I+1}$ subquiver respectively. However, $\tilde{k}_I=\tilde{k}_{I+1}$ implies there is a zero in the matrix
\be
\kappa= \mathbf{v}-B=\sum_I \tilde{k}_I H_I={\rm diag}(\kappa_1,...,\kappa_N)~,
\ee
and hence $p_{I+1}=v_{I+1}$. Therefore, these terms will themselves cancel and thus should not appear in the term $(x+x^{-1}-2)$Tr$_V~x^{K}$. Therefore, without loss of generality, we will consider the generic case $\tilde{k}_I\neq \tilde{k}_J$.

By careful analysis (see Appendix \ref{app:AppG}) of the boundary cases where $\Sigma_I,\Sigma_{I+1}$ joins to $\Gamma_{I,I+1}$, one can show that the contribution to the character equation will be of the form
\begin{align}\label{eq:carefulanalysis}
\begin{split}
&(x+x^{-1}-2){\rm Tr}_V~x^{K}=\\
&=1-x^{\tilde{k}_1}-x^{(p_N-p_1 -\tilde{k}_{N-1})}+x^{(p_N-p_1)}+\sum_{I=1}^{N-2}\left( x^{(p_{I+1}-p_1)}-x^{ (p_{I+1}-p_1+\tilde{k}_{I+1}-\tilde{k}_I)}\right)~.
\end{split}\end{align}
Now we can fix the overall factor (that is by shifting $K$ so that it is traceless) by multiplying by a factor of $x^{p_1}$, we find the contribution to be
\begin{align}\begin{split}
&(x+x^{-1}-2){\rm Tr}_V~x^{K}=\\
&=x^{p_1}-x^{(p_1+\tilde{k}_1)}-x^{(p_N-\tilde{k}_{N-1})}+x^{p_N}+\sum_{I=1}^{N-2}\left( x^{p_{I+1}}-x^{(p_{I+1}+\tilde{k}_{I+1}-\tilde{k}_I)}\right)~.
\end{split}\end{align}
Now by using the identity
\be
\tilde{k}_I=(h^I,\kappa)=\half\left( \sum_{J=1}^I \kappa_J-\sum_{J=I+1}^N \kappa_J\right)\quad\Longrightarrow\quad \tilde{k}_{I+1}-\tilde{k}_I=\kappa_{I+1}=v_{I+1}-p_{I+1}~,
\ee
we can see that $p_{I+1}+\tilde{k}_{I+1}-\tilde{k}_I=v_{I+1}.$ Therefore, we see that the contribution to the character equation determined by the brane configuration can in fact be reduced
\begin{align}\begin{split}
(x+x^{-1}-2){\rm Tr}_V~x^{ K}&=x^{p_1}-x^{ v_1}-x^{v_{N}}+x^{p_N}+\sum_{I=1}^{N-2}\left( x^{p_{I+1}}-x^{v_{I+1}}\right)~,\\
&=\sum_{I=1}^N\left(x^{p_I}- x^{v_I}\right)={\rm Tr}_{N} x^{B}-{\rm Tr}_N x^{ \mathbf{v}}~,
\end{split}\end{align}
and thus solves the character equation.

\section*{Acknowledgements}
We thank M. Bullimore, S. Cherkis, D. Gaiotto, Hee-Cheol Kim, H. Nakajima, A. Neitzke, T. Okuda, W. Peelaers, and
A. Royston for discussion and
correspondence on related issues. A.D. thanks the organizers of ``Pollica Summer Workshop
on Dualities in Superconformal Field Theories" and the organizers of the ``Simons
Summer Workshop", where part of this work was done. G.M. thanks  the Aspen Center for Physics (partially supported by NSF Grant No. PHY1066293)
for hospitality while this work was being done. The authors acknowledge support by the DOE under grant DOE-SC0010008 to Rutgers University.
\appendix
\section{$5d,\,\N=1^*$ Instanton Partition function and Witten index of ADHM QM} \label{WittenIndex}

Consider a 4d $\N=2^*$ Lagrangian theory with gauge group $G$ (and maximal torus $T_G$) on $\C^2$. With the same data, one can define a 5d $\N=1^*$ SYM  on $\C^2 \times S^1_\beta$ with the same gauge group and matter content. One can now define a supersymmetric index in 5d w.r.t to the supercharges $Q \equiv \overline{Q}^1_{\dot{1}}=-\overline{Q}^{1\dot{2}},\, Q^\dagger \equiv \overline{Q}^{2}_{\dot{2}}=\overline{Q}^{2\dot{1}}$ (where $Q^A_{{\alpha}}, \overline{Q}^A_{\dot{\alpha}}$ are the supercharges of the 4d/5d theory):
\begin{equation}\label{5dIndex-A}
Z_{5d}(\epsilon_{1}, \epsilon_{2}, a_i, m)= \Tr_{\cH_{QFT}(\C^2)} (-1)^F e^{-\beta\{Q,Q^\dagger\}}e^{-(\epsilon_1 (J_1+J_R)+\epsilon_2 (J_2+J_R)+2m\,J_f + \sum_{i} a_i O_i)},
\end{equation}
where the trace is over the Hilbert space $\cH_{QFT}(\C^2)$ getting contributions only from states which are invariant under $Q$-supersymmetry. Additionally, $J_1,J_2$ are the Cartans of the spatial $SO(4)$ rotating two orthogonal $\C\cong\R^2$s which we denote as $SO(4)_1$. Writing $so(4)_1 \cong su(2)_l \oplus su(2)_r$, the Cartan generators of $SU(2)_l$ and $SU(2)_r$ are given in terms of $J_1,J_2$ as: $J_l=\frac{J_1- J_2}{2}, J_r=\frac{J_1+J_2}{2}$. Another $SO(4)$ symmetry arises as the subgroup of the $SO(5)$ R-symmetry which is unbroken by a single non-zero scalar vev (see below), which we denote as $SO(4)_2$. Writing $so(4)_2 \cong su(2)_R \oplus su(2)_f$, we denote the Cartan generators of $SU(2)_R$ and $SU(2)_f$ as $J_R$ and $J_f$ respectively. $\{O_i\}$ denotes the Cartan generators
of the gauge group.\\

 Geometrically, the twists introduced by $J_1,J_2$ in the definition of the index above can be realized by replacing the flat 5d spacetime by a $\C^2$ bundle over $S^1$, i.e. $\C^2 \times \R $ coordinatized by $(z_1,z_2, \tau) \in \C^2 \times S^1$, with the following identification (Melvin identification):
\begin{equation}
(z_1,z_2, \tau) \sim (e^{\epsilon_1}z_1,e^{\epsilon_2}z_2, \tau+\beta)~,
\end{equation}
so that we can take $0\leq \tau < \beta$.
The metric on the fiber bundle is chosen such that the monodromy along $S^1$ is an element $(g, \, r) \in SO(4)_1 \times SU(2)_R$. Explicitly, parametrizing $\R^4$ as a circle fibration over $\R^3$ and defining $\epsilon_{\pm} =\frac{\epsilon_1\pm \epsilon_2}{2}$, the 5d metric is
\begin{equation}
\de s^2_5(\Omega) =\frac{1}{4r}\Big(\de r^2 + r^2 \de \theta^2 + r^2 \sin^2{\theta} (\de\phi +  \widetilde V^\phi \de\tau)^2\Big) + \frac{r}{4} (\de \psi+ \omega + \widetilde V^\psi \de \tau)^2 + \de \tau^2~,
\end{equation}
where the vector field $\widetilde V$ is given as follows:
\begin{subequations}\label{V-1}
\begin{empheq}{align}
& \widetilde V^{r} = \widetilde V^{\theta} =0\quad,\qquad \widetilde V^\phi =-\frac{2\I\epsilon_+}{\beta}\quad,\qquad  \widetilde V^\psi = -\frac{2\I\epsilon_-}{\beta}~.
\end{empheq}
\end{subequations}
The resulting space-time is called an $\Omega$--background. Note that the $\Omega$-deformed action for the $4d, \N=2^*$ theory can be obtained by using this metric to write the 5d theory on the bundle and then dimensionally reducing along the circle (which amounts to setting the Lie derivatives of all fields along the circle to zero).\\

The index can be written as a  path integral with the following boundary condition at the infinity of $\R^4$ :
\begin{equation}
\begin{split}
 F^{(4)} \longrightarrow 0\quad, \qquad (A_\tau + \I Y) \longrightarrow \,a~, \quad \,a \in \Big(\mathfrak{t}_G \otimes {\C}\Big)/\Lambda_{\rm cr}~.
\end{split}
\end{equation}
The standard 5d $\N=1^*$ SYM action has to be deformed to accommodate the various twists in the index.
For generic values of the parameters $\epsilon_1, \epsilon_2$, and appropriate background fields turned on,
the $\Omega$-deformed theory preserves a supercharge $Q$, which squares to a $U(1)^2_{\epsilon_1, \epsilon_2} \times T_G \times U(1)_m$-
transformation on the fields.
The Q-fixed locus of the path integral consists of a set of isolated fixed points
on the moduli space of $G$-instantons on $\R^4$ under the combined $U(1)^2_{\epsilon_1, \epsilon_2} \times T_G \times U(1)_m$ action \cite{Nekrasov:2002qd, Nekrasov:2003rj}.
For $G=SU(N)$, these fixed points are labelled by $N$-tuples of Young diagrams consisting of $k$ boxes, where $k$ is the instanton number.\\

The path integral can then be evaluated from the one-loop determinant arising from fluctuations of fields around these fixed points. The universal part
of the determinant is denoted as $Z_{\rm 1-loop}$, while the part dependent on the fixed points is denoted as $Z_{\rm inst}$. The localized 5d index can
therefore be written as \footnote{The function $(x; y,z)_\infty$ is defined as $(x; y,z)_\infty=\prod^\infty_{i,j=0}(1-x\, y^i \, z^j)$.}
\begin{align}\begin{split}
& Z_{5d} =Z_{\rm 1-loop}.\, Z_{\rm inst}~,\\
& Z_{\rm 1-loop} = Z^{\rm vec}_{\rm 1-loop}.\, Z^{\rm adj. hyper}_{\rm 1-loop} \nonumber \\
&=\Big((uv;u,v)^{{\rm rank}(G)}_\infty \prod_{\alpha \in {\rm roots}} (uv\, e^{\alpha(a)};u,v)_\infty \Big) \times \Big(\prod_{\al \in {\rm roots}} (\sqrt{uv}\, e^{\al(a) +m};u,v)^{-1}_\infty \Big)~,\\
&Z_{\rm inst} = \sum^{\infty}_{k=0} \fq^k Z_{k}^{\rm inst},\quad (u = e^{-\epsilon_1}, \, v= e^{-\epsilon_2}, \, \fq=e^{-\frac{8\pi^2 \beta}{g^2_{5d}}})~.
\end{split}\end{align}
where $\{\mu_i\}$ are chemical potentials associated to the global symmetry of the theory. \\

Now consider the instanton part of the 5d path integral. A saddle point of the path integral at a given $\tau$-slice corresponds to a 4d instanton localized at the origin. These saddle points can therefore be visualized as $k$-instantons whose parameters slowly vary with $\tau$. This implies that one can approximate the path integral with that of a quantum mechanical particle moving in the moduli space of instantons -- this is called the moduli space approximation, and it becomes exact in computing certain quantities in theories with supersymmetry. Using the moduli space approximation, the instanton part of the 5d index can be written in terms of the Witten index of a $(4,4)$ supersymmetric quantum mechanics (SQM):
\begin{equation}
Z^{\rm inst}_k(\epsilon_{1,2}, a_i, m)= \Tr_{\cH^{SQM}_{k}} (-1)^F e^{-\beta\{Q,Q^\dagger\}} e^{-(\epsilon_1 (J_1+J_R)+\epsilon_2 (J_2 +J_R)+2 m\,J_f + \sum_{i} a_i O_i)}~,
\end{equation}
where $\cH^{SQM}_{k}$ is the Hilbert space of the supersymmetric quantum mechanics on $k$-instanton moduli space. The bosonic part of $\cH^{SQM}_{k}$ has  complex dimension $2h^{\vee}(G) k$ (where $h^{\vee}(G)$ is the dual Coxeter number) which is the dimension of $k$-instanton moduli space $\cM^k_{inst}$. The fermionic part also has complex dimension $2h^{\vee}(G) $
-- this is the dimension of a fiber of the vector bundle $\V(R_{\rm adj})$ on $\cM^k_{inst}$ associated with fermionic zero modes from the
adjoint hypermultiplet.

The natural action of $U(1)_{\epsilon_1} \times U(1)_{\epsilon_2} $ on $\C^2$ induces an action on $\cM^k_{inst}$. Similarly, there are
natural actions of $T_G$ and $U(1)_m$.  Therefore, the Witten index is given by a $U(1)^2 \times T_G \times U(1)_m$ equivariant integral over  $\cM^k_{inst}$ with an appropriate characteristic class on the manifold as integrand (such integrals were first considered in \cite{Moore:1997dj} and then shown to be related to the instanton partition function in \cite{Nekrasov:2002qd}).  If $\cM^k_{inst}$ were a smooth compact space with isolated $U(1)^2 \times T_G \times U(1)_m$ fixed points, the integral would be well-defined and then one could use a generalization of the Atiyah-Bott localization formula to write the integral formally as a sum over fixed points.  However, $\cM^k_{inst}$ is noncompact and has singularities due to small instantons, and therefore one has to be careful in defining such equivariant integrals.
A standard alternative is to replace $\cM^k_{inst}$ by the smooth space $\cM^k_{ADHM}$ via the ADHM construction with a non-zero real stability/FI parameter and regularize the infinite volume with a moment map \cite{Moore:1997dj}.
The ADHM construction has a clear interpretation in the string theory embedding, where the SQM is realized as a world volume gauge theory on a stack
of D0-branes probing a stack of D4-branes which engineers the 5d gauge theory. The group action as well as the characteristic classes can be extended to $\cM^k_{ADHM}$ and the equivariant integral is well-defined. In the case of the $\N=2^*$ theory studied in this paper the resulting Witten index is
independent of the FI parameter. For more general hypermultiplet representations this will not be the case.

We discuss some basic properties of the (4,4) Witten index in appendix \ref{App-WI}. We review the related
equivariant integral in appendix \ref{equiv structure}.

\section{Basic Properties of the Witten Index}\label{App-WI}

In this section, we will focus on ADHM SQMs associated with instantons in an $\N=1^*$ $SU(N)$ theory
on $S^1 \times \C^2/\Z_n$. Consider a (4,4) SQM living on a circle of radius $\beta$ with a gauge group $G_{\rm gauge}$ and a flavor symmetry group
$G_{\rm flavor}$.  These are quiver gauge theories with $G_{\rm gauge} = \prod_{i=1}^n U(k_i)$, where $\sum_i k_i=k$,
with fundamental and bifundamental matter. For $n=1$, we have a single $U(k)$ gauge group with a single adjoint hypermultiplet
and fundamental matter.\\

The full global symmetry of the theory, including the R-symmetry, is
$$G_{\rm global} = SU(2)_l \times SU(2)_r \times SU(2)_R \times SU(2)_f \times G_{\rm flavor}~,$$
where $SU(2)_l \times SU(2)_r \times SU(2)_R$ is the R-symmetry associated with (4,4) supersymmetry.
Let $J_l, J_r$,$J_R$, $J_f$ be the Cartan generators of $SU(2)_l$, $SU(2)_r$,$SU(2)_R$ and $SU(2)_f$ respectively, while the flavor symmetry generators are collectively labelled as $\{O_j\}$.  The Witten index of the theory is then formally written as
\begin{align}\label{WI-def}
Z_{SQM}(\epsilon_1, \epsilon_2, a_i, m)= \Tr_{\cH^{SQM}} (-1)^F e^{-\beta\{Q,Q^\dagger\}} e^{-(\epsilon_1 (J_1+J_R)+\epsilon_2 (J_2 +J_R)+2 m\,J_f + \sum_{i} a_i O_i)}~,
\end{align}
where the generators $J_1, J_2$ are related to $J_l, J_r$ as $J_{l,r} = \frac{J_1\mp J_2}{2}$.\\

\subsection{(4,4) multiplets in terms of (0,2) multiplets}\label{multiplet}
Let us first list the (4,4) multiplets and their global symmetries, which can be effectively read off from a Type IIA description.
Recall that the ADHM SQMs are realized as D0 world volume theories in a D0-D4-brane system
where the D4-branes wrap the orbifold $\C^2/\Z_n$. The massless modes of the open string spectrum in the D0-D4-brane system
can be assembled in $(4,4)$ multiplets on the D0-brane as follows:
\begin{align}\begin{split}
  \textrm{D0-D0}:& {\rm vector~ multiplet}\hspace{29mm}(A_t,\varphi,
  {\varphi_{Aa}}),
  (\bar\lambda^A_{\dot\alpha},
  {\bar\lambda^a_{\dot\alpha}})\\
  & {\rm adjoint/bifundamental ~hyper} \quad(a_{\alpha\dot\beta}),(\lambda^A_\alpha,
 {\lambda^a_\alpha})\\
  \textrm{D0-D4}:& \textrm{fundamental ~hyper}\hspace{23mm}
  (q_{\dot\alpha}),(\psi^A,{\psi^a}).
\end{split}\end{align}
where the indices correspond to the different global symmetries : $\alpha, \dot\alpha, A, a \in \{1,2\}$ label the indices of $SU(2)_l, SU(2)_r$, $SU(2)_R$ and $SU(2)_f$ respectively. Note that we have suppressed all gauge indices for the fields listed above.\\

Here $\varphi$ is a ``real'' scalar in the sense that it is valued in the Lie algebra of the
compact gauge group, while $\varphi_{A a}$ and $a_{\alpha\dot\beta}$ are complex
scalars satisfying a natural reality constraint, namely, they define quaternions. \\

The localization formula for the Witten index is given in terms of (0,2) supermultiplets (see below). Therefore, we need to write the various (4,4) supermultiplets in our theory in terms of (0,2) supermultiplets. To do this, it is convenient to first split up (4,4) multiplets into (0,4) multiplets, and then split them further into (0,2) constituents. This is summarized in Table \ref{Tab:multiplets}. We refer the reader to \cite{Hwang:2014uwa} for more details.\\
\begin{center}
\begin{table}[htbp]
\begin{tabular}{|m{80pt}|m{150pt}|m{150pt}|m{100pt}|}
\hline
(4,4) multiplets & (0,4) constituents  &  (0,2) constituents & $(r_+,r_-, f)$  \\
\hline \hline
Vector & \begin{tabular}{@{}c@{}} Vector ($A_t,\varphi, \bar\lambda^A_{\dot\alpha}$) \\  Twisted adj. hyper (${\varphi_{Aa}}, \bar\lambda^a_{\dot\alpha}$) \end{tabular} &  \begin{tabular}{@{}c@{}} Vector + Adj. Fermi \\  (Adj. +$\overline{\rm Adj.}$) Chiral\end{tabular} &  \begin{tabular}{@{}c@{}} (0,0,0) + (1,0,0) \\
(\half, 0, \half) + (-\half, 0, \half) \end{tabular}\\
\hline
Adj./Bif.Hyper& \begin{tabular}{@{}c@{}} Adj./Bif Hyper ($a_{\alpha\dot\beta},\lambda^A_\alpha$) \\  Adj./Bif.  Fermi (${\lambda^a_\alpha}$)\end{tabular} & \begin{tabular}{@{}c@{}} (Adj./Bif.+ $\overline{\rm Adj./Bif.}$) Chiral \\ (Adj./Bif. + $\overline{\rm Adj./Bif.}$) Fermi \end{tabular} & \begin{tabular}{@{}c@{}} (\half,\half,0) + (\half, -\half, 0) \\
(0,\half,\half) + (0,-\half,\half) \end{tabular}\\
\hline
Fund. Hyper & \begin{tabular}{@{}c@{}} Fund. Hyper ($q_{\dot\alpha}, \psi^A$)\\ Fund. Fermi ($\psi^a$)\end{tabular} & \begin{tabular}{@{}c@{}}  (Fund. + $\overline{\rm Fund.}$) Chiral \\ (Fund. + $\overline{\rm Fund.}$) Fermi \end{tabular}  & \begin{tabular}{@{}c@{}} (\half,0,0) + (\half,0,0) \\
(0, 0, \half) + (0, 0, \half) \end{tabular}\\
\hline
\end{tabular}
\caption{(4,4) mutiplets in terms of (0,4) and (0,2) multiplets. The last column lists the charges $(r_+,r_-, f)$ for the various chiral and fermi multiplets
that constitute the (4,4) multiplets. Note that the $m$-dependent terms in \eref{D0quiver-basic} and \eref{D0quiver-orbifold} arise from the fields charged under the $SU(2)_f$.}
\label{Tab:multiplets}
\end{table}
\end{center}

\subsection{Localization Formula} \label{Loc-app}
For the index to be computable using standard localization techniques, the space of supersymmetric vacua should not have any flat directions. The global symmetry twists in the definition of the Witten index ensure that the flat directions coming from various hypermultiplet scalars are lifted. However, one of the adjoint scalars $\varphi$, which lives in a $(0,2)$ vector multiplet inside the $(4,4)$ vector multiplet, is neutral under these symmetries and therefore flat directions associated with it cannot be lifted by the above twists. For unitary gauge groups, one can turn on FI parameters which lift the flat directions for $\varphi$.
In this paper, we will only consider SQMs which arise as ADHM QM of instantons associated with 5d $\N=1^*$ $SU(N)$ gauge theories. The gauge groups $G_{\rm gauge}$ for these ADHM QM are products of unitary factors so that one can always turn on appropriate FI parameters. Following the approach in \cite{Hori:2014tda, Hwang:2014uwa}, we will only turn on real FI parameters $\{\zeta_i\}$. For our study of Witten indices associated with 5d instanton partition functions, it will be sufficient to take $\zeta_i =\zeta$, for all $i$.\\

The path integral associated with the index can then be computed in the weak gauge coupling limit $e^2 \beta^3 \to 0$ using standard localization techniques \cite{Hori:2014tda, Hwang:2014uwa}. The answer generically depends on the FI parameter $\zeta$. In the present problem, we are interested in computing a 5d instanton partition function, which is given by an equivariant integral of trigonometric characteristic classes over the Higgs branch of the SQM.
Therefore, we should compute the associated Witten index in a region
of the parameter space of $\zeta$ such that the index has support only on the Higgs branch.
The relevant limit of the Witten index is the {\it Higgs scaling limit} \cite{Hori:2014tda} where we take $e^2 \beta^3 \to 0$
holding $\zeta'= \beta^2 e^2 \zeta$ fixed to a non-zero value. In this limit, the vector multiplet and the chiral adjoint multiplet
become massive with  a mass of the order of $M_H=e\sqrt{|\zeta|}$, and can be integrated out
so that the low energy effective theory is well approximated by the theory on the Higgs branch.
The Witten index computed in the Higgs scaling limit is piecewise constant in $\zeta'$,
and undergoes wall-crossing at $\zeta'=0$ where the effective Higgs masses $M_H$ vanish. \\

We now present the localization formula for a (4,4) quiver ADHM SQM (associated with instantons in an $\N=1^*$ $SU(N)$ theory
on $S^1 \times \C^2/\Z_n$) with $G_{\rm gauge} = \prod^n_{i=1} U(k_i)$ with $\sum^n_{i=1} k_i=k$
- we refer the reader to \cite{Hori:2014tda, Hwang:2014uwa} for details. The Witten index can be written in terms of the (0,2) multiplets, i.e. (0,2) vector multiplets and (0,2) chiral and fermi multiplets transforming in a representation $R$ of $G_{\rm gauge} \times G_{\rm flavor}$.
The path integral in \eref{WI-def} can be reduced to an integral over the space $\mathfrak{M}$ of bosonic zero modes from the
vector multiplets, given by the holonomy of the gauge field around $S^1$ and the adjoint scalar $\varphi$ (neutral under the global symmetry twists), which by constant gauge transformations can be put in the Cartan subalgebra of the SQM gauge group. Given the eigenvalues $\varphi^i_I$ and $A^i_{\tau\, I}$ (such that $A^i_{\tau\, I}+ 2\pi \sim A^i_{\tau\, I}$), with $I=1,2,\ldots, k_i$ and $i=1,2,\ldots, n$, the $k$ variables
$\phi^i_I=\varphi^i_I + \I A^i_{\tau\, I}$ define complex coordinates on $\mathfrak{M}$. Therefore, the space of bosonic zero modes can be
identified as $\mathfrak{M}=\mathfrak{t}_{G_{\rm gauge}} \otimes \C/\Lambda_{\rm coroot} \cong (\C^\star)^k$.

The integral on $\mathfrak{M}$ can be further reduced to a contour integral over $k$ complex variables $\phi^i_I$.
In the Higgs scaling limit, the contour integral is explicitly given as
\begin{align}\begin{split}\label{WI-(0,4)-(0,2)}
& Z^{(4,4)}_{SQM} (\mu, \epsilon_\pm; \zeta')=\frac{1}{\prod^n_{i=1} k_i !} \oint_{JK(\zeta')} \prod^n_{i=1} \prod^{k_i}_{I=1}\Big[\frac{\de \phi^i_I}{2\pi \I}\Big] \, Z_{\rm 1-loop} ~,\\
& Z_{\rm 1-loop}:= Z^{(0,2)}_{\text{vector}} \cdot Z^{(0,2)}_{\text{chiral}} \cdot Z^{(0,2)}_{\text{fermi}}~,
\end{split}\end{align}
The various contributions to $Z_{\rm 1-loop}$ are given as:
\begin{align}\begin{split}
& Z^{(0,2)}_{\rm vector}= \prod_{\al \in {\rm roots}} 2\sinh{\frac{\al(\phi)}{2}}~,\\
& Z^{(0,2)}_{\rm chiral}= \prod_{\rho \in {\rm weights(R_{\rm chiral})}} \Big({2\sinh{\frac{\rho (\phi, a) + 2\epsilon_{+}\,r^+ +2\epsilon_{-}\, r^- + 2m\,f }{2}}}\Big)^{-1}~,\\
& Z^{(0,2)}_{\rm fermi}= \prod_{\rho \in {\rm weights(R_{\rm fermi})}} {2\sinh{\frac{\rho (\phi, a) + 2\epsilon_{+}\,r^+ +2\epsilon_{-}\,r^- + 2m\,f}{2}}}~,
\end{split}\end{align}
where $r^+, r^-, f$ denote the charges of the respective fields under the Cartan generators $J_r + J_R, J_l, J_f$ respectively.\\

The integrand diverges along certain hyperplanes $H_i$ in $\mathfrak{M}$,
where non-zero modes arising from chiral multiplets become massless. Such a hyperplane is of the form:
\be
H_i =\{\phi \in \mathfrak{M} |  Q_i (\phi) + 2{r^+_i} \epsilon_+ + 2 {r^-_i} \epsilon_- + 2 m\,f + Q^F_i (a)=0\}~,
\ee
where $Q_i \in \mathfrak{t}^*_{G_{\rm gauge}}, Q^F_i \in \mathfrak{t}^*_{G_{\rm flavor}}$ are charge covectors associated to the
gauge and flavor symmetry respectively. Let $\mathfrak{M}_{\rm sing}$ be a collection of points in $\mathfrak{M}$ where at least
$k$ such linearly independent hyperplanes intersect. Following \cite{Hori:2014tda, Hwang:2014uwa}, the integral in \eref{WI-(0,4)-(0,2)}
should be evaluated on a compact contour which is a given by a collection of infinitesimal compact contours around a certain
subset of points in $\mathfrak{M}_{\rm sing}$. The appropriate subset and the resultant sum of residues can be conveniently
stated using the Jeffrey-Kirwan residue prescription \cite{Jeffrey:1995rs, Szenes:2003pl} which we will describe momentarily.\\

Let $\{Q_l \in \mathfrak{t}^*\}$ be a collection of charge covectors, with $l=1,\ldots,L$, for some $L$,
such that $\{H_l\}$ defines a collection of $L$ hyperplanes in $\mathfrak{M}$ intersecting at $\phi=\phi_{*}$, i.e.
\be
H_l = \Big\{\phi \in \mathfrak{M} | Q_l(\phi - \phi_{*})=0 \Big\}~.
\ee
For notational simplicity, let us take $\phi_{*}=0$ -- for generic $\phi_{*}$ one has to shift the variables $\phi_I$ appropriately.
The contour integral of $k$ complex variables has a pole at $\phi=0$ if $L\geq k$ hyperplanes intersect at that point.
This hyperplane arrangement is called projective \cite{Benini:2013xpa, Szenes:2003pl} when the $L$ charge covectors are contained in a
half-space of $\mathfrak{t}^*$.  In all ADHM SQMs associated with instantons in 5d $\N=1^*$ SYM, the projective condition is satisfied.

Now, let us compute the JK residue of the above integrand at $\phi=\phi_{*}=0$.
On Laurent-expanding the integrand around $\phi=0$, the non-zero residues are obtained from simple poles.
Near the singularity, the relevant denominator takes the form:
$$ \frac{1}{Q_{l_1}(\phi)\ldots Q_{l_{k}}(\phi)}~, $$
where $Q_{l_1}, \cdots, Q_{l_{k}}(\phi)$ are $k$ independent covectors.
The definition of the JK residue also depends on a covector $\eta \in \mathfrak{t}^*$. For a projective arrangement, the JK residue at
$\phi=0$ is then defined as
\be \label{JK-def}
{\rm JK-Res}(\{Q_l\}, \eta)\frac{\de \phi_1\cdots \de \phi_k}{Q_{l_1}(\phi)\ldots Q_{l_{k}}(\phi)} :=\begin{cases}
   |{\rm det} (Q_{l_1}\ldots Q_{l_{k}})|^{-1},& \text{if } \eta \in {\rm Cone}(Q_{l_1}\ldots Q_{l_{k}})\\
    0,              & \text{otherwise,}
\end{cases}
\ee
where $\eta \in {\rm Cone}(Q_{l_1}\ldots Q_{l_{k}})$ if $\eta = \sum^k_{i=1} a_i Q_i$ with strictly positive coefficients $a_i$
($\eta$ should be in the interior of the cone).
Finally, to complete the contour prescription given in \eref{WI-(0,4)-(0,2)}, we set $\eta =\zeta' (1, \ldots, 1)$. It was shown in
\cite{Hori:2014tda, Hwang:2014uwa}, that this choice sets the residues of all poles coming from the asymptotic region of
$\mathfrak{M}$ to zero. We denote this contour prescription as $JK(\zeta)$ in \eref{WI-(0,4)-(0,2)}. \\

As an illustrative example, consider the case of an Abelian quiver gauge theory. The singular hyperplanes are of the form:
\be
H^{\rm Abelian}_i =\{\phi \in \mathfrak{M} |  Q_i \phi + 2{r^+_i} \epsilon_+ + 2 {r^-_i} \epsilon_- + 2m\,f + Q^F_i (a)=0\}~,
\ee
In this case, one can choose $r^+_i > 0$, for all chiral multiplets, using shifts by gauge and/or flavor charges. Therefore,
a given pole $\phi=\phi_*$, can either correspond to a set of singular hyperplanes with $Q_i > 0$ or a set with $Q_i < 0$,
but never both. Let $\Delta^{(\pm)}$ denote the set of poles of the contour integral corresponding to singular hyperplanes with
$Q_i > 0$ for all $i$, and $Q_i < 0$ with all $i$ respectively. Then, applying the definition \eref{JK-def} for $r=1$ to the formula \eref{WI-(0,4)-(0,2)}, we
get \cite{Benini:2013xpa}:
\begin{align} \label{JK-Abelian}
Z^{(4,4)}_{\rm SQM} (a,m, \epsilon_\pm; \zeta')=& \oint_{JK(\zeta')} \Big[\frac{\de \phi}{2\pi \I}\Big] \, Z_{\rm 1-loop} \nonumber \\
= & \begin{cases} \sum_{\phi_* \in \Delta^+} {\rm Res}_{\phi=\phi_*}\Big[Z_{\rm 1-loop} \frac{\de \phi}{2\pi \I}\Big] \, & \text{if} \, \zeta' > 0~, \\
-\sum_{\phi_* \in \Delta^-} {\rm Res}_{\phi=\phi_*}\Big[Z_{\rm 1-loop} \frac{\de \phi}{2\pi \I}\Big] & \text{if}\, \zeta' < 0~.
\end{cases}
\end{align}

\subsection{ADHM SQM for 5d $\N=1^*$ $SU(N)$ SYM on $S^1 \times \C^2$}\label{WI-N=2*}

As an illustrative example, consider the Witten index for the (4,4) ADHM SQM associated with $k$-instantons in a 5d $U(N)$ or $SU(N)$ $\N=1^*$ SYM on $S^1 \times \C^2$ -- this corresponds to the $n=1$ quiver in the notation of appendix \ref{Loc-app}. The SQM consists of a single $U(k)$ vector multiplet with a single adjoint
hyper and $N$ fundamental hypers.
The Witten index for this theory could be written from the general equation \eref{WI-(0,4)-(0,2)} and Table \ref{Tab:multiplets} as follows:
\footnote{In all Witten index formulae, we adopt the notation: $2 \sinh{(x \pm y)} = 2\sinh{(x+y)} \, 2\sinh{(x-y)}$.}

\begin{empheq}{align}\label{D0quiver-basic}
Z^{(4,4)}_{SQM}(a,m,\epsilon_{1,2}; \zeta') = \frac{1}{k!}\oint_{JK(\zeta')} \left[\frac{d\phi_I}{2\pi i}\right]  {Z}_{k,\, (4,4)}^{\rm vector}(\phi,m,\epsilon_{1,2}) \cdot  {Z}_{k, \, (4,4)}^{\rm adj}(\phi,m,\epsilon_{1,2}) \cdot {Z}_{k, \,(4,4)}^{\rm fund}(\phi,a,\epsilon_{1,2})~,
\end{empheq}
where the contribution of different (4,4) multiplets to the index
\begin{align}\begin{split}
& {Z}^{\rm vec}_{k, \,(4,4)}(\phi,a,\epsilon_{1,2}) = \prod_{I,J=1}^k \frac{2\sinh \half(\phi_{IJ}+2\epsilon_+)}{2\sinh{ \half(\phi_{IJ} + m \pm \epsilon_+)}} \times
\prod_{I \neq J}^k 2\sinh \frac{\phi_{IJ}}{2}~ , \\
& {Z}^{\rm adj}_{k, \,(4,4)}(\phi,a,m,\epsilon_{1,2}) =\prod_{I,J=1}^k\frac{2\sinh{\half(\phi_{IJ} + m \pm \epsilon_-)}}{2\sinh \half(\phi_{IJ}+\epsilon_1)\, 2\sinh \half(\phi_{IJ}+\epsilon_2)}~, \\
& {Z}_{k, \,(4,4)}^{\rm fund}(\phi,a,m,\epsilon_{1,2})=\prod_{I=1}^k\prod_{i=1}^N\frac{2\sinh{\half(\pm (\phi_I-a_i) + m)}}{2\sinh \half(\pm(\phi_I-a_i ) + \epsilon_+)}~.
\end{split}\end{align}
can be computed from the decomposition of (4,4) multiplets into (0,2) multiplets, and then using the prescription in \eref{WI-(0,4)-(0,2)}.

The contour integral should be evaluated using the JK prescription.
Let us write down the formula \eref{D0quiver-basic} explicitly in the chamber $\zeta' > 0$. The $\zeta' < 0$ formula can be
worked out in an analogous fashion. For the (4,4) ADHM SQM under consideration, it was explicitly shown \cite{Hwang:2014uwa}
that the JK prescription leads to the Young diagram formula, such that the poles of the above contour integral are labelled
by $N$-tuples of Young diagrams $ (Y_1,Y_2,\ldots,Y_N)$ with the total number of boxes $|\vec Y|=\sum^N_{\alpha=1} |\vec Y_\alpha| =k$.
The resultant Witten index, which is usually written in terms of the 5d $\N=1^\ast$ vector- and adjoint hyper-multiplets, 
can be expressed as:
\begin{align}\label{basic-Nek1}
& Z^{(4,4)}_{k\, SQM}(a,m,\epsilon_{1,2}; \zeta'>0) =\sum_{\vec Y} z_{\vec Y,\, 5d}^\text{vec}(\epsilon_1,\epsilon_2, a) z_{\vec Y,\, 5d}^\text{adj}(\epsilon_1,\epsilon_2, a, m)~,
\end{align}
where $z_{\vec Y,\, 5d}^\text{vec}$ and $z_{\vec Y,\, 5d}^{\text{adj}}$ are contributions of the 5d vector multiplet and the 5d hypermultiplet at the pole labelled by
$\vec Y$ -- the explicit expressions are discussed below.
In order to write these we note that for a given $\vec Y$, each box in a given N-tuple is labelled by a $\phi_I$ for some $I$ (we choose a rule where the count of $I$ starts at the box at the leftmost corner of the first non-empty Young diagram) and the corresponding poles in $\phi_I$ are given by
\footnote{It is a special feature of 5d N=1* U(N) partition function that the residues arising from the other poles (i.e. the ones which depend
on the adjoint mass $m$) are zero. This was already noted in the original paper of Nekrasov \cite{Nekrasov:2002qd} and proved carefully in later papers -- we refer the reader to section 3.1 of \cite{Hwang:2014uwa} for a detailed proof.}
\begin{align}
\phi_I=\phi_s:= a_\alpha + \epsilon_+ - i_\al \epsilon_1 -j_\al \epsilon_2~,
\end{align}
where $I=1,\ldots, k$ and $\al=1,\ldots, N$, with $s=(i_\al, j_\al)$ denoting a box in the $\al$-th Young diagram in $\vec Y$ \footnote{Our convention for Young diagrams is to draw them in the first quadrant with $i$ and $j$ labelling the horizontal and vertical axes respectively, with $i$ and $j$ increasing away from the origin.}.  \\

The 5d vector multiplet contribution to the residue at $\vec Y$ is
\begin{align}\begin{split}\label{basic-vec}
&z_{\vec Y,\, 5d}^\text{vec}(\epsilon_1,\epsilon_2, \vec a) = \frac{1}{\prod_{(\alpha,\beta,s \in Y_\alpha)}\sinh{\half E_{\alpha \beta}(s)}  \sinh{\half(-2\epsilon_+ + E_{\alpha \beta}(s))}}~, \\
&E_{\alpha \beta}(s):= E(a_\alpha-a_\beta, Y_\alpha,Y_\beta,s)= a_\alpha-a_\beta -\epsilon_1 L_{Y_\beta}(s) + \epsilon_2(A_{Y_\al} (s)+1)~,
\end{split}\end{align}
where $L_{Y_\al}(s)$ is the distance of the box $s$ from the rightmost edge of the Young diagram in the same row, and $A_{Y_\al} (s)$ is the distance of the box $s$ from the bottom of the diagram in the same row.\\

The 5d adjoint hypermultiplet contributes as follows:
\begin{align}
&z_{\vec Y,\, 5d}^\text{adj}(\epsilon_1,\epsilon_2, \vec a,m)= \prod_{(\alpha,\beta,s \in Y_\alpha)}\sinh{\half(E_{\al\beta}(s) +m -\epsilon_+)} \times \sinh{\half( E_{\al,\beta}(s)-m-\epsilon_+)}~. \label{basic-adj}
\end{align}

Combining all the residues, the $k$ instanton partition for a given $N$-tuple Young diagrams $\vec{Y}=\{Y_1,Y_2,\cdots,Y_N\}$ is
\begin{align}
	Z^{\rm inst}_k = & \sum_{|\vec{Y}|=k}\prod_{\al,\beta=1}^N\prod_{s\in Y_\al}\frac{\sinh \half (E_{\al,\beta}(s)+m-\epsilon_+)\sinh{\half(E_{\al,\beta}(s)-m-\epsilon_+)}}{\sinh{\half E_{\al,\beta}(s)} \sinh {\half (E_{\al,\beta}(s)-2\epsilon_+)}}~.
\end{align}

\section{Computation of $Z_{\rm mono}$ from the defect SQM} \label{Okuda-2*}
In this section, we compute explicit expressions for $Z_{\rm mono}$ associated with 't Hooft operators in $\N=2^*$ $SU(2)$ SYM
using the Witten index formula \eref{D0quiver-orbifold}-\eref{explicitfunctions} of the related SQMs discussed in section \ref{WI}.
The function $Z_{\rm mono}$ is labelled by the following defect data:
\begin{align}
B= \half {\rm diag} (p, \, -p)\quad, \qquad {\bf v}= \half {\rm diag} (v, \, -v)~,
\end{align}
where $p$ is a positive integer, and $v=p, p-2, p-4,\ldots, -p$. We will compute $Z_{\rm mono}$ for a few small values of $p$ and $v$ below
-- the SQMs, along with the defect data and the instanton data, associated with $Z_{\rm mono}$ in these examples are listed in Table \ref{Tab:SQM1}.
\begin{center}
\begin{table}[htbp]
\begin{tabular}{|m{80pt}|m{200pt}|m{200pt}|}
\hline
Defect Data $(B, \mathbf{v})$ & KN Data $(\vec k, \vec w)$ & Quiver SQM \\
\hline \hline
$B=\half(3,-3)$, \newline $v= \half(1,-1)$. & $\vec k= (0,1,1,0,\ldots,0)$, \newline  $\vec w=(0,1,1,0,\ldots,0)$. & \includegraphics[scale=0.5]{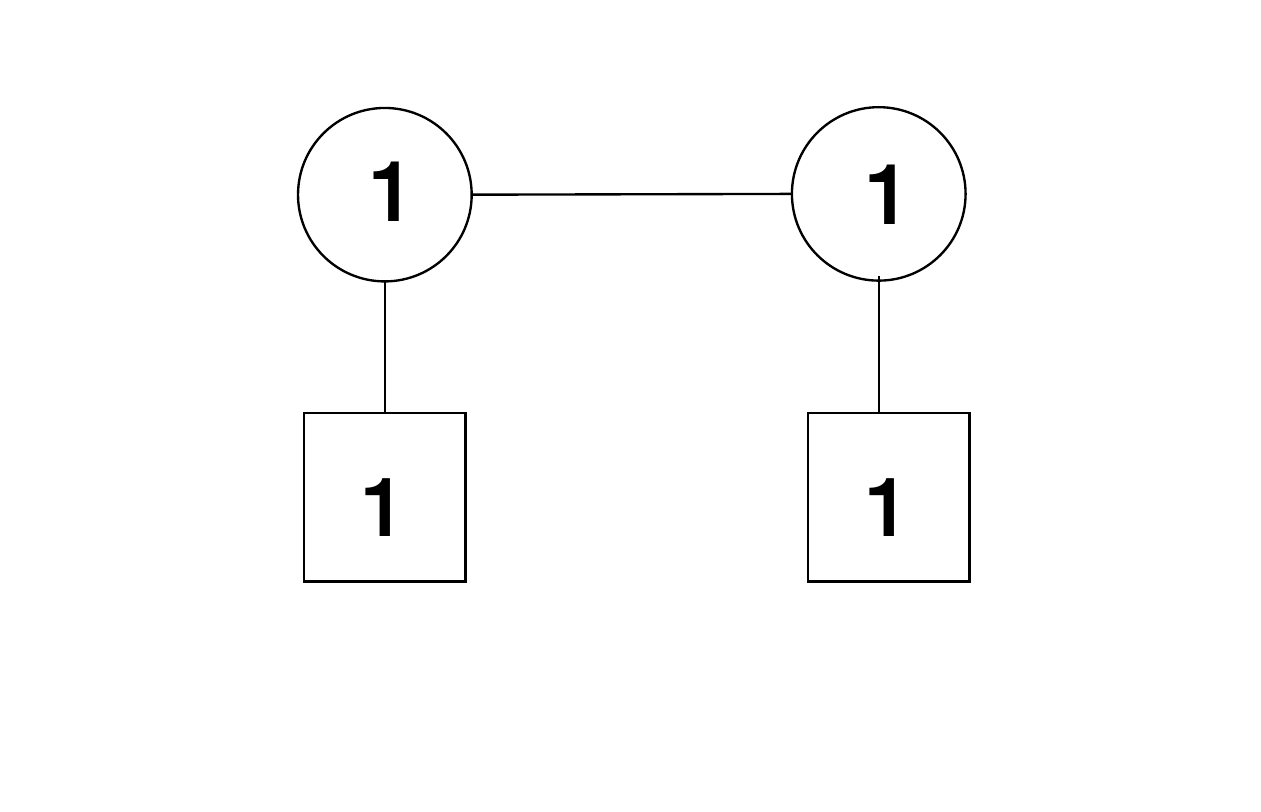} \\
\hline
$B=\half(4,-4)$,  $v= \half(0,0)$. & $\vec k= (0,1,2,1,\ldots,0)$, \newline $\vec w=(0,0,2,0,\ldots,0)$. & \includegraphics[scale=0.5]{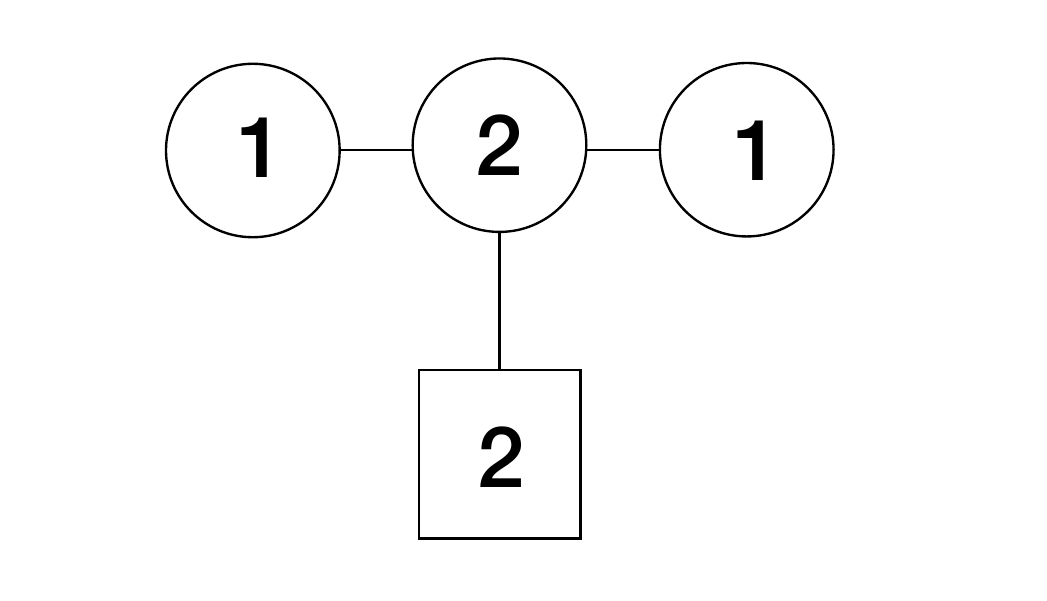} \\
\hline
$B=\half(4,-4)$,  $v= \half(2,-2)$.   & $\vec k= (0,1,1,1,\ldots,0)$, \newline $\vec w=(0,1,0,1,0\ldots,0)$. & \includegraphics[scale=0.5]{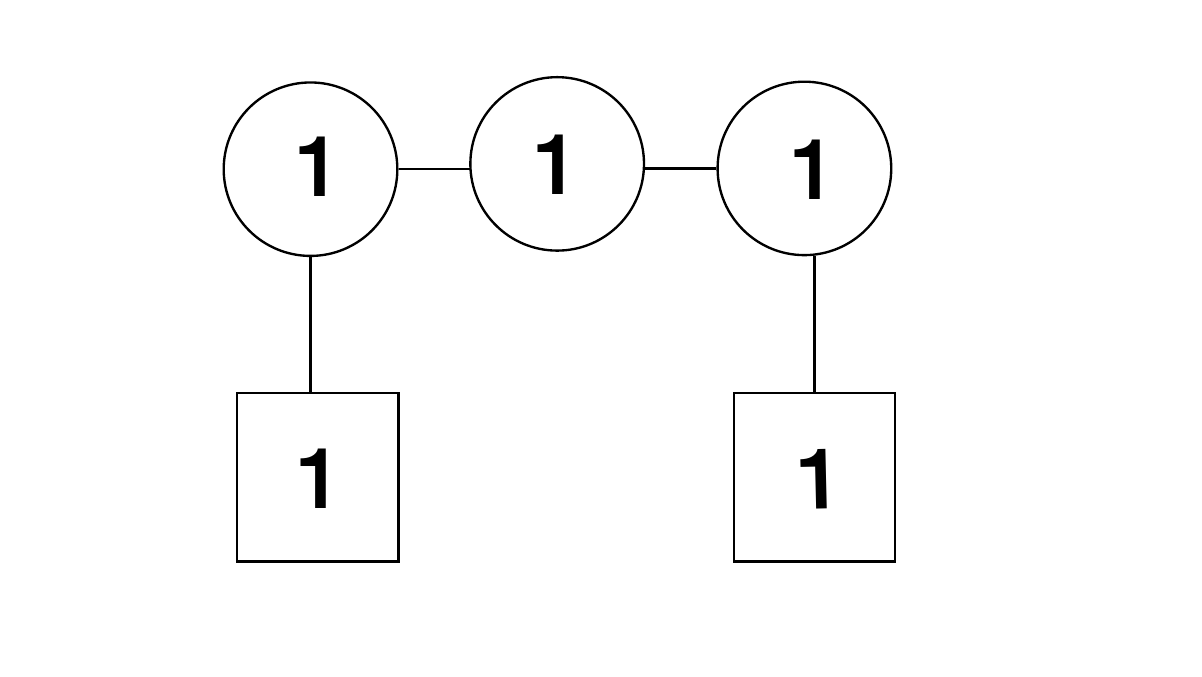}\\
\hline
\end{tabular}
\caption{Summary table for examples of quiver SQMs associated with the monopole bubbling indices of 't Hooft operators in $\N=2^*$ $SU(2)$ SYM.}
\label{Tab:SQM1}
\end{table}
\end{center}
The resultant
expressions are identical to those assembled from the IOT expressions summarized in \eref{mono-vec}-\eref{mono-adj}, if we identify the equivariant parameters in the
following fashion\footnote{Here $a$ a complex number.}:
\begin{align}\label{repr-4d/5d-A}
a:= 2\I \pi \fa\quad, \quad m:= 2\I \pi \fm\quad, \quad \epsilon_+:=\I \pi {\lambda}\quad, \quad \epsilon_-:=0~.
\end{align}
The Witten indices are even functions of $\epsilon_+$ indicating that they are invariant under wall-crossing w.r.t the FI parameters.
We will use the pole prescription corresponding to the chamber $\zeta' >0$ to evaluate them.

\begin{itemize}
\item $\mathbf{(p=3,v=1)}$ :\quad The defect data is given as $B=\text{diag}\half(3,\,-3)$ and $\mathbf{v}=\text{diag}\half(1,-1)$.
From equation \eref{char-eqn-main}, we obtain $K=\text{diag}(-\half, \half)$.
Using the shift transformation as discussed in section \ref{Ex-WI}, we have $K=\text{diag}(1,2)$ and $\mathbf{v}=\text{diag}(2,1)$, which leads to
the following $U(2)$ instanton data on a $\C^2/\Z_n$ orbifold ($n > 3$):
\be
\vec k=(0,1,1,0,\ldots,0) \hspace{10mm} \vec w=(0,1,1,0,\ldots,0)  \hspace{10mm} \mathbf{v}=\text{diag}(2,1).
\ee
The associated $(4,4)$ SQM has a $U(1) \times U(1)$ gauge group with one bifundamental hyper and one fundamental hyper at each node, as given in
Table \ref{Tab:SQM1}. The Witten index in the Higgs scaling limit can be written as
\begin{align}
	& Z_{\rm SQM}(a,m,\epsilon_{1,2}) = \oint \left[\frac{d\phi_I}{2\pi i}\right] \,  {Z}^{\rm vec}_{\vec k} \cdot  {Z}^{\rm bif}_{\vec k}  \cdot  {Z}^{\rm fund}_{\vec k, \vec w} ~ ,\cr
	&  {Z}^{\rm vec}_{\vec k}(\phi,m,\epsilon_{1,2}) =  \Big(\frac{2\sinh(\epsilon_+)}{2\sinh{\half(m \pm \epsilon_+)}}\Big)^2 ~, \cr
	&  {Z}^{\rm bif}_{\vec k}(\phi,m,\epsilon_{1,2}) =\frac{\sinh{\half(\phi^{2}_1 -\phi^1_1 + m + \epsilon_-)} \sinh{\half(\phi^{1}_1 -\phi^2_1 + m - \epsilon_-)}}{\sinh{\half(\phi^{2}_1-\phi^1_1+\epsilon_+ +\epsilon_-)}\sinh{\half(\phi^{1}_1-\phi^2_1+\epsilon_+ - \epsilon_-)}}   \ , \cr
	&  {Z}^{\rm fund}_{\vec k, \vec w}(\phi,a,m,\epsilon_{1,2})=\frac{\sinh{\half(\phi^1_1-a_2 + m)} \sinh{\half(-\phi^1_1 + a_2 + m)}}{\sinh{\half(\phi^1_1-a_2+ \epsilon_+)} \sinh{\half(-\phi^1_1+ a_2+ \epsilon_+)}}  \cr
 & \qquad \qquad \qquad \qquad \times \frac{\sinh{\half(\phi^2_1-a_1 + m)} \sinh{\half(-\phi^2_1 + a_1 + m)}}{\sinh\half(\phi^2_1-a_1+ \epsilon_+) \sinh \half(-\phi^2_1+ a_1+ \epsilon_+)}~.
\end{align}

 From the residue prescription following \eref{D0quiver-orbifold}-\eref{explicitfunctions} in section \ref{WI}, the poles are labelled by the following doublets of colored Young diagrams:
\begin{align}\begin{split}
\text{I. }\left( \text{ } \begin{Young} \Yvcentermath1 1 \cr  2 \cr \end{Young} \text{ , } \emptyset \text{ } \right)\hspace{5mm}\text{II. }\left(\text{ } \begin{Young} 2 \cr \end{Young}\text{ , }\begin{Young}   1\cr \end{Young}\text{ }\right)\hspace{5mm}\text{III. }\left(\text{ }\emptyset \text{ , } \begin{Young}  1 & 2 \cr  \end{Young}\text{ }\right)
\end{split}\end{align}
where the $(i,j)$-th box in the $\ell$-th Young diagram is labelled by its $\Z_n$-charge, i.e $s=r_\ell+i-j$, such that the total number of boxes with charge $s$ is $k_s$.
The pole associated with a box with coordinates $(i,j)$ and  $\Z_n$-charge $s$ in the $\ell^{th}$ Young diagram of a given doublet is:
\be
\phi^{s}_I=a_\ell+\epsilon_+-i \epsilon_1-j\epsilon_2~.
\ee
Explicitly, the poles of the contour integral are given by our residue prescription as follows:
\begin{align}\begin{split}
& \text{I. }  \phi^1_1= a_1- 2\epsilon_+ + \epsilon_- ,\, \phi^2_1= a_1-\epsilon_+~,\\
& \text{II. } \phi^1_1= a_2-\epsilon_+,\, \phi^2_1= a_1 - \epsilon_+~,\\
& \text{III. } \phi^1_1= a_2-\epsilon_+,\, \phi^2_1= a_2-2 \epsilon_+ - \epsilon_-~.
\end{split}\end{align}

Computing residues at the three poles, with $a_1=-a_2=a$, we obtain
\begin{align}
Z_{\rm mono}(a,m,\epsilon_{+}; p&=3,v=1) =Z_{\rm SQM}(a,m;\epsilon_{+}, \epsilon_-)|_{\epsilon_{-}=0} \nonumber\\ =&\frac{\sinh{\half(2a+m -2\epsilon_+)} \sinh{\half(2a-m -2\epsilon_+)}}{\sinh{\half(2a-\epsilon_+)}\sinh{\half(2a-3\epsilon_+)}} \nonumber \\
+& \frac{\sinh{\half(2a+m +2\epsilon_+)} \sinh{\half(2a-m +2\epsilon_+)}}{\sinh{\half(2a+\epsilon_+)}\sinh{\half(2a+3\epsilon_+)}} \nonumber\\
+ & \frac{\sinh{\half(2a+m)} \sinh{\half(2a-m )}}{\sinh{\half(2a+\epsilon_+)}\sinh{\half(2a-\epsilon_+)}}.
\end{align}

\item $\mathbf{(p=4,v=0)}$ :\quad The defect data, after the usual shift, is given as:
\be
B= {\rm diag}(4,0) \quad,\qquad\mathbf{v}= {\rm diag}(2,2)~,
\ee
with the associated KN data :
\be
\vec{k}=(0,1,2,1,0,...)\quad,\qquad\vec{w}=(0,0,2,0,0,...) \quad,\qquad {r}=\mathbf{v}= {\rm diag}(2,2)~.
\ee
The defect SQM is given by the $\CN=(4,4)$ quiver in Table \ref{Tab:SQM1}.
This has contributions from the vector, fundamental chiral, and bifundamental chiral multiplets:
\begin{align}\begin{split}\label{eq:z40}
&Z^{\rm vec}_{\vec k}=\left(\frac{2\sinh(\epsilon_+)}{2\sinh \half(m \pm \epsilon_+)}\right)^4\frac{\sinh \half(\pm\phi^2_{12})\sinh \half(\pm \phi^2_{12}+2\epsilon_+)}{\sinh \half(\phi^2_{12}+m\pm \epsilon_+)\sinh \half(\phi^2_{21}+m\pm \epsilon_+)}~,\\
&Z^{\rm fund}_{\vec k, \vec w}=\prod_{i=1}^2\frac{\sinh \half(\phi^2_1-a_i+m)\sinh \half(\phi^2_2-a_i+m)\sinh \half(-\phi^2_1+a_i+m)\sinh \half(-\phi^2_2+a_i+m)}{\sinh(\phi^2_1-a_i+\epsilon_+)\sinh \half(\phi^2_2-a_i+\epsilon_+)\sinh \half(-\phi^2_1+a_i+\epsilon_+)\sinh \half(-\phi^2_2+a_i+m)}~,\\
&Z^{\rm bif}_{\vec k}=\left(\frac{\sinh \half(\phi^{21}_{11}+m+\epsilon_-)\sinh \half(\phi^{12}_{11}+m-\epsilon_-) \sinh \half(\phi^{21}_{21}+m+\epsilon_-)\sinh \half(\phi^{12}_{12}+m-\epsilon_-)}{\sinh \half(\phi^{21}_{11}+\epsilon_1)\sinh \half(\phi^{12}_{11}+\epsilon_2)\sinh \half(\phi^{21}_{21}+\epsilon_1) \sinh \half(\phi^{12}_{12}+\epsilon_2)}\right)\\
&\times\left(\frac{\sinh \half(\phi^{32}_{11}+m+\epsilon_-)\sinh \half(\phi^{23}_{11}+m-\epsilon_-) \sinh \half(\phi^{32}_{12}+m+\epsilon_-)\sinh \half(\phi^{23}_{21}+m-\epsilon_-)}{\sinh \half(\phi^{32}_{11}+\epsilon_1)\sinh \half(\phi^{23}_{11}+\epsilon_2)\sinh \half(\phi^{32}_{12} +\epsilon_1)\sinh \half(\phi^{23}_{21}+\epsilon_2)}\right)~,
\end{split}\end{align}
where $\phi^i_{IJ}=\phi^i_I-\phi^i_J$ and $\phi^{ij}_{IJ}=\phi^i_I-\phi^j_J$.\\

From the residue prescription following \eref{D0quiver-orbifold}-\eref{explicitfunctions} in section \ref{WI}, the poles are labelled by the following doublets of colored Young diagrams and their symmetric pairs (i.e. doublets of Young diagrams I, II and III, with $Y_1 \leftrightarrow Y_2$):
\begin{align}\label{N2*ex2-YD}
\begin{split}
\text{I. }\left(\text{ }\begin{Young} 1 & 2 \cr 2 & 3 \cr \end{Young}\text{ },\text{ }\emptyset\right)\hspace{5mm}\text{II. }\left(\text{ } \begin{Young}  1 \cr 2 & 3 \cr \end{Young}\text{ , }\begin{Young} 2 \cr  \end{Young}\text{ }\right)\hspace{5mm}\text{III. }\left(\text{ }\begin{Young} 2 & 3 \cr \end{Young}\text{ , } \begin{Young} 1 \cr 2 \cr \end{Young}\text{ }\right)
\end{split}\end{align}
where the $(i,j)$-th box in the $\ell$-th Young diagram is labelled by its $\Z_n$-charge, i.e $s=r_\ell+i-j$, such that the total number of boxes with charge $s$ is $k_s$.
The pole associated with a box with coordinates $(i,j)$ and  $\Z_n$-charge $s$ in the $\ell^{th}$ Young diagram of a given doublet is:
\be
\phi^{s}_I=a_\ell+\epsilon_+-i \epsilon_1-j\epsilon_2.
\ee
Explicitly the poles can be listed as follows:
\begin{align}\label{N2*ex2-poles}
\begin{split}
\begin{array}{rlcrlcrl}
\text{I.)}&\phi^1_1=a+\epsilon_+-\epsilon_1-2\epsilon_2&&
\text{II.)}&\phi^1_1=a+\epsilon_+ - \epsilon_1 -2\epsilon_2&&
\text{III.)}&\phi^1_1=-a+\epsilon_+-\epsilon_1-2\epsilon_2\\
&\phi^2_1=a+\epsilon_+-\epsilon_1-\epsilon_2&&&
\phi^2_1=a+\epsilon_+-\epsilon_1-\epsilon_2&&&
\phi^2_1=a+\epsilon_+-\epsilon_1-\epsilon_2\\
&\phi^2_2=a+\epsilon_+-2\epsilon_1-2\epsilon_2&&&
\phi^2_2=-a+\epsilon_+-\epsilon_1-\epsilon_2&&&
\phi^2_2=-a+\epsilon_+-\epsilon_1-\epsilon_2\\
&\phi^3_1=a+\epsilon_+-2\epsilon_1-\epsilon_2&&&
\phi^3_1=a+\epsilon_+-2\epsilon_1-\epsilon_2&&&
\phi^3_1=a+\epsilon_+ -2\epsilon_1-\epsilon_2
\end{array}
\end{split}\end{align}

Plugging these into \ref{eq:z40}, we get the contributions:
\begin{align}\begin{split}
\text{I.}&\hspace{5mm}\frac{\sinh \half(2a\pm m-\epsilon_+)\sinh \half(2a\pm m-3\epsilon_+)}{\sinh(a)\sinh^2(a-\epsilon_+)\sinh(a-2\epsilon_+)}~,\\
\text{II.}&\hspace{5mm}\frac{\sinh^2\half(2a\pm m-\epsilon_+)}{\sinh^2(a)\sinh^2(a-\epsilon_+)}~,\\
\text{III.}&\hspace{5mm}\frac{\sinh \half(2a\pm m+\epsilon_+)\sinh \half(2a\pm m-\epsilon_+)}{\sinh(a\pm \epsilon_+)\sinh^2(a)}~.
\end{split}\end{align}
The symmetric pair for each of these diagrams leads to poles for which the contribution to $Z_{\rm mono}$ is given by I, II or III,
with $ a \to - a$. This gives us the final result:
\begin{align}\begin{split}
&Z_{\rm mono}(a,m,\epsilon_+; p=4,v=0)= \Big[\frac{\sinh \half(2a\pm m-\epsilon_+)\sinh \half(2a\pm m-3\epsilon_+)}{\sinh(a)\sinh^2(a-\epsilon_+)\sinh(a-2\epsilon_+)}\\
+& \frac{\sinh^2 \half(2a\pm m-\epsilon_+)}{\sinh^2(a)\sinh^2(a-\epsilon_+)}
+\frac{\sinh \half(2a\pm m+\epsilon_+)\sinh \half(2a\pm m-\epsilon_+)}{\sinh(a\pm \epsilon_+)\sinh^2(a)}\Big] \\ & + \Big[a \to - a \Big]~.
\end{split}\end{align}

\item $\mathbf{(p=4,v=2)}$ :\quad The KN data of this contribution is described by the vectors:
\be
\vec{k}=(0,1,1,1,0,...)\quad,\qquad\vec{w}=(0,1,0,1,0,...) \quad,\qquad {r}=\mathbf{v}= {\rm diag}(3,1).
\ee
The SQM is given by the $\CN=(4,4)$ quiver given in Table \ref{Tab:SQM1}.
This has contributions from the vector, fundamental chiral, and bifundamental chiral multiplets:
\begin{align}\begin{split}\label{eq:z42}
&Z^{\rm vec}_{\vec k}=\Big(\frac{2\sinh(\epsilon_+)}{2\sinh \half(m \pm \epsilon_+)}\Big)^3~,\\
&Z^{\rm bif}_{\vec k}=\frac{\sinh \half(\phi^{21}_{11}+m+\epsilon_-)\sinh \half(\phi^{12}_{11}+m-\epsilon_-) \sinh \half(\phi^{32}_{11}+m+\epsilon_-) \sinh \half(\phi^{23}_{11}+m-\epsilon_-)}{\sinh \half(\phi^{21}_{11}+\epsilon_1)\sinh \half(\phi^{12}_{11}+\epsilon_2) \sinh \half(\phi^{32}_{11}+\epsilon_1)\sinh \half(\phi^{23}_{11}+\epsilon_2)}~,\\
&Z^{\rm fund}_{\vec k, \vec w}=\frac{\sinh \half(\phi^1_1-a_2+m)\sinh \half(-\phi^1_1+a_2+m)\sinh \half(\phi^3_1-a_1+m)\sinh \half(-\phi^3_1+a_1+m)}{\sinh \half(\phi^1_1-a_2+\epsilon_+)\sinh \half(-\phi^1_1+a_2+\epsilon_+)\sinh \half(\phi^3_1-a_1+\epsilon_+)\sinh \half(-\phi^3_1+a_1+\epsilon_+)}~,
\end{split}\end{align}
where $\phi^{ij}_{IJ}=\phi^i_I-\phi^j_J$.\\

The poles for the contour integral are labelled by the following doublets of Young diagrams:
\begin{align}\label{N2*ex3-YD}
\begin{split}
\text{I. }\left(\text{ }\begin{Young} 1 \cr 2 \cr  3 \cr \end{Young}\text{ },\text{ }\emptyset\right)\hspace{5mm}\text{II. }\left(\text{ } \begin{Young} 2 \cr 3 \cr \end{Young}\text{ , }\begin{Young} 1 \cr  \end{Young}\text{ }\right)\hspace{5mm}\text{III. }\left(\text{ }\begin{Young} 3 \cr \end{Young}\text{ , } \begin{Young} 1 & 2 \cr \end{Young}\text{ }\right)
\hspace{5mm}\text{IV. }\left(\text{ }\emptyset \text{ , }\begin{Young} 1 & 2 & 3 \cr \end{Young}\text{ }\right)
\end{split}\end{align}
Explicitly, these poles are of the form:
\begin{align}\label{N2*ex3-poles}
\begin{split}
&\begin{array}{rlcrlcrl}
\text{I.}&\phi_1=a+\epsilon_+-\epsilon_1-3\epsilon_2&&
\text{II.}&\phi_1=-a+\epsilon_+-\epsilon_1-\epsilon_2&&
\text{III.}&\phi_1=-a+\epsilon_+-\epsilon_1-\epsilon_2\\
&\phi_2=a+\epsilon_+-\epsilon_1-2\epsilon_2&&&
\phi_2=a+\epsilon_+-\epsilon_1-2\epsilon_2&&&
\phi_2=-a+\epsilon_+-2\epsilon_1-\epsilon_2\\
&\phi_3=a+\epsilon_+-\epsilon_1-\epsilon_2&&&
\phi_3=a+\epsilon_+-\epsilon_1-\epsilon_2&&&
\phi_3=a+\epsilon_+-\epsilon_1-\epsilon_2
\end{array}\\
&\begin{array}{rl}
\text{IV.}&\phi_1=-a+\epsilon_+-\epsilon_1-\epsilon_2\\
&\phi_2=-a+\epsilon_+-2\epsilon_1-\epsilon_2\\
&\phi_3=-a+\epsilon_+-3\epsilon_1-\epsilon_2
\end{array}
\end{split}\end{align}

From \eref{eq:z42}, we get the following residues (setting $\epsilon_-=0$):
\begin{align}\begin{split}
\text{I.}&\hspace{5mm}\frac{\sinh \half(2a\pm m-3\epsilon_+)}{\sinh (a- \epsilon_+)\sinh (a-2\epsilon_+)}\quad,\qquad \text{III.}\hspace{5mm}\frac{\sinh \half(2a\pm m+\epsilon_+)}{\sinh(a)\sinh(a+\epsilon_+)}~,\\
\text{II.}&\hspace{5mm}\frac{\sinh \half(2a\pm m-\epsilon_+)}{\sinh(a)\sinh(a-\epsilon_+)}\qquad\qquad,\qquad
\text{IV.}\hspace{5mm}\frac{\sinh \half(2a\pm m+3\epsilon_+)}{\sinh(a+\epsilon_+)\sinh(a+2\epsilon_+)}~.
\end{split}\end{align}
This leads to the result
\begin{align}\begin{split}
Z_{\rm mono}(a,m, &\epsilon_+; p=4, v=2)=\frac{\sinh \half(2a\pm m-3\epsilon_+)}{\sinh(a-\epsilon_+)\sinh(a-2\epsilon_+)}+\frac{\sinh \half (2a\pm m-\epsilon_+)}{\sinh(a)\sinh(a-\epsilon_+)}\\&
+\frac{\sinh \half(2a\pm m+\epsilon_+)}{\sinh(a)\sinh(a+\epsilon_+)}+\frac{\sinh \half(2a\pm m+3\epsilon_+)}{\sinh(a+\epsilon_+)\sinh(a+2\epsilon_+)}~.
\end{split}\end{align}

\end{itemize}

\section{Bubbling index as an equivariant integral} \label{equiv structure}
The five dimensional instanton partition function of a 5d $\N=1$ theory on $S^1 \times \R^4$ is given by an equivariant integral of certain trigonometric characteristic classes over the moduli space of instantons on $\R^4$ \cite{Nekrasov:2002qd}. Similarly, the instanton partition function of a 5d $\N=1$ theory on $S^1 \times \R^4/\Z_n$ is given by an equivariant integral with the same characteristic classes as above, and the domain of integration is an appropriate
KN moduli space. Since instanton moduli spaces on $\R^4$ as well as KN moduli spaces have small instanton singularities, these
equivariant integrals are not well-defined in general. However, in both cases, there exist resolutions of the moduli spaces obtained by introducing
suitable stability parameters (FI parameters). The group action lifts naturally such that the equivariant characteristic classes
can be extended to these resolved spaces, and therefore one can unambiguously define these integrals.

In both cases, the equivariant integral may be reduced to a contour integral. For instanton partition functions on $\R^4$ and $S^1 \times \R^4$,
such contour integrals were studied in detail by Nekrasov and Shadchin \cite{Shadchin:2004yx, Nekrasov:2004vw, Shadchin:2005mx}. In the 5d case, these contour integrals coincide with the
Witten index of the ADHM quiver SQM in the Higgs scaling limit, i.e. in the limit of $e^2 \to 0$ with the FI parameter $|\zeta| \to \infty$
such that $\zeta' =e^2 \zeta$ is held fixed \cite{Kim:2011mv, Hwang:2014uwa}. The instanton partition function then depends only on the sign of the FI parameter.
In a pure $\N=2$ or $\N=2^*$ $SU(N)$ SYM, the instanton partition function is completely independent of the FI parameter,
but this is not true if we include hypermultiplets in general representations.\\

In section \ref{char-classes-S1xR4}, we discuss the equivariant integral formula for 5d instanton partition functions on $S^1 \times \R^4$ together with the relevant characteristic classes. In section \ref{KN-equiv}, we write down the analogous expressions for $S^1 \times \R^4/\Z_n$.

\subsection{Equivariant integrals for 5d instanton partition function on $S^1 \times \R^4$}\label{char-classes-S1xR4}

 \subsubsection{4d partition function}
 Let us first review the equivariant integral formula for a 4d instanton partition function of a pure $\N=2$ $U(N)$ SYM on $\R^4$ and how it reduces to a
 contour integral. Let $\cM^k$ be the affine space of ADHM data, and $\cM^k_{\rm ADHM}$ is the ADHM moduli space with
fixed framing at infinity (i.e. choice of a basis of the vector space $W$) obtained as a non-compact
hyperk\"ahler quotient $\cM^k////U(k)$ implemented via
the ADHM equations:
$$\mu_{\mathbb C} \equiv  [B_1,B_2]+IJ=0\quad,\quad
\mu_{\mathbb R} \equiv  [B_1^\dagger, B_1]+[B_2^\dagger, B_2]+II^\dagger-J^\dagger J=0~.
$$\\
Note that by splitting the moment maps into real and complex, we are implicitly choosing a complex structure on $\R^4$. Let
$\omega$ be the symplectic (1,1) form w.r.t. the chosen complex structure. As discussed in \cite{Nekrasov:2002qd, Moore:1997dj}, 
the 4d instanton partition function involves computation of a $T$-equivariant volume, associated with the
 torus action of $T=U(1)^2|_{\epsilon_1,\epsilon_2} \times U(1)^N|_{\vec a}$, i.e.
\begin{align}
Z_{\rm inst}(\epsilon_{1},\epsilon_{2}, \vec a)= \sum_{k \geq 0} \fq^k \int_{\cM^k_{\rm ADHM}=\cM^k////U(k)} e^{\omega + \mu_T}~,
\end{align}
where $\mu_T$ is the $T$-moment map so that we have an equivariant 2-form.

The smooth locus of the moduli space $\cM^k_{\rm ADHM}$ is metrically incomplete as a \hk manifold and this can be addressed by adding point/ideal
instantons (in the Uhlenbeck compactification):
\begin{align}
\overline{\cM}^k= \cM^k_{\rm ADHM} \cup \big(\cM^{k-1}_{\rm ADHM} \times \R^4 \big)\cup \big(\cM^{k-2}_{\rm ADHM} \times Sym^2(\R^4) \big)\cup \ldots \cup Sym^k(\R^4)~.
\end{align}
The resultant space $\overline{\cM}^k$ is a singular manifold and one cannot apply the standard theorems of localization directly
to such spaces. However, the Uhlenbeck compactification $\overline{\cM}^k$ admits a smooth resolution $\widetilde{\cM}^k_{ADHM}(\zeta)$,
which is the moduli space of torsion free sheaves on $\mathbb{C} \mathbb{P}^2$ with fixed framing of the line at infinity,
with rank $N$ and second Chern class $c_2=k$ \cite{1995alg.geom..7012N, Nakajima:2003pg, Nakajima:2005fg}. $\widetilde{\cM}^k_{ADHM}(\zeta)$
is a \hk manifold and can be shown to be isomorphic to the \hk quotient \cite{Nakajima:2003pg}:
\begin{eqnarray}\label{res-adhm}
\widetilde{\cM}^k_{ADHM}(\zeta) \cong
\left   \{
 (B_1,B_2,I,J)
 \left|
\begin{array}{ccc}
 \mu_{\C}&=&0\\
  \mu_{\R}&=&\zeta \cdot \, \mathbb{I}_k\\
 \end{array}
\right.
\right\}
{\Big /}
U(k,\mathbb C)~ ,
 \end{eqnarray}
where $\zeta$ is a fixed positive real number. In terms of the string theory picture of Dp-D(p+4) branes, where the ADHM construction can be
understood as the Higgs branch of the Dp world volume gauge theory, this amounts to turning on an FI parameter for the $U(1)$ factor of the
$U(k)$ gauge group. \\

In addition, the T-action lifts to $\widetilde{\cM}^k_{ADHM}(\zeta)$, so that one can now unambiguously define the equivariant volume of the resolved moduli space. As explained in \cite{Nekrasov:2002qd, Martens:2006hu}, the equivariant volume relevant for the original gauge theory problem of instanton counting is the one computed with respect to the pull back of the symplectic 2-form on the Uhlenbeck compactified moduli space $\overline{\cM}^k$. The resulting 2-form on $\widetilde{\cM}^k_{ADHM}(\zeta)$ vanishes on the exceptional set $\widetilde{\cM}^k_{ADHM}(\zeta) \to \overline{\cM}^k$ and reduces to the original 2-form on $\cM^k_{\rm ADHM} \subset \widetilde{\cM}^k_{ADHM}(\zeta)$. Thinking of the equivariant integral as an
integral of a function with respect to a volume form, and noting that $\widetilde{\cM}^k_{ADHM}(\zeta) \setminus \cM^k_{\rm ADHM}$ has measure zero,
one can attempt to define the singular integral on $\cM^k_{\rm ADHM}$ by:
\begin{align}
\int_{\cM^k_{\rm ADHM}} e^{\omega + \mu_T} := \int_{\widetilde{\cM}^k_{\rm ADHM}(\zeta)} e^{\omega + \mu_T}~,
\end{align}
where we have used the same symbol for the symplectic (1,1)-form and its pull back. Of course this definition only makes sense if the right hand side is $\zeta$-independent. \\

Integrals of equivariant characteristic classes
over $\cM^k_{\rm ADHM}$ can be similarly written as integrals over the resolved space with pulled back equivariant classes as integrands.
For the special case of a pure $\N=2$ SYM (and $\N=2^*$ SYM), it turns out that the volume integral/instanton partition function defined above is $\zeta$-independent.\\

Computing the integral $\int_{\widetilde{\cM}^k_{\rm ADHM}} e^{\omega + \mu_T}$ can be done in two steps : firstly, consider the integral on the level set $\mu^{-1}_{\C}(0) \cap \mu^{-1}_{\R}(\zeta)$ and write it as an integral over $\cM^k$ which can be computed using the Duistarmaat-Heckman theorem of equivariant localization for a non-compact space \footnote{The extension of the Duistarmaat-Heckman theorem to non-compact \hk quotients was derived
in \cite{Moore:1997dj}. More rigorous treatment of the problem can be found in \cite{1993alg.geom..7005P, Martens:2006hu}.}.
Finally, integrate over the group $G=U(k)$. \\

Schematically, one has
 \begin{align}
 \int_{\widetilde{\cM}^k_{\rm ADHM}} e^{\omega + \mu_T} = \int \frac{D\phi}{Vol(G)} \int_{\cM^k} e^{\omega + \mu_T + \mu_{T_G}} = \int \frac{D\phi}{Vol(G)} \sum_F \int_F
 \frac{e^{\omega + \mu_T + \mu_{T_G}}}{e_{T \times T_G} (\nu_F)}~,
 \end{align}
 where $\phi$ lives in the Cartan subalgebra of $G=U(k)$ and $T_G = U(1)^k$. $F$ denotes the fixed point set under the $T \times T_G$-action on $\cM^k$, and
 $e_{T \times T_G} (\nu_F)$ is the equivariant Euler class of the normal bundle at $F$. Since $\cM^k$ is non-compact, there is an additional restriction on the quantity on the RHS of the last equality i.e. the equivariant parameters lie in a open cone $C$ -- this is precisely the set of all parameters for which the RHS converges \cite{1993alg.geom..7005P, Martens:2006hu}. The choice of this cone $C$ depends on the sign of the FI parameter. \\

 We specialize to the case relevant for the Nekrasov partition function, where $F$ consists of a single point since only the origin is preserved under the full $T \times T_G$-action, and the denominator then is a product of weights of the  $T \times T_G$-action on the tangent space at the origin.
 It is useful to describe the integral over $\cM^k$ in the cohomological QFT approach of \cite{Shadchin:2005mx, Moore:1997dj}  (see \cite{Cordes:1994fc}
 for more background)
 where the above integral is written in terms of the ADHM variables $\{B_1,B_2,I,J\}$ (and their superpartners) as well as certain auxiliary multiplets $(\chi_\R,H_\R)$
 and $(\chi_\C,H_\C)$ (with $\chi$ fermionic and $H$ bosonic) which implement restriction of the fields to the level set $\mu^{-1}_{\C}(0) \cap \mu^{-1}_{\R}(\zeta)$. In this language, the above integral can be packaged into a contour integral, i.e.
 \begin{align}\begin{split}
  \int_{\widetilde{\cM}^k_{\rm ADHM}} &e^{\omega + \mu_T} = \oint_{JK(\zeta)} \frac{\prod^k_{i=1}\de \phi_i}{k!} \prod_{\Phi}\Big(\frac{1}{\prod_{P} w_P ^{T\times T_G}(\Phi)} \Big)^{\epsilon_{\Phi}}|_{\Phi \in \{B_1,B_2,I,J, \chi_{\R}, \chi_{\C}\}}~,\\
  =& \oint_{JK(\zeta)} \frac{\prod^k_{i=1}\de \phi_i}{k!} \Big(\frac{\prod_{i,j} w_i ^{T\times T_G}(\chi_{\R}) w_j ^{T\times T_G}(\chi_{\C})}{  \prod_{k,l,m,n} w_k ^{T\times T_G}(B_1) w_l^{T\times T_G}(B_2) w_m ^{T\times T_G}(I) w_n ^{T\times T_G}(J)} \Big)~,
\end{split} \end{align}
 where the integrand involves the weights $w_P ^{T \times T_G}(\Phi)$ of the ADHM variables and constraints under the torus action $T \times T_G$ at the origin, with $P$ labelling the individual weights of an ADHM variable $\Phi$ under the torus action (see equation \eref{equiv.wts.} below). Also, $\epsilon_{\Phi} \in \{\pm 1\}$ denotes the fermionic parity, and in writing the second equality we have used the fact that $\{B_1,B_2,I,J\}$ are bosonic while $\{\chi_\R, \chi_\C\}$ are
 fermionic. In the second equality, the indices $i,j,k,l,m,n$ run over the non-zero weights of the respective ADHM fields and constraints as indicated.\\

 The residues of the contour integral should be computed using the Jeffrey-Kirwan (JK) prescription
 \cite{1996alg.geom..8029J} (reviewed in appendix \ref{Loc-app})-- this is inherited from the restriction of the set of equivariant parameters to a cone $C$ \cite{1993alg.geom..7005P, Martens:2006hu}. It can be shown that the JK prescription is equivalent to the standard Young diagram rule for computing these integrals \cite{Hwang:2014uwa}.\\

Finally, one needs to compute the weights $w_P ^{T \times T_G} (\Phi)$.  Given $u =e^{\I a_l T^l}\in T_{U(N)}$ and $g = e^{\I \phi_i H^i} \in T_G$,
the action of $T \times T_G$ on the ADHM variables and constraints is\footnote{ The following is that action on the ADHM variables defining $\cM^k_{ADHM}(\zeta)$. In order
to define the resolution of singularities one uses geometric invariant theory and hence the lifting of the
$T$-action under the resolution of singularities is not simple in terms of ADHM variables
defining the hyperkahler quotient.}
 \begin{align}\begin{split}\label{equiv.trans.}
 & B_1 \to e^{\I \epsilon_1} g \,B_1\, g^{-1}\quad, \qquad I \to e^{\I \epsilon_+} g \,I \,u^{-1}\quad,\qquad \chi_{\C} \to e^{\I(\epsilon_1+ \epsilon_2)} g\, \chi_{\C} \,g^{-1},\\
 & B_2 \to e^{\I \epsilon_2} g \, B_2\, g^{-1}\quad, \qquad J \to  e^{\I \epsilon_+} u\, J \,g^{-1}\quad,\qquad  \chi_{\R} \to g \,\chi_{\R} \, g^{-1}~,
\end{split} \end{align}
and the weights $w_P ^{T \times T_G} (\Phi)$ can be read off as follows:

\begin{align}\begin{split}\label{equiv.wts.}
B_s  :& \quad \epsilon_s + \phi_i -\phi_j , \,\, \forall i,j, \,\, (s=1,2) ,\\
I :& \quad \epsilon_+ - a_l + \phi_i  , \,\, \forall i, \, \,\, \forall l\quad, \qquad \chi_{\C}  : \quad \epsilon_1+ \epsilon_2+ \phi_i- \phi_j , \,\, \forall i,j~,\\
J  :&\quad \epsilon_+ + a_l -\phi_i, \,\, \forall i, \, \,\, \forall l\quad,\qquad
\chi_{\R}  : \quad \phi_i- \phi_j, \,\, \forall i \neq j.
\end{split}\end{align}

 Putting everything together, we get the final expression for the $T$-equivariant volume
 \begin{align}\begin{split}
 Z^{4d}_{k\,{\rm inst}}:&=  \int_{\widetilde{\cM}^k_{\rm ADHM}} e^{\omega + \mu_T} \\
  =  \oint_{JK(\zeta)}& \frac{\prod^k_{i=1}\de \phi_i}{k!}\frac{\prod_{i\neq j} (\phi_i- \phi_j) \prod_{i, j} (\phi_i- \phi_j +\epsilon_1+ \epsilon_2)}{\prod_{i,j} (\phi_i -\phi_j + \epsilon_1) (\phi_i -\phi_j + \epsilon_2)\prod_{i,l} (\phi_i - a_l + \epsilon_+) (\epsilon_+ + a_l -\phi_i)}~,
\end{split} \end{align}
 where the residues are given by the JK prescription, or equivalently by the Young diagram rule. \\

 \subsubsection{5d partition function}
Instanton partition functions of theories (with or without matter) on $S^1 \times \C^2$ are given by integrals of $T$-equivariant characteristic classes, and can be similarly expressed as contour integrals. Consider a $T$-equivariant characteristic class $F_{T}(T\cM^k_{\rm ADHM})$ given as a function of Chern roots $\{x_1,\ldots, x_d\}$, i.e. $F_T(T\cM^k_{\rm ADHM}) =\prod^d_{i=1} F(x_i)$, where $\mbox{ch}_T(T\cM^k_{\rm ADHM}) =\sum^d_{i=1} e^{x_i}$ and $d= \mbox{dim} (\widetilde{\cM}^k_{\rm ADHM})$. Proceeding in the same fashion as before, the corresponding contour integral is of the following form:
 \begin{align}\begin{split} \label{eqv2con-general}
 \int_{\widetilde{\cM}^k_{\rm ADHM}(\zeta)}& e^{\omega + \mu_T} F_T(T\widetilde\cM^k_{\rm ADHM}(\zeta)) \\
 &= \oint_{JK(\zeta)} \frac{\prod^k_{i=1}\de \phi_i}{k!} \prod_{\Phi} \Big(\frac{\prod_{P} F(w_P ^{T\times T_G}(\Phi))}{\prod_{P} w_P ^{T\times T_G}(\Phi)} \Big)^{\epsilon_{\Phi}}\Big{|}_{\Phi \in \{B_1,B_2,I,J, \chi_{\R}, \chi_{\C}\}}~,
\end{split} \end{align}
where, as before, the statistics of the field has to be taken into account while unpacking the integrand. \\

In an $\N=1$ theory on $S^1 \times \C^2$ with hypermultiplets in a representation $\cR$, the BPS equations of the $\Omega$-deformed path integral consists of the self-duality equation for the gauge fields on $\R^4$ as well as a Dirac equation in the instanton background, where the connection transforms in the representation $\cR$ of the gauge group.
Let $\V({\cR})$ be a vector bundle over $\widetilde{\cM}^k_{\rm ADHM}$  such that the fiber at a given point $m \in \widetilde{\cM}^k_{\rm ADHM}$ is the index of the Dirac operator in the instanton background (labelled by $m$) with the connection transforming in the representation $\cR$.
The instanton partition function therefore involves equivariant characteristic classes of these vector bundles $\V({\cR})$, and we will write the corresponding equivariant integral momentarily. The weights of the torus action on these bundles can be read off from the equivariant index of the
Dirac operator, which in turn can be computed from the Chern character of the universal bundle -- we refer the reader to the papers
\cite{Shadchin:2004yx, Shadchin:2005mx, Losev:2003py, Losev:1997tp, Lossev:1997bz} for details.\\

The equivariant integral formula for the partition function is most conveniently read off from the contour integral formula of the Witten index of the associated SQM. For a hyper in an arbitrary representation $\cR$ of a $U(N)$ gauge group, we have
\begin{align}\begin{split}\label{chclass-genrep}
&Z^{5d}_{\rm inst}(\epsilon_{1,2}, \vec a, \vec m; \zeta) = \sum_{k \geq 0} \fq^k \int _{\widetilde{\cM}^k_{\rm ADHM}} e^{\omega + \mu_T} \widehat{A}_{T } (T\widetilde{\cM}^k_{\rm ADHM})
\cdot C_{T  \times T_F} (\V(\cR))~, \\
&= \sum_{k \geq 0} \fq^k \oint_{JK(\zeta)} \frac{\prod^k_{i=1}\de \phi_i}{k!} \prod_{\Phi} \Big(\frac{\prod_{P} A(w_P ^{T\times T_G}(\Phi))}{\prod_{P} w_P ^{T\times T_G}(\Phi)} \Big)^{\epsilon_{\Phi}}
\times \prod_{K} \Big(\prod_{P'} \sinh{w_{P'}^{T \times T_G \times T_F}(K)} \Big)^{\epsilon_{K}} ~,
\end{split}\end{align}
where $\Phi \in \{B_1,B_2,I,J, \chi_{\R}, \chi_{\C} \}$ is the set of ADHM variables and constraints for a pure SYM, $K$ denotes ADHM variables
which parametrize the hypermultiplet zero modes, and $\epsilon_{\Phi} \in \{\pm 1\}$,  $\epsilon_{K} \in \{\pm 1\}$ denote the fermionic parity of the
set of fields $\{\Phi\}$ and $\{K\}$ respectively. Additionally, $T_F$ indicates that we also work equivariantly with respect to flavor symmetry.

The equivariant characteristic classes
$\widehat{A}_{T } (T\widetilde{\cM}^k_{\rm ADHM})$ and $C_{T \times T_F} (\V(\cR))$, and the function $A(x)$, are defined as
\begin{align}\begin{split}
& \widehat{A}_{T } (T\widetilde{\cM}^k_{\rm ADHM}) = \prod^d_{i=1} \frac{ x_i}{e^{ x_i/2} - e^{- x_i/2}}:= \prod^d_{i=1} A(x_i)\quad, \quad  \mbox{ch}_{T }(T\widetilde{\cM}^k_{\rm ADHM}) =\sum^{d}_{i=1} e^{x_i}~,\\
& C_{T  \times T_F} (\V(\cR))=  \prod^{{d}_{\cR}}_{i=1} 2 \sinh{\frac{\xi_i}{2}}, \quad \mbox{ch}_{T  \times T_F}(\V(\cR)) =\sum^{{d}_{\cR}}_{i=1} e^{\xi_i}~,
\end{split}\end{align}
where $T_F$ is the maximal torus of the flavor symmetry group associated with the hypermultiplet, $d= {\rm dim}(T\widetilde{\cM}^k_{\rm ADHM})$
and ${d}_{\cR}= {\rm dim}(\V(\cR))$.\\

For a  pure 5d $\N=1$ SYM, the integral involves the $T$-equivariant A-roof genus:
\begin{align}
Z^{5d}_{\rm inst}(\epsilon_{1,2}, \vec a; \zeta)= \sum_{k \geq 0} \fq^k \int _{\widetilde{\cM}^k_{\rm ADHM}} e^{\omega + \mu_T} \widehat{A}_T (T\widetilde{\cM}^k_{\rm ADHM})~,
\end{align}

From the general formula (\ref{eqv2con-general}), we have
\begin{align}\begin{split}
& \int _{\widetilde{\cM}^k_{ADHM}} e^{\omega + \mu_T} \widehat{A}_T (T\widetilde{\cM}^k_{ADHM}) \\
=&   \oint_{JK(\zeta)} \frac{\prod^k_{i=1}\de \phi_i}{k!} \prod_{\Phi}\Big(\frac{1}{\prod_{P} w_P ^{T\times T_G}(\Phi)} \Big)^{\epsilon_{\Phi}} \prod_{P} \Big(\frac{ w^{T\times T_G}_P(\Phi)}{e^{ \half w_P ^{T\times T_G}(\Phi)} - e^{-\half w_P ^{T\times T_G}(\Phi)} }\Big)^{\epsilon_{\Phi}}\Big{|}_{\Phi \in \{B_1,B_2,I,J, \chi_{\R}, \chi_{\C}\}} \nonumber \\
=& \oint_{JK(\zeta)} \frac{\prod^k_{i=1}\de \phi_i}{k!} \prod_{\Phi} \prod_{P} \Big(\frac{1}{e^{ \half w_P ^{T\times T_G}(\Phi)} - e^{-\half w_P ^{T\times T_G}(\Phi)} }\Big)^{\epsilon_{\Phi}}\Big{|}_{\Phi \in \{B_1,B_2,I,J, \chi_{\R}, \chi_{\C}\}} \nonumber \\
= & \oint_{JK(\zeta)} \frac{\prod^k_{i=1}\de \phi_i}{k!} \frac{\prod_{i\neq j} \sinh \half (\phi_i- \phi_j) \prod_{i, j} \sinh \half(\phi_i- \phi_j +\epsilon_1+ \epsilon_2)}{\prod_{i,j} \sinh \half(\phi_i -\phi_j + \epsilon_1) \sinh \half(\phi_i -\phi_j + \epsilon_2)\prod_{i,l} \sinh \half(\pm(\phi_i - a_l)+\epsilon_+ )}~.
\end{split}\end{align}

 For matter multiplets, one can read off the weights $w_{P'}^{T \times T_G \times T_F}(K)$ from the Chern character $\mbox{ch}_{T  \times T_G \times T_F}(\V(\cR))$, and these were computed for various representations and gauge groups
 in \cite{Shadchin:2005mx}. For example, in $\N=1^*$ theory, one can show that
 \be\label{adj-char}
 \begin{split}
 & \mbox{ch}_{T  \times T_F}(\V(\cR={\rm adj})) =e^{m}\mbox{ch}_{T }(T\widetilde{\cM}^k_{\rm ADHM})
  = \sum_i e^{(x_i +m)} ~,\\
 \implies & C_{T  \times T_F} (\V(\cR={\rm adj})) =  \prod^d_{i=1} 2\sinh{\frac{(x_i +m)}{2}},
 \quad \text{where}\,\, \mbox{ch}_{T }(T\widetilde{\cM}^k_{\rm ADHM})= \sum_i e^{x_i}~.
  \end{split}
  \ee
 Therefore the integrand in \eref{chclass-genrep} can be combined to give $T$-equivariant $\widehat{\chi}_y$ genus
 \begin{align}\begin{split}\label{chiygenus1}
& Z^{5d}_{\rm inst}(\epsilon_{1,2}, \vec a, m; \zeta)=\sum_{k \geq 0} q^k \int _{\widetilde{\cM}^k_{\rm ADHM}} e^{\omega + \mu_T} \widehat{\chi}_{y,\,T} (T\widetilde{\cM}^k_{\rm ADHM})~, \\
& \widehat{\chi}_{y,\,T} (T\widetilde{\cM}^k_{\rm ADHM})=  \prod^d_{i=1} \frac{(y e^{x_i/2}- y^{-1} e^{-x_i/2} )  x_i}{e^{x_i/2} - e^{- x_i/2}}~,
\end{split}\end{align}
where $y =e^{m/2}$ and $\{x_1,\ldots, x_d\}$ are Chern roots as before.
Again using the general formula \eref{eqv2con-general}, we get
 \begin{align}\label{eqv-adj}
& \int _{\cM^k_{ADHM}} e^{\omega + \mu_T} \widehat{\chi}_{y,\,T} (T\cM^k_{ADHM}) \nonumber \\
=& \oint_{JK(\zeta)} \frac{\prod^k_{i=1}\de \phi_i}{k!} \prod_{\Phi}  \prod_{P} \Big(\frac{ (e^{ \half (w_P ^{T\times T_G}(\Phi)+m)} - e^{-\half (w_P ^{T\times T_G}(\Phi)+m)})}{e^{ \half w_P ^{T\times T_G}(\Phi)} - e^{-\half w_P ^{T\times T_G}(\Phi)}}\Big)^{\epsilon_{\Phi}}|_{\Phi \in \{B_1,B_2,I,J, \chi_{\R}, \chi_{\C}\}} \nonumber \\
= &  \oint_{JK(\zeta)} \frac{\prod^k_{i=1}\de \phi_i}{k!} \frac{\prod_{i\neq j} 2 \sinh \half(\phi_i- \phi_j) \prod_{i, j} 2\sinh \half(\phi_i- \phi_j +\epsilon_1+ \epsilon_2)}{\prod_{i,j} 2 \sinh \half(\phi_i -\phi_j + \epsilon_1) 2\sinh \half(\phi_i -\phi_j + \epsilon_2)\prod_{i,l} 2 \sinh \half(\pm(\phi_i - a_l)+\epsilon_+)} \nonumber \\
 \times &\frac{\prod_{i,j} 2 \sinh \half(\phi_i -\phi_j +m + \epsilon_1) 2\sinh \half(\phi_i -\phi_j +m + \epsilon_2)\prod_{i,l} 2\sinh \half(\pm(\phi_i - a_l)+m+\epsilon_+)}{\prod_{i,j} 2\sinh \half(\phi_i- \phi_j+m ) \prod_{i, j} 2\sinh \half(\phi_i- \phi_j +m+\epsilon_1 + \epsilon_2)}~,
\end{align}
The expression matches with \eref{D0quiver-basic} after a redefinition of the adjoint mass $m \to m - \epsilon_+$.\\

 \subsubsection{Transformation of the equivariant integrals under $\zeta \to -\zeta$ and wall-crossing}\label{WC-2}
 We now describe how the contour integral expressions for 4d/5d instanton partition function change
 under a change in the sign of the real FI parameter $\zeta$ in \eref{res-adhm}. The moment maps in the
 ADHM construction are then given as:
 \begin{align}\begin{split}
 \mu_{\mathbb C} &\equiv  [B_1,B_2]+IJ=0~,\\
\mu_{\mathbb R} &\equiv  [B_1^\dagger, B_1]+[B_2^\dagger, B_2]+II^\dagger-J^\dagger J= -\zeta\quad,\quad\zeta >0~.
\end{split}\end{align}

 Define a new set of ADHM variables : $\Big(\widetilde{I}, \widetilde{J}, \widetilde{B_1}, \widetilde{B_2}\Big)$, such
 that
 \begin{align}\begin{split}
 & \widetilde{B_1}= B^\dagger_2\quad,\qquad \widetilde{I} =J^\dagger~, \\
 & \widetilde{B_2}= B^\dagger_1\quad,\qquad  \widetilde{J}=J^\dagger~.
\end{split} \end{align}
 In terms of the variables $\Big(\widetilde{I}, \widetilde{J}, \widetilde{B_1}, \widetilde{B_2}\Big)$,
 the moment maps can be written as:
\begin{align}\begin{split}
\widetilde{\mu}_\C&= \mu^\dagger_{\mathbb C} \equiv  [\widetilde B_1, \widetilde B_2]+\widetilde I \widetilde J=0~,\\
\widetilde{\mu}_\R &= - \mu_{\mathbb R} \equiv  [\widetilde B_1^\dagger, \widetilde B_1]+[\widetilde B_2^\dagger, \widetilde B_2]+\widetilde I \, \widetilde I^\dagger- \widetilde J^\dagger \widetilde J= \zeta~.
\end{split}\end{align}

The $T \times T_G$ group action on the ADHM variables and constraints is then given as:
 \begin{align}\begin{split}\label{equiv.trans.neg.}
 & \widetilde B_1 \to e^{- \I \epsilon_2} g \,\widetilde B_1\, g^{-1}\quad,\qquad
\widetilde I \to e^{-\I \epsilon_+} g \,\widetilde I \,u^{-1}\quad,\qquad
 \widetilde \chi_{\C} \to e^{-\I(\epsilon_1+ \epsilon_2)} g\, \chi_{\C} \,g^{-1}~,\\
 & \widetilde B_2 \to e^{-\I \epsilon_1} g \, \widetilde B_2\, g^{-1}\quad,\qquad
 \widetilde J \to  e^{-\I \epsilon_+} u\, \widetilde J \,g^{-1}\quad,\qquad
  \widetilde \chi_{\R} \to g \,\chi_{\R} \, g^{-1}~.
\end{split} \end{align}

Comparison with \eref{equiv.trans.} shows that the group action above is identical, with $\epsilon_+ \to - \epsilon_+$.
Therefore, a change of sign in $\zeta$ in the ADHM moduli space \eref{res-adhm} leads to exactly the same manifold
with an almost identical group action -- the only difference being a change of sign in the equivariant parameter $\epsilon_+$.
The equivariant weights of the ADHM variables can be obtained from those in \eref{equiv.wts.}
after the transformation $\epsilon_+ \to - \epsilon_+$. The equivariant weights associated to the matter multiplets
can be read off from the original ones after substituting $\epsilon_+ \to - \epsilon_+$.\\

The integrand of the contour integral for a 5d partition function in the $-\zeta$-chamber can be obtained from
the $\zeta$-chamber integrand by substituting $\epsilon_+ \to - \epsilon_+$, while  the JK-residue should be
taken w.r.t. $\zeta$ (and not $-\zeta$). As an example, consider the 5d instanton partition function for a pure $\N=1$ $SU(N)$ SYM:
\begin{align}
& Z^{5d}_{\rm k-inst}(\epsilon_{+}, \epsilon_{-}, \vec a; - \zeta) =\int _{\widetilde{\cM}^k_{ADHM}(-\zeta)} e^{\omega + \mu_T} \widehat{A}_T (T\widetilde{\cM}^k_{ADHM}) \nonumber \\
= & \oint_{JK(\zeta)} \frac{\prod^k_{i=1}\de \phi_i}{k!} \frac{\prod_{i\neq j} \sinh \half (\phi_i- \phi_j) \prod_{i, j} \sinh \half(\phi_i- \phi_j - \epsilon_1 - \epsilon_2)}{\prod_{i,j} \sinh \half(\phi_i -\phi_j - \epsilon_2) \sinh \half(\phi_i -\phi_j - \epsilon_1)\prod_{i,l} \sinh \half(\pm(\phi_i - a_l) -\epsilon_+ )} \nonumber \\
=&  Z^{5d}_{\rm k-inst}(- \epsilon_{+}, \epsilon_{-}, \vec a; \zeta)~.
\end{align}

On evaluating the contour integral, one can check that $Z^{5d}_{\rm k-inst}$ is an even function of $\epsilon_+$, i.e.
\be
Z^{5d}_{\rm k-inst}(- \epsilon_{+}, \epsilon_{-}, \vec a; \zeta) = Z^{5d}_{\rm k-inst}(\epsilon_{+}, \epsilon_{-}, \vec a; \zeta)~,
\ee
which implies that it is wall-crossing invariant.\\

For an $\N=1^*$ $SU(N)$ theory, equation \eref{adj-char} implies that the equivariant weights associated with the adjoint hypermultiplet
are related to those of the vector multiplet by an overall shift of the adjoint mass $m$. As discussed above, the partition function is then
obtained from \eref{eqv-adj} after shifting the adjoint mass $m$: $m \to m -\epsilon_+$.
Under a transformation $\zeta \to -\zeta$, the instanton partition function is given as:
\begin{align}\begin{split}
 Z^{5d}_{\rm k-inst}(\epsilon_{+}, \epsilon_{-}, \vec a, m ; - \zeta) = & \int _{\widetilde{\cM}^k_{ADHM}(-\zeta)} e^{\omega + \mu_T}\widehat{\chi}_{y,\,T} (T\widetilde{\cM}^k_{\rm ADHM})\\
 =& Z^{5d}_{\rm k-inst}(-\epsilon_{+}, \epsilon_{-}, \vec a, m ; \zeta)~.
\end{split}\end{align}
As before, on computing the contour integral explicitly, one can check that $Z^{5d}_{\rm k-inst}$ for $\N=1^*$ $SU(N)$ theory is invariant under wall-crossing.\\

Wall-crossing invariance of the 5d instanton partition function for the $\N=1^*$ theory and the pure $\N=1$ SYM
can be checked (without actually performing the contour integrals) as follows.
Consider first the $\N=1^*$ theory in the instanton sector $k=1$ which is associated
with an Abelian SQM. In this case, the wall-crossing formula of the Witten index can be read off from \eref{JK-Abelian}:
\be
\begin{split}
&Z^{(4,4)}_{\rm SQM} (a,m, \epsilon_\pm; \zeta' < 0) - Z^{(4,4)}_{\rm SQM} (a,m, \epsilon_\pm; \zeta' > 0)\\
=& -\sum_{\phi_* \in \Delta^-} {\rm Res}_{\phi=\phi_*}\Big[Z_{\rm 1-loop} \frac{\de \phi}{2\pi \I}\Big]
- \sum_{\phi_* \in \Delta^+} {\rm Res}_{\phi=\phi_*}\Big[Z_{\rm 1-loop} \frac{\de \phi}{2\pi \I}\Big]\\
=& R_{-\infty} + R_{\infty}~,
\end{split}
\ee
where $R_{\pm \infty}$ are the residues of $\Big[Z_{\rm 1-loop} \frac{\de \phi}{2\pi \I}\Big]$ at $\phi=\pm \infty$.
The sum $(R_{-\infty} + R_{\infty})$ vanishes for the (4,4) ADHM SQMs associated
with instanton particles in 5d $\N=1^*$ $SU(N)$ SYM on $S^1 \times \C^2$ or $S^1 \times \C^2/\Z_n$,
which can be directly checked from the Abelian version of \eref{D0quiver-basic} and \eref{D0quiver-orbifold} respectively.
For generic $k$, the change in the Witten index as $\zeta \to -\zeta$ is similarly given by a sum over the various asympototic residues
(i.e. when one or more of the $\phi_I$s or $\phi^i_I$s go to $\pm \infty$). However, from equation \eref{D0quiver-basic} and \eref{D0quiver-orbifold},
one can directly check that the residues for a given $\phi_I$ (or $\phi^i_I$) from $\pm \infty$ (with other integration variables generic) cancel
against each other. Therefore, the sum over the asymptotic residues vanish as in the case of $k=1$ leading to a wall-crossing invariant Witten index.

For the pure $\N=1$ SYM, in the instanton sector k=1, the residues  $R_{-\infty}$ and $R_{\infty}$ vanish individually.
For generic $k$, the asymptotic residues also vanish individually since the residues associated with any $\phi_I \to \pm \infty$
or $\phi^i_I \to \pm \infty$, with other integration variables generic, is zero. Therefore, we also have a wall-crossing invariant Witten index
in this case.

 \subsection{Equivariant integrals for $Z_{\rm mono}$ and 5d instantons on $S^1 \times \C^2/\Z_n$}\label{KN-equiv}
Let us review the equivariant integral formula for the 4d instanton partition function of a pure $\N=2$ $U(N)$ SYM on an orbifold $\C^2/\Z_n$
and show how it reduces to a contour integral using the cohomological QFT approach \cite{Moore:1997dj}. The moduli space of instantons
on $\C^2/\Z_n$ can be constructed as a \hk quotient of the $\Z_n$-invariant ADHM data, as reviewed in section \ref{KN-review}.
As in the case of the ADHM construction of instanton moduli space on $\C^2$, the Uhlenbeck compactification of the moduli space $\cM_{KN}$
is singular. The smooth resolution in this case is the moduli space of $\Z_n$-equivariant torsion free sheaves on $\mathbb{C} \mathbb{P}^2$
with fixed framing at the line at infinity \cite{Fujii:2005dk, Bruzzo:2014jza, Nakajima:2005fg}. The resolved space $\widetilde{\cM}_{KN}(\zeta^i_{\R})$ can again
be described as a \hk quotient after introducing stability/FI parameters which deform the real moment map as follows:
\begin{eqnarray}\label{res-adhm-2}
 \widetilde{\cM}_{KN}(\zeta^i_{\R})
 \cong
\left   \{
 (B_1,B_2,I,J)_{\Z_n}
 \left|
\begin{array}{cccc}
 \mu_{\C}&=&0& \\
  \mu_{\R}&=&\zeta_{\R} \\
 \end{array}
\right.
\right\}
{\Big /}
U(k,\mathbb C)~ ,
\end{eqnarray}
where we only consider $\zeta_\R$ in the set\footnote{The cone associated to the other chamber for $\zeta_\R$ discussed above corresponds to taking $\zeta^i_\R>0$, $\forall i$.}
\begin{align}
C_0 := \{\zeta_{\R} = (\zeta^i_{\R}) \in \R^n | \zeta^i_{\R} < 0 ,\,\, \forall i=1,\ldots,n \}~.
\end{align}
Given the above definition, equivariant integrals on
$\widetilde{\cM}_{KN}(\zeta^i_{\R})$ can be written as $\Z_n$--invariant projections of equivariant integrals on the moduli
space of instantons on $\C^2$. Similar to the case of instantons on $\C^2$, such equivariant integrals may be written as
contour integrals using the cohomological QFT prescription discussed before. These contour integrals coincide with the
Witten index formula for the ADHM SQM in the Higgs scaling limit, i.e.  $e^2 \to 0$ and $|\zeta| \to \infty$
(after setting the gauge couplings $e_i=e$, and $\zeta^i_\R = \zeta$, for all $i$) holding $\zeta'= e^2 \zeta$ fixed.
The instanton partition function therefore depends only on the sign of $\zeta$ or $\zeta'$.

The contour integral can be constructed using the orbifold-invariant ADHM variables
\begin{align}\begin{split}\label{ADHM-inv data}
 & B_1  \in \bigoplus^{q_{max}-1}_{j=q_{min}} \text{Hom}(V_{j+1}, V_{j})\quad~~~,\qquad  I  \in  \bigoplus^{q_{max}}_{j=q_{min}} \text{Hom}(V_{j}, W_{j} )~,\\
 & B_2  \in  \bigoplus^{q_{max}}_{j=q_{min}+1} \text{Hom}(V_{j-1}, V_{j})\quad,\qquad
  J  \in  \bigoplus^{q_{max}}_{j=q_{min}} \text{Hom}(W_{j}, V_{j})~.
\end{split}\end{align}
and the fields imposing the moment map equations
\begin{align}
(\chi_\R,\chi_\C) \in   \bigoplus^{q_{max}}_{j=q_{min}} \Big(\text{Hom}(V_{j}, V_{j}), \text{Hom}(V_{j}, V_{j})\Big)~.
\end{align}
The generating function for 5d instanton partition functions on $S^1 \times \C^2/\Z_n$ with a monodromy vector $\vec w$ at spatial infinity can be written as
\begin{align}
\widetilde{Z}^{S^1 \times \C^2/\Z_n}_{\rm inst}(\epsilon_{1,2}, \vec a, \vec m; \,\zeta\, |\, \vec w) = \sum_{\vec k} \fq^k \prod^{q_{max}}_{j=q_{min}} u_j^{\beta_j} \,\, Z^{S^1 \times \C^2/\Z_n}_{\rm inst}(\vec k, \vec w; \vec a, \vec m; \,\zeta\,)~,
\end{align}
where the sum is over $\vec{k}$ such that $ \sum^{q_{max}}_{j=q_{min}} k_j=k$, $\beta_j = w_j + k_{j-1} +k_{j+1} - 2k_j$ (these $\beta_j$s are the beta functions of $j$-th gauge
node of the quiver), $\fq$ is the fugacity associated with the instanton number, and $u_j$ are fugacities associated with the second Chern class
of the instanton bundle (see \eref{cc}).
The instanton partition function labelled by the KN vector $\vec k$ and the monodromy vector $\vec w$ is
\begin{align}\begin{split}\label{chclass-genrep-KN}
& Z^{S^1 \times \C^2/\Z_n}_{\rm inst}(\vec k, \vec w; \vec a, \vec m; \,\zeta\,) = \int _{\widetilde{\cM}_{KN}(\zeta')} e^{\omega + \mu_T} \widehat{A}_{T } (T\widetilde{\cM}_{KN}(\zeta^i_{\R}))
\cdot C_{T  \times T_F} (\V_{KN}(\cR)) \\
&=  \oint_{JK(\zeta)} \frac{1}{\prod^{q_{max}}_{i=q_{min}}k_i!} \prod^{q_{max}}_{i=q_{min}} \prod^{k_i}_{I=1} \frac{d\phi^i_I}{2\pi i}\prod_{\Phi} \Big(\frac{\prod_{P} A(w_P ^{T\times T_G}(\Phi))}{\prod_{P} w_P ^{T\times T_G}(\Phi)} \Big)
\times \prod_{K} \Big(\prod_{P'} \sinh{ \half w_{P'}^{T \times T_G \times T_F}(K)} \Big)~,
\end{split}\end{align}
where, as before, $\{\Phi\}$ runs over the $\Z_n$ invariant ADHM variables $\{B_1,B_2, I, J\}$ while $\{K\}$ parametrizes the
$\Z_n$ invariant zero modes arising from the hypermutiplets in representation $R$ in the ADHM construction.
$\widehat{A}_{T } (T\widetilde{\cM}_{KN}(\zeta^i_{\R}))$ is the A-roof genus and $C_{T  \times T_F} (\V_{KN}(\cR))$ is the characteristic class
associated with the matter bundle:
\begin{align}\begin{split}
& \widehat{A}_{T } (T\widetilde{\cM}_{KN}(\zeta^i_{\R})) = \prod^d_{i=1} \frac{ x_i}{e^{ x_i/2} - e^{- x_i/2}}:= \prod^d_{i=1} A(x_i)\quad, \qquad  \mbox{ch}_{T }(T\widetilde{\cM}_{KN}(\zeta^i_{\R})) =\sum_i e^{x_i}~,\\
& C_{T  \times T_F} (\V_{KN}(\cR))=  \prod^{{d}_{\cR}}_{i=1} \sinh{\half \xi_i}, \quad \mbox{ch}_{T  \times T_F}(\V_{KN}(\cR)) =\sum_i e^{\xi_i}~,
\end{split}\end{align}
where $T_F$ is the maximal torus of the flavor symmetry group associated with the hypermultiplet, $d= {\rm dim}(T\widetilde{\cM}_{KN}(\zeta^i_{\R}))$
and ${d}_{\cR}= {\rm dim}(\V_{KN}(\cR))$. In particular, for the case of a 5d $\N=1^*$ theory where $\cR$ is adjoint, the characteristic classes in the integrand
can be combined to give a $T$-equivariant $ {\chi}_y$-genus of the KN moduli space, which can be written as a contour integral, i.e.
 \begin{align}\label{chiy genus-KN}
&Z^{S^1 \times \C^2/\Z_n}_{\rm inst}(\vec k, \vec w; \vec a, \vec m; \,\zeta\,)  = \int _{\widetilde{\cM}_{KN}(\zeta')} e^{\omega + \mu_T}  {\chi}_y  \nonumber \\
&= \oint_{JK(\zeta)} \frac{1}{\prod^{q_{max}}_{i=q_{min}}k_i!} \prod^{q_{max}}_{i=q_{min}} \prod^{k_i}_{I=1} \frac{d\phi^i_I}{2\pi i} \prod_{\Phi}  \prod_{P} \frac{ (e^{ \half (w_P ^{T\times T_G}(\Phi)+m)} - e^{-\half (w_P ^{T\times T_G}(\Phi)+m)})}{e^{ \half w_P ^{T\times T_G}(\Phi)} - e^{-\half w_P ^{T\times T_G}(\Phi)}}|_{\Phi \in \{B_1,B_2,I,J, \chi_{\R}, \chi_{\C}\}_{\Z_n}} \nonumber \\
&\stackrel{m \to m- \epsilon_+}{=}  \oint_{JK(\zeta)} \frac{1}{\prod^{q_{max}}_{i=q_{min}}k_i!} \prod^{q_{max}}_{i=q_{min}} \prod^{k_i}_{I=1} \frac{d\phi^i_I}{2\pi i} \, \prod^{q_{max}}_{i=q_{min}}\prod_{I,J=1}^{k_i} \frac{2\sinh \half(\phi^i_{IJ}+2\epsilon_+)}{2\sinh{\half(\phi^i_{IJ} + m  \pm \epsilon_+)}} \times \prod_{I \neq J}^{k_i} \sinh \half\phi^i_{IJ} \nonumber \\
& \times \prod^{q_{max} -1}_{j=q_{min}}\prod_{I=1}^{k_{j+1}} \prod_{J=1}^{k_j} \frac{2\sinh{ \half(\phi^{j+1}_{I}-\phi^j_{J} + m + \epsilon_-)} 2\sinh{ \half(\phi^{j}_{J}-\phi^{j+1}_{I} +m -\epsilon_-)}}{2\sinh \half(\phi^{j+1}_{I}-\phi^j_{J}+\epsilon_+ + \epsilon_-) 2\sinh \half(\phi^j_{J}-\phi^{j+1}_{I}+\epsilon_+ - \epsilon_-)}   \nonumber \\
& \times \prod^{q_{max}}_{j=q_{min}}\prod_{I=1}^{k_i}\prod_{l=1}^{w_i}\frac{2\sinh{ \half(\phi^i_I-a_l + m)} 2\sinh{ \half(-\phi^i_I + a_l + m)}}{2\sinh \half(\phi^i_I-a_l + \epsilon_+) 2\sinh \half(-\phi^i_I+a_l + \epsilon_+)} ~.
\end{align}
The last line of the above formula is precisely the same as equation \eref{D0quiver-orbifold}-\eref{explicitfunctions} above. From \eref{4d-5d}, we can
therefore write down a formula for $Z^{\R^3 \times S^1}_{\rm mono}$ as an equivariant integral on a resolved KN moduli space:
\be \label{chi-y-mono}
\boxed{Z^{\R^3 \times S^1}_{\rm mono}(B,\mathbf{v}; \fa,\fm, \lambda| G=SU(N)) = \int _{\widetilde{\cM}_{KN}(\zeta')} e^{\omega + \mu_T}  {\chi}_y|_{\vec k = \vec k(B,\mathbf{v}), \vec w=\vec w(B,\mathbf{v})}}
\ee
where the equivariant parameters on the two sides of the equation are related as in \eref{eqpar-map}.

The formula for these contour integrals under a change of sign of all the FI parameters, i.e. $\zeta \to -\zeta$ (or $\zeta' \to -\zeta'$),
can be obtained in a similar fashion as discussed in appendix \ref{WC-2} in the context of partition functions on $S^1 \times \R^4$.
The resultant contour integral can be obtained from the original one by substituting $\epsilon_+ \to -\epsilon_+$. One can check
that the expressions in \eref{chiy genus-KN} and \eref{chi-y-mono} are even functions of $\epsilon_+$ and therefore
invariant under $\zeta \to -\zeta$.

\section{Character Equation Analysis}
\label{app:AppG}

In this appendix we will derive equation \ref{eq:carefulanalysis}. Let us introduce the notation
\be
K={\rm diag}(K_1,K_2,...,K_k)\quad,\qquad x=e^{2\pi \I \nu}~,
\ee
where the entries can be repeated.
 In the character equation we will want to reduce the term
\be
(x+x^{-1}-2){\rm Tr}_V~x^K=(x+x^{-1}-2) \sum_{s=1}^{k}  x^{K_s}=(x+x^{-1}-2)\sum_{s=1}^{n-1}k_s x^s~,
\ee
where $\vec{k}=(k_0,...,k_{n-1})$. Note that generically
\be
k_i=k_{i+1}\pm 1\quad\text{or} \quad k_i=k_{i+1}~.
\ee
This means that the factor of $(x+x^{-1}-2)$ will actually eliminate most of the terms. Consider two sequence of $k's$: (a) $(k_s-1, k_s, k_s+1)$ and (b) $(k_s,k_s,k_s)$. In the case of (a), we have the terms of degree $x^n$ will cancel:
\begin{align}\begin{split}
x \cdot \Big[(k_s-1) x^{s-1}\Big] -2\cdot \Big[( k_s) x^s\Big]+x^{-1} \cdot \Big[(k_s+1) x^{s+1}\Big]\qquad\\
= (k_s-1) x^s -2 k_s x^s +(k_s+1) x^s =0~.
\end{split}\end{align}
Similarly for the case of (b) the terms of degree $x^s$ will cancel:
\begin{align}\begin{split}
x \cdot \Big[(k_s) x^{s-1}\Big] -2\cdot \Big[( k_s) x^s\Big]+x^{-1} \cdot \Big[(k_s) x^{s+1}\Big]\qquad\\
= (k_s) x^s -2 k_s x^s +(k_s) x^s =0~.
\end{split}\end{align}
This means that the product $(x+x^{-1}-2)$Tr$_k~ x^K$ will cancel order by order along the sequences of purely increasing, decreasing, or constant $k_s$'s respectively. Therefore, the only sequences where there is not a complete cancellation is at the connection between the quivers of type $\Sigma_i$ and $\Gamma_{j,j+1}$.

Now let us compute the terms which contribute to the character equation. There are 4 such sequences
\begin{align}\begin{split}
{\rm I}.)\quad (k_s-1,k_s, k_s)\qquad, \qquad {\rm II}.)&\quad (k_s,k_s,k_s-1)~,\\
 {\rm III}.) \quad (k_s+1, k_s,k_s)\qquad,~ ~\quad {\rm IV}.)&\quad  (k_s,k_s,k_s+1)~ ,
\end{split}\end{align}
where we have taken the middle term to be the $s$-th term in the vector $\vec k$.

Computing the terms of degree $s$ we see
\begin{align}\begin{split}\label{eq:cont}
{\rm I}.) &\quad (k_s-1) x^s -2 k_s x^s +k_s x^s=- x^s ~,\\
{\rm II}.)&\quad k_sx^s -2 k_s x^s + (k_s-1) x^s=-x^s~,\\
{\rm III}.) &\quad (k_s+1) x^s -2 k_s x^s +k_s x^s =x^s~,\\
{\rm IV}.) &\quad k_s  x^s -2 k_s x^s +(k_s+1)x^s=x^s~.
\end{split}\end{align}
Note that each term is $(+1)$ or $(-1)$ times a simple power of $x$. Therefore, we see that there will be a sum of monomials with positive or negative coefficient whose degree is the position along the full quiver of the beginning and end nodes of the $\Sigma_i$ subquivers.

Now to determine the contribution to the character equation, we must determine the generic positions of all of the $\Sigma_i$ quivers. Let us use the notation
\begin{align}\begin{split}
&B={\rm diag}(p_1,...,p_N)\qquad,\quad \mathbf{v}={\rm diag}(v_1,...,v_N)~,\\
&\kappa=\mathbf{v}-B=\sum_I \tilde{k}_I H_I={\rm diag}(\kappa_1,...,\kappa_N)~,
\end{split}\end{align}
where the $p_I$ and $v_I$ are non-decreasing.

First, note that in the case $\tilde{k}_I=\tilde{k}_{I+1}$, we have that there will be no $\Gamma_{I,I+1}$ subquiver, and consequently there will be no contribution from the pair of edges connection $\Sigma_I$ to $\Sigma_{I+1}$. This is okay though, because it means that there is a zero in the matrix $\kappa$ and hence there is a value of $p_i=v_i$ and hence the terms drop from the character equation. So therefore we will consider the generic case where $\tilde{k}_I\neq \tilde{k}_J$ for $I\neq J$.

Second, it is particularly insightful to consider the contributions from the terms surrounding a given $\Gamma_{I,I+1}$ for $I\neq 0,N-1$:

\begin{center}
\begin{tikzpicture}[node distance = 2.2cm, auto]
\node (2) [right of = 1] {};
\node (3) [hexs,right of =2,xshift=-0.5cm] {$\Sigma_{I}$};
\node (4) [hex,right of= 3] {$\Gamma_{I,I+1}$};
\node (5) [hexs,right of=4] {$\Sigma_{I+1}$};
\node (6) [right of=5,xshift=-0.5cm] {};
\draw[-] (2) --  (3);
\draw[-] (3) --  (4);
\draw[-] (4) --  (5);
\draw[-] (5) --  (6);
\end{tikzpicture}
\end{center}
Note that the length of $\Gamma_{I,I+1}$ and $\Sigma_I$ (denoted $|\Gamma_{I,I+1}|$ and $|\Sigma_I|$ respectively) are given by
\be
|\Gamma_{I,I+1}|=|\tilde{k}_{I+1}-\tilde{k}_I| -1 \quad,\qquad |\Sigma_I|=n_I +1 -|\tilde{k}_{I+1}-\tilde{k}_I|\omega_{I,I+1}-|\tilde{k}_I-\tilde{k}_{I-1}|\omega_{I,I-1}~,
\ee
where again
\be
\omega_{I,J}=\begin{cases}
0&\tilde{k}_I \leq \tilde{k}_J\\
1&\tilde{k}_I > \tilde{k}_J
\end{cases}
\ee
Let us assume for simplicity that $\tilde{k}_I<\tilde{k}_{I-1}$ and $\tilde{k}_{I+1}<\tilde{k}_{I+2}$. Additionally let us assume that the first node of $\Sigma_I$ is at the position $m+1$ in the vector $\vec k$. Then using \ref{eq:cont}, we see that the terms contributing from the above subquivers is given by
\be
x^m+(-1)^{\omega_{I,I+1}} x^{m+ n_I-|\tilde{k}_{I +1}-\tilde{k}_I|\omega_{I,I+1}}+(-1)^{\omega_{I+1,I}} x^{m+n_I+ |\tilde{k}_{I+1}-\tilde{k}_{I}|\omega_{I+1,I}}+x^{m+n_I+n_{I+1}}
\ee
So, no matter what the sign of $(\tilde{k}_I-\tilde{k}_{I+1})$ is, there will always be the contribution of the form\footnote{Note that this also holds for the special cases $\ell^{(m)}_i=0$ and $|\tilde{k}_i-\tilde{k}_{i+1}|=1$. }
\be
x^m +x^{m+n_I}- x^{m+n_I+(\tilde{k}_{I+1}-\tilde{k}_I)} + x^{m+n_I+n_{I+1}}~.
\ee

Now once we solve the beginning (and end) couple contributions, we can iterate on the above formula, and compute the entire contribution to the character equation. Using the fact that $|\Gamma_{0,1}|=|\tilde{k}_1|-`$, $|\Gamma_{N-1,N}|=|\tilde{k}_N|-`$, we have that the first two contributions are of the form
\be
\quad 1-x^{|\tilde{k}_1|}~.
\ee
Now by iterating, we see that the full contribution to the character equation is of the form
\be
1- x^{\tilde{k}_1}-x^{p_N-p_1-\tilde{k}_{N-1}}+ x^{p_N-p_1} +\sum_{I=1}^{N-2} \left(x^{p_{I+1}-p_1}- x^{p_{I+1}-p_1 +(\tilde{k}_{I+1}-\tilde{k}_I)}\right)~.
\ee
Here we used the relations
\be
n_I=p_{I+1}-p_I\quad,\qquad \sum_{J=1}^I n_J=p_{I+1}-p_1~,
\ee
where here $I=1,...,N-1$ and we extend the definition of $n_I$ to $n_N=0$.

\section{Q-fixed point equations and 't Hooft defect}\label{BPS-4d-App}
In this section, we discuss the Q-fixed locus of the 4d path integral associated with an 't Hooft defect. For the sake of brevity, we focus on vector multiplets -- including hypermultiplets in an arbitrary representation will involve an obvious generalization of the procedure presented here. We choose to write the Q-fixed equations in Minkowskian signature, with the metric $\de s^2= \sum^3_{i=1} (\de x^i)^2 - (\de \tau)^2$ on $\R^3 \times S^1$, to match conventions of recent papers \cite{Moore:2015szp, Moore:2014gua, Moore:2014jfa} on monopole moduli spaces.
The Euclidean versions of these equations can be obtained by Wick rotating appropriate bosonic fields.\\

The bosonic part of an $\N=2$ vector multiplet in four dimensions consists of a gauge field $A=(A_\tau, A_i)$, with $i=1,2,3$, and a complex scalar field $\varphi$
(or a pair of real scalars $X,Y$), while the fermionic part consists of a pair of Weyl spinor doublets $\psi_{\al \, A}$, $\bar{\psi}^{\dot{\al}}_A$, with $A=1,2$ being the $SU(2)_R$ index, and $(\al, \dot{\al})$ labelling the $SU(2)_l \times SU(2)_r$ Lorentz spinor indices respectively. The Weyl spinor doublets obey reality conditions : $(\psi_{\al \, A})^*=-\bar\psi_{\dot\al}^{ A}$. We adopt the following convention for the $\sigma$-matrices:
\begin{align}
\sigma_a = (I, \vec \sigma)\quad, \qquad \bar\sigma_a = (I, -\vec \sigma)\quad, \quad a=0,1,2,3~,
\end{align}
where $I$ is the $2 \times 2$ unit matrix and $\vec \sigma$ are the Pauli matrices. While writing multilinear expressions in terms of the scalar fields, we will often suppress the Lorentz spinor indices -- the undotted indices will be contracted in the ``northwest to southeast" convention while the dotted ones will follow the ``southwest to northeast" convention.\\

\subsection{Q-fixed point equations of the undeformed 4d path integral}

Let us first discuss the Q-fixed point equations for an 't Hooft defect on the undeformed space $\R^3 \times S^1$ (i.e. when $\lambda=0$ in \eref{def-dyonic}).
Given the field content described above, the action of an $\N=2$ vector multiplet with an 't Hooft defect at the origin, is
\begin{align}\begin{split}
& S= S_{\rm vector} + S_{\rm boundary}~, \\
& S_{\rm vector}=\frac{1}{g^2} \int_{\R^3 \times S^1} \de^4 x \Tr{\Big(\half F^{\mu \nu} F_{\mu \nu} + D_\mu \varphi D^\mu \bar \varphi - \qtr [\varphi,\bar\varphi]^2 \Big)}\\
& + \frac{1}{g^2} \int_{\R^3 \times S^1} \de^4 x \Tr{\Big(-2 \I  \bar{\psi}^A \bar{\sigma}^\mu D_\mu {\psi}_A - \I {\psi}^A [\bar\varphi,{\psi}_A]  + \I {\bar\psi}^A [\varphi,{\bar\psi}_A]  \Big)}\\
& + \frac{\vartheta}{8 \pi ^2} \int_{\R^3 \times S^1} \Tr \Big(F \wedge F \Big)~,\\
& S_{\rm boundary}=\frac{-\I}{g^2} \int_{\Sigma_3=\{x^\mu\,|\,r=\delta\}} \Tr ((\varphi - \bar\varphi)\, F + (\varphi + \bar\varphi) \star^{(4)} F) \wedge \de \tau~,
\end{split}\end{align}
where $S_{\rm vector}$ is the standard action for an $N=2$ vector multiplet, and $S_{\rm boundary}$ is a boundary term
\footnote{The boundary term as written in \cite{Moore:2015szp} is dependent on the complex structure $\zeta$ associated with the line operator
$L_\zeta$. Here we have chosen $\zeta=1$.}
necessary to regularize the the classical action in the 't Hooft background \cite{Moore:2015szp}. The $\N=2$ supersymmetry transformations for the vector multiplet fields are generated by the parameters $\xi_{\al \, A}, \bar{\xi}^{\dot{\al}}_A$ (we take these to be bosonic) which are solutions of the Killing spinor equations:
\begin{align}\begin{split}
&\nabla_\mu \xi _ {A} := \Big(\partial_\mu  + \qtr \omega^{a\,b}_\mu \sigma_{a\,b} \Big)\xi_{A} =0~,\\
& \nabla_\mu \bar \xi_{A} := \Big(\partial_\mu  + \qtr \omega^{a\,b}_\mu \bar\sigma_{a\,b}\Big) \bar\xi_{A}=0~,
\end{split}\end{align}
where we have suppressed the Lorentz spinor indices. In the case of undeformed $\R^3 \times S^1$, we have $\omega^{a\,b}_\mu=0$, which implies that the supersymmetry parameters $\xi_{\al \, A}, \bar{\xi}^{\dot{\al}}_A$ are constants. \\

Explicitly, the supersymmetry transformation rules for the bosonic fields are
\begin{align}\begin{split}
&\delta A_i= \xi^A \sigma_i \bar\psi_A + \bar\xi^A \bar\sigma_i \psi_A\quad,\qquad  \delta \phi =2 \xi^A \psi_A~,\\
&\delta A_\tau=  \xi^A \sigma_0 \bar\psi_A +\bar\xi^A \bar\sigma_0 \psi_A\quad,\qquad
 \delta \bar \phi=2 \bar\xi^A \bar\psi_A~,
\end{split}\end{align}
while variation of the fermionic fields are
\begin{align}
& \delta \psi_A=- \I \sigma^{\mu \nu} F_{\mu \nu} \xi_A + \I \sigma^\mu D_\mu \varphi \bar\xi_A +\frac{\I}{2} \xi_A [\varphi, \bar\varphi] ~,\\
& \delta \bar\psi_A=\I \bar\sigma^{\mu \nu} F_{\mu \nu} \bar\xi_A -  \I\bar\sigma^\mu D_\mu \bar\varphi \xi_A + \frac{\I}{2} \bar\xi_A [\varphi, \bar\varphi]~.
\end{align}

For treating line defects, it is more convenient to work with the following redefined fields:
\begin{align}
& Y= \half (\varphi + \bar\varphi)\quad,\qquad\rho_A=\half(\psi_A + \sigma^0 \bar\psi_A) ~,\\
& X= \frac{1}{2\I} (\varphi - \bar\varphi)\quad,\qquad
 \lambda_A =\frac{1}{2\I}(\psi_A - \sigma^0 \bar\psi_A)~,
\end{align}
where $X,Y$ are real scalar fields and
$\rho_A, \lambda_A$ are symplectic Majorana Weyl spinors -- $\bar\rho^A=\bar\sigma^0 \rho^A,\, \bar\lambda^A=\bar\sigma^0 \lambda^A$.
Similarly, one redefines the supersymmetry parameters in the following fashion:
\begin{align}\begin{split}
& \varepsilon_A=\half(\xi_A + \sigma^0 \bar\xi_A) ~,\\
& \eta_A=\frac{1}{2\I}(\xi_A - \sigma^0 \bar\xi_A)~.
\end{split}\end{align}
where $ \varepsilon_A, \eta_A$ are symplectic Majorana Weyl spinors.
Supersymmetry transformation generated by the parameter $\varepsilon_A$, generating $\cR$-supersymmetry \cite{Moore:2015szp}, may be explicitly written as
\begin{align}\begin{split}
&\delta A_i= 2 \varepsilon_A \sigma^0 \bar{\sigma}_i \rho^A\quad,\qquad\delta Y=2 \varepsilon^A \rho_A,\\
&\delta A_\tau=-2 \I \varepsilon_A \lambda^A \quad~~,\qquad
\delta X=2  \varepsilon^A \lambda_A~ ,\\
&\delta \rho^A= [-(D_0 X -[Y,X])
 + \I \sigma^0 \bar{\sigma}^i  (E_i - D_i Y)]\varepsilon^A~,\\
&\delta \lambda^A=[D_0 Y   + \I \sigma^0 \bar{\sigma}^i  (B_i -D_i X)]\varepsilon^A~,
\end{split}\end{align}
while supersymmetry generated by the parameter $\eta_A$, generating $\cT$-supersymmetry, has the following form:
\begin{align}\begin{split}
&\delta A_i= 2 \eta_A \sigma^0 \bar{\sigma}_i \lambda^A\quad,\qquad
\delta X=2  \eta_A \lambda^A~,\\
&\delta A_\tau=2 \I \eta_A \rho^A \qquad~~,\qquad
\delta Y=-2  \eta_A \rho^A~,\\
&\delta \rho^A= [ D_0 Y
 - \I \sigma^0 \bar{\sigma}^i (B_i + D_i X)] \eta^A~,\\
&\delta \lambda^A=[( D_0 X + [Y,X]) + \I \sigma^0 \bar{\sigma}^i \varepsilon^A (E_i + D_i Y)]\eta^A~.
\end{split}\end{align}

Various derivatives of vector multiplet fields appearing in the above equations are defined as follows:
\begin{align}
 &D_i X= \partial_i X + [A_i, X]\quad,\qquad B_i= \half \epsilon_{ijk} F^{jk}~, \\
 &D_i Y= \partial_i Y + [A_i, Y]\quad,\qquad E_i=F_{i\tau}=\partial_i A_\tau - \partial_\tau A_i + [A_i, A_\tau]~.\\
\end{align}
In the undeformed background $\R^3 \times S^1$ i.e. for $\lambda=0$, an 't Hooft operator insertion at the origin, specified by the boundary condition \eqref{'t Hooft b.c.}, only preserves four supercharges generated by $\varepsilon_A$, with $\eta_A=0$. Therefore, setting $\delta \rho^A=0$ and $\delta \lambda^A=0$ for a generic symplectic-Majorana-Weyl spinor $\varepsilon^A$, the BPS equations for the undeformed background with a line defect are
\begin{empheq}[box=\widefbox]{align}\begin{split}
& B_i -D_i X=0\quad,\qquad  D_\tau X -[Y,X]=0~,\\
& E_i - D_i Y=0,\quad,\qquad
 D_\tau Y=0~,
\end{split}\end{empheq}
of which the last three equations impose $Q^2$--invariance on the bosonic fields. Note that
the Dirac monopole configuration in \eref{'t Hooft b.c.} is an exact solution of the above equations.\\

\subsection{Q-fixed point equations of the deformed 4d path integral}

Now consider the $\Omega$-deformed background with $\lambda \neq 0$. The metric in terms of the local coordinates is given as
\begin{align}\begin{split}
\de s^2 &= \de r^2 + r^2 \de \theta^2 + r^2 \sin^2 \theta (\de \phi + \frac{\lambda}{R} \de \tau)^2 - \de \tau^2,\\
&= \sum^3_{i=1} (\de x^i + V^i \de \tau)^2 - \de \tau ^2, \quad V^1 = \frac{\lambda}{R} x^2, V^2 = -\frac{\lambda}{R} x^1, V^3=0,
\end{split}\end{align}
while all the fields in the theory are understood to be periodic under $\tau$-direction.\\

One can choose the following orthonormal basis (and its inverse):
\begin{align}
e^{\;\;a}_{\mu} =\left(
\begin{array}{cccc}
1 & 0 & 0 & 0  \\
 0 &  1  & 0 & 0  \\
 0 & 0 & 1 & 0  \\
 V^1 &  V^2 & 0 & 1
\end{array}
\right)\quad,\qquad
E_{a}^{\;\;\;\; \mu} =\left(
\begin{array}{cccc}
1 & 0 & 0 & 0  \\
 0 &  1  & 0 & 0  \\
 0 & 0 & 1 & 0  \\
- V^1 & - V^2 & 0 & 1
\end{array}
\right)~.
\end{align}

Let us comment on the supersymmetry preserved by the line defect in this deformed background. Preserving part of the supersymmetry of the undeformed background requires turning on a background gauge field which lives in the Cartan subalgebra of the $SU(2)_R$ symmetry. The supersymmetry parameters are solutions of a more general Killing spinor equation \footnote{For the most general form, see \cite{Hama:2012bg}.}:
\begin{align}\begin{split}
&D_\mu \xi _ {A} := \Big(\partial_\mu  + \qtr \omega^{a\,b}_\mu \sigma_{a\,b} \Big)\xi_{A} + \I V^B_{\mu \, A} \xi_B=0~,\\
& D_\mu \bar \xi_{A} := \Big(\partial_\mu  + \qtr \omega^{a\,b}_\mu \bar\sigma_{a\,b}\Big) \bar\xi_{A}+ \I V^B_{\mu \, A} \bar \xi_B=0~,
\end{split}\end{align}
where $V^B_{\mu \, A}$ is the background $SU(2)_R$ gauge field.\\

It is convenient to write the supersymmetry in the Donaldson-Witten twisted form, i.e. let $SU(2)_r \cong SU(2)_R$, which implies that the supersymmetry parameters may be written as
\begin{align}\begin{split}
& \bar \xi^{\dot\al}_{A} \to \bar \xi^{B}_{A} = \delta^B_{\,A} \bar\xi + (\bar{\sigma}_{ab})^B_A \bar{\xi}^{ab}~, \\
& \xi_{\al\, A} \to \xi_{\al\, A} = (\sigma_a)_{\al\, A} \xi^a~.
\end{split}\end{align}
where $a,b=0,1,2,3$ label the vierbeins.\\

The $\cR$ and $\cT$ supersymmetry parameters can also be written in terms of the twisted supersymmetry parameters:
\begin{align}\begin{split}
& \varepsilon_{\al \, A}= (\sigma_0)_{\al\, A}\bar\xi + (\sigma_a)_{\al\, A} \xi^a + (\sigma_0)_{\al\, B} (\bar{\sigma}_{ab})^B_{\, A} \bar{\xi}^{ab}~,\\
& \eta_A= (\sigma_0)_{\al\, A}\bar\xi - (\sigma_a)_{\al\, A} \xi^a - (\sigma_0)_{\al\, B} (\bar{\sigma}_{ab})^B_{\, A} \bar{\xi}^{ab}~.
\end{split}\end{align}
Setting the background $SU(2)_R$ gauge field to cancel the self-dual part of the spin connection, i.e.
\begin{align}
\I V^B_{\mu \, A} + \qtr \omega^{a\,b}_\mu (\bar\sigma_{a\,b})^B_{\, A}=0,
\end{align}
one obtains the following solution of the Killing spinor equations in the deformed background:
\begin{align}
\xi^i=\bar{\xi}^{ab}=0\quad,\qquad \partial_\mu \bar\xi=\partial_\mu \xi^0=0\quad \implies \quad \bar\xi, \xi^0= {\rm constant}~,
\end{align}
The deformed background therefore preserves only two supercharges, with associated parameters $ \bar\xi$ and $\xi^0$. In terms of the $\cR$ and $\cT$ supersymmetry parameters, we have
\begin{align}
& \varepsilon_{\al \, A} = (\sigma^0)_{\al\, A} (- \xi^0 + \bar\xi)\quad,\qquad
 \eta_{\al \, A}=-(\sigma^0)_{\al\, A} (\xi^0 + \bar\xi)~.
\end{align}

Now,  a line defect in this deformed background, specified by the boundary conditions at $r \to 0$, preserves a single supercharge :  the condition $\eta_A=0$ sets a linear combination of $ \bar\xi$ and $\xi^0$ to zero. More explicitly,
\begin{align}
& \eta_{\al \, A}=0 \implies \xi^0 + \bar\xi=0\quad,\qquad
 \varepsilon_{\al \, A} = 2 (\sigma^0)_{\al\, A} \bar\xi.~.
\end{align}

The transformation of the bosonic fields under this supercharge are:
\begin{align}\begin{split}
&\delta X=2 \I \varepsilon_A \rho^A\qquad,\qquad \delta A_i= 2 \varepsilon_A \sigma^0 \bar{\sigma}_i \rho^A~,\\
&\delta Y=-2 \I \varepsilon_A \lambda^A\quad,\qquad \delta A_\tau=-2 \I \varepsilon_A \lambda^A + 2 V^i \varepsilon_A \sigma^0 \bar{\sigma}_i \rho^A~.
\end{split}\end{align}
Note that the supersymmetry preserves a Wilson loop at the origin where $V^i=0$, so that $\delta_{\rm susy} (A_\tau - Y)=0$.
The fermionic fields transform as :
\begin{align}\begin{split}
&\delta \rho^A= [-(D_\tau X -[Y,X]- V^i D_i X)
 + \I \sigma^0 \bar{\sigma}^i  (E_i - D_i Y- V^j F_{ij})]\varepsilon^A,\\
&\delta \lambda^A=[(D_\tau Y - V^i D_i Y) + \I \sigma^0 \bar{\sigma}^i  (B_i -D_i X)]\varepsilon^A~,
\end{split}\end{align}

Therefore, BPS equations in the deformed background with the 't Hooft operator insertion are then given as follows:
\begin{empheq}[box=\widefbox]{align}\begin{split}\label{4d BPS def}
& B_i -D_i X=0\qquad\qquad,\qquad D_\tau X -[Y,X] - V^i D_i X =0~,\\
&E_i - D_i Y- V^j F_{ij} =0\quad,\qquad
 D_\tau Y - V^i D_i Y=0~.\\
\end{split}\end{empheq}
Note that the last three equations give the $Q^2$--invariance of the fields $(X,Y,A_i)$ \footnote{$Q^2$--invariance of $A_\tau$ is obtained as a linear combination of the other equations and is therefore identically satisfied. Explicitly, $\delta^2 A_\tau = -V^i F_{i \tau} + D_\tau Y=0$, using \eqref{4d BPS def}.}, where $Q^2 = \cL_{G} + \rm gauge \, transformation$, with $\cL_{G}$ being a covariant Lie derivative w.r.t a vector field $G$. In the vierbein basis, the vector field is defined as
\begin{align}
G^a = \bar\xi^A \bar\sigma^a \xi_A = (1,0,0,0)\quad, \quad a=0,1,2,3.
\end{align}
where $\bar\xi$ is appropriately normalized. Therefore, in the coordinate basis, $G$ is given as
\begin{align}
G^\mu = E^\mu_0 G^0 \quad \implies \quad G^\tau =1\quad, \quad G^i = -V^i\quad, \quad i=1,2,3~,
\end{align}
leading to the above $Q^2$--invariance equations. Therefore, $Q^2$ generates the following group action
\begin{align}
Q^2 A_i  = {\tau-{\rm translation}} + {\rm rotation}  + {\rm gauge\, transformation}~.
\end{align}
The BPS equations imply that the 4d path integral localizes on a sublocus of the
moduli space of singular monopoles on $\R^3$ which is invariant under the group
action generated by $Q^2$. Kronheimer's correspondence \cite{KronCorr} states that moduli space
of singular monopoles on $\R^3$ is isomorphic to the moduli space of $U(1)$--invariant
instantons on a Taub-NUT space. IOT/GOP argued that, for studying the monopole bubbling locus,
it is sufficient to consider instantons localized at the tip of of the Taub-NUT which is locally $\R^4$.
In addition, the group action generated by $Q^2$ can be lifted to an appropriate group action on the
moduli space of instantons. Therefore, the Q-fixed locus of the 4d path integral can also be thought
of as a sublocus of the moduli space of $U(1)$--invariant instantons on $\R^4$, which is invariant under the
above group action. In analogy to Nekrasov's original computation \cite{Nekrasov:2002qd}, the Q-fixed locus is given by a set
of isolated fixed points on the $U(1)$--invariant instanton moduli space.

\section{IOT result: $Z_{\rm mono}$ from 5d instanton partition function} \label{IOT}
 In this subsection, we show that IOT formula \cite{Ito:2011ea} for $Z_{\rm mono}$ for pure 't Hooft operators on $S^1 \times \R^3$ may be derived from Nekrasov's partition function for instantons on $S^1 \times \R^4$ by imposing the constraint of $U(1)_K$ invariance. In a 4d $\N=2$ $SU(N)$ SYM with matter in
 representation $R$, the monopole bubbling contribution $Z_{\rm mono}$ for an 't Hooft defect labelled by $B$, in the bubbling sector labelled by $\mathbf{v}$, has the following form:
 \begin{equation}\label{mono-total}
 Z_\text{mono}(\fa,\fm_f,\lambda;B,\mathbf{v})
=\sum_{\vec Y'} z_{\vec Y'}^\text{vec}(\fa,\lambda;B,\mathbf{v})
 z_{\vec Y'}^R(\fa,\fm_f;B,\mathbf{v})~,
\end{equation}
The sum in \eref{mono-total} is over a $U(1)_K$-constrained set of fixed points on the moduli space of instantons on $\R^4$, which are labelled by
$U(1)_K$-invariant $N$-tuples of Young diagrams $\vec Y'$. The one-loop determinants $z_{\vec Y'}^\text{vec}, z_{\vec Y'}^R$ at a given fixed point are obtained
by restricting to $U(1)_K$--invariant weights, as we discuss below. We would like to emphasize that the above formula gives the complete
answer for $Z_\text{mono}$ only for an $\N=2^*$ $SU(N)$ theory.\\

One can derive the above formula using two standard ingredients: the ADHM construction of $U(1)_K$ invariant instantons on $\C^2$ \cite{Kapustin:2006pk} and Nekrasov's formula for the instanton partition function of 5d $\N=1$ theories on $S^1 \times \C^2$ \cite{Nekrasov:2002qd}.

\subsection{$Z_{\rm mono}$ from 5d instanton partition function}\label{IOT-Zmono}
The Q-fixed locus of the 5d $G=SU(N)$ instanton partition function on $S^1 \times \C^2$ (defined as the non-perturbative part of the 5d supersymmetric
index in \eref{5dIndex-A}) is given by a finite set of fixed points on the moduli space
of $SU(N)$ instantons on $\C^2$ under the $U(1)_{\epsilon_1} \times U(1)_{\epsilon_2} \times T_G$ equivariant action
\footnote{The structure of fixed points remains the same for $SU(N)$ theory with hypermultiplets in arbitrary representation $R$. The one-loop determinant at a
given fixed point is obtained from the weights of the $U(1)_{\epsilon_1} \times U(1)_{\epsilon_2} \times T_G \times T_F$ action ($T_F$ being the
maximal torus of the flavor symmetry group) on the vector bundle $\V(R)$ on the instanton moduli space, associated with fermion zero modes of the hypermultiplet.}.

Using the standard ADHM description
of a $k$-instanton moduli space, the sub-locus invariant under the $U(1)_{\epsilon_1} \times U(1)_{\epsilon_2} \times T_G$ action is given by the ADHM data
$(B_1,B_2,I,J)$ that satisfy
\begin{empheq}{align}\begin{split}\label{inst-fixed-point-eq}
    &\epsilon_1 B_1+[\phi, B_1]=0\quad,\qquad \phi I-I a=0~, \\
    &\epsilon_2 B_2+[\phi, B_2]=0\quad,\qquad (\epsilon_1+\epsilon_2) J+ a J- J\phi=0~,
\end{split}\end{empheq}
for generic equivariant parameters $(\epsilon_1,\epsilon_2,a)$ (where $a$ is an element of the Cartan subalgebra of $SU(N)$), and
for some $\phi=\text{diag}(\phi_1,\ldots,\phi_k)$ parametrizing the Cartan subalgebra of $U(k)$.
The invariant sub-locus consists of a finite set of isolated points
if the above equations are satisfied only for discrete choices of $\phi$, which turns out to be the case \cite{Nekrasov:2002qd}. A fixed point
is then labelled by a particular value of $\phi$, which in turn could be read off from an $N$-tuple of Young diagrams
$\vec Y$ consisting of a total of $k$ boxes.
Explicitly, the solution for $\phi$ associated with a fixed point labelled by a given $N$-tuple of Young diagrams is:
\begin{equation}\label{phi-soln1}
\phi_s = a_\al + \epsilon_+ + \epsilon_1 (i_{s,Y_\al}-1) + \epsilon_2 (j_{s,Y_{\al}}-1)\quad, \quad s=1,\ldots,k\quad, \quad \al=1,\ldots,N~,
\end{equation}
where $(i_{s,Y_\al}, j_{s,Y_\al})$ denotes the $s$-th box (out of the total $k$) which belongs to the diagram $Y_\al$.\\

Now, consider the case of $U(1)_K$--invariant instantons as discussed in section \ref{ADHM-KN}. For $e^{2\pi \I \nu} \in U(1)_K$, the
$U(1)_K$--invariance imposes a set of constraints on the ADHM variables -- summarized in \eref{U(1)K action}.
Invariance under an infinitesimal $U(1)_K$ transformation therefore leads to the following constraints on the ADHM variables:
\begin{empheq}{align}\begin{split}\label{U(1)K-inv-eq}
& - B_1+ [K,B_1]=0\quad,\qquad KI-I\mathbf{v}=0~,\\
&B_2+[K,B_2]=0\quad,\qquad
\mathbf{v} J-J K=0~,
\end{split}\end{empheq}
where $K$ is a cocharacter which is determined by the defect data $(B, \mathbf{v})$ via \eref{char-eqn-main}.

To derive  the $U(1)_K$--invariant fixed points we proceed as follows.
We multiply the equations \eref{U(1)K-inv-eq} by $\nu$ and add them to the corresponding
equation in the set \eref{inst-fixed-point-eq}, which leads to
 \begin{empheq}{align}\begin{split}\label{inst-fixed-point-eq-K}
    &\widetilde\epsilon_1 B_1+[\widetilde\phi, B_1]=0\quad,\qquad \widetilde\phi I-I \widetilde{a}=0~,\\
    &\widetilde\epsilon_2 B_2+[\widetilde\phi, B_2]=0\quad,\qquad
(\widetilde\epsilon_1+\widetilde\epsilon_2) J+ \widetilde{a} J- J\widetilde\phi=0~,
\end{split}\end{empheq}
where the new parameters are simply
\begin{empheq}{align}\begin{split}\label{K-fixedpoints}
&\widetilde\phi_s = \phi_s + K_s \nu \quad,\qquad \widetilde{\epsilon_1}=\epsilon_1 - \nu~,\\
&\widetilde{a}_{\al}=a_{\al} + \mathbf{v}_{\al} \nu\quad,\qquad
 \widetilde\epsilon_2=\epsilon_2 + \nu~.
\end{split}\end{empheq}
Since the equations \eref{inst-fixed-point-eq-K} are of the same form as the equations \eref{inst-fixed-point-eq}, the solution for $\widetilde\phi$
is given by equation \eref{phi-soln1} with the equivariant parameters $(\epsilon_1, \epsilon_2, a_\al)$ replaced by
$(\widetilde\epsilon_1, \widetilde\epsilon_2, \widetilde{a}_{\al})$, i.e.
\begin{align}\label{phi-soln2}
& \widetilde\phi_s = \widetilde{a}_\al + \epsilon_+ + \widetilde\epsilon_1 (i_{s,Y_\al}-1) + \widetilde\epsilon_2 (j_{s,Y_{\al}}-1) ,\\
 \implies &  \phi_s = a_\al + \epsilon_+ + \epsilon_1 (i_{s,Y_\al}-1) + \epsilon_2 (j_{s,Y_{\al}}-1) + \Big(-K_s + \mathbf{v}_\al +(j_{s,Y_{\al}} - i_{s,Y_{\al}})\Big)\nu.
 \end{align}

The $U(1)_K$-invariant fixed points must be independent of $\nu$, and therefore correspond to the following $N$-tuple of Young diagrams
\begin{equation}\begin{split}
\vec Y = (Y_1,Y_2,\ldots,Y_N) \quad \text{ such that} \quad
K_s=\mathbf{v}_\al +(j_{s,Y_{\al}} - i_{s,Y_{\al}}) ~,
\label{vec-Y-cond}
\end{split}\end{equation}
up to a permutation of $s\in \{1,\ldots, k\}$, with $\al=1,\ldots,N$ and $(i_{s,Y_\al}, j_{s,Y_\al})$ representing $s$-th box in the $\al$-th Young diagram.
This gives a clear recipe for determining the fixed points on the $U(1)_K$ invariant instanton moduli space under the $U(1)_{\epsilon_1} \times U(1)_{\epsilon_2} \times T_G$ action.\\

For computing the one-loop determinants in equation \eref{mono-total}, one should restrict to $U(1)_{\epsilon_1} \times U(1)_{\epsilon_2} \times T_G \times T_F$  weights ($T_F$ being the maximal torus of the flavor symmetry group) that contribute to the index at a given fixed point are the ones that are $U(1)_K$-invariant.
Consider the vector multiplet contribution to the instanton partition function in the standard case \cite{Nekrasov:2002qd}
\footnote{We adopt the notation $$2\I \sin(x\pm y) = 2\I\sin(x+y)\, 2\I\sin(x-y)~.$$}:
\footnote{The arm and leg-lengths of a given Young diagram w.r.t. a box $s=(i,j)$ (not necessarily inside the diagram) are defined as
 $ A_Y(s)=\lambda_i-j\,,\quad L_Y(s)=\lambda^T_j-i,$
where $\lambda_i$ and $\lambda^T_i$ are the numbers of boxes in the $i$-th row
and column of $Y$, respectively. Note that $A_Y, L_Y$ can be negative if $s$ is outside the diagram.}:
\begin{equation}
  \begin{aligned}
z^\text{vec}_{\vec Y,\, {\rm Nek.}}=& \prod_{(\alpha, \beta, s)}  \left(2\sinh\left[\half \left(a_\alpha- a_\beta+ (A_{Y_\alpha}(s)-L_{Y_\beta}(s)\pm 1
)\epsilon_+ - (A_{Y_\alpha}(s)+L_{Y_\beta}(s)+ 1) \epsilon_- \right) \right]\right)^{-1}~,
  \end{aligned}
\end{equation}
 where the products are over the triples $(\alpha, \beta, s)$ with $s \in Y_\al$. In the present case, we should only include in the product those triples $(\al,\beta,s)$
 in the above product for which the argument of the sinh function is invariant under the transformation of the equivariant parameters
 $(a,\epsilon_1,\epsilon_2) \to (\widetilde{a}, \widetilde{\epsilon_1}, \widetilde{\epsilon_2})$ , with $(\widetilde{a}, \widetilde{\epsilon_1}, \widetilde{\epsilon_2})$ given in \eref{K-fixedpoints}. From \eref{K-fixedpoints}, the argument of the $\sinh$ function transforms as
 \be
 \begin{split}
& (a_\alpha- a_\beta+ (A_{Y_\alpha}(s)-L_{Y_\beta}(s)\pm 1)\epsilon_+ - (A_{Y_\alpha}(s)+L_{Y_\beta}(s)+ 1) \epsilon_-)\\
\to & (a_\alpha + \mathbf{v}_\al \nu - a_\beta -\mathbf{v}_\beta \nu + (A_{Y_\alpha}(s)-L_{Y_\beta}(s)\pm 1)\epsilon_+ - (A_{Y_\alpha}(s)+L_{Y_\beta}(s)+ 1) (\epsilon_- -\nu)) \\
= & (a_\alpha- a_\beta+ (A_{Y_\alpha}(s)-L_{Y_\beta}(s)\pm 1)\epsilon_+ - (A_{Y_\alpha}(s)+L_{Y_\beta}(s)+ 1) \epsilon_-)\\
 & + (\mathbf{v}_\al - \mathbf{v}_\beta + A_{Y_\alpha}(s)+L_{Y_\beta}(s)+ 1)\nu~,
 \end{split}
 \ee
which implies that the argument is invariant under the $U(1)_K$-action for a triple $(\al,\beta,s)$ if
\be \label{alpha-beta-s-cond}
\mathbf{v}_\al - \mathbf{v}_\beta + A_{Y_\alpha}(s)+L_{Y_\beta}(s)+ 1= 0~.
\ee

 Therefore, using the identification $a_\al = 2\I \pi \fa_\al$, $\epsilon_+ = {\I \pi \lambda}$, and $\epsilon_-=0$, the function $z^\text{vec}_{\vec Y}$ in the $U(1)_K$-invariant case is
\begin{equation}\label{mono-vec}
\boxed{z^\text{vec}_{\vec Y}= \prod_{(\alpha, \beta, s)} \left(2\I\sin\left[ \pi \left(\fa_\alpha-\fa_\beta+\half(A_{Y_\alpha}(s)-L_{Y_\beta}(s)\pm 1
)\lambda \right) \right]\right)^{-1}}
\end{equation}
where the products are over the triples $(\alpha, \beta, s)$, with $s \in Y_\al$, satisfying \eref{alpha-beta-s-cond}.

This reproduces the IOT formula for a vector multiplet \footnote{The formula for $z^\text{vec}_{\vec Y}$ is identical to equation 5.25 in IOT up to some overall factors
of $\I$. These factors of $\I$ are needed to produce the correct overall sign of $Z_{\rm mono}$, which IOT ignored in their expressions. See discussion after equation 6.11 in \cite{Ito:2011ea}. The same is true for $z^\text{adj}_{\vec Y}$ and $z_{\vec Y}^\text{fund}$.
}.\\

Similarly, proceeding as above and defining $m=2 \I \pi \fm$, contribution of the adjoint hyper is given as:
\begin{equation} \label{mono-adj}
\boxed{z^\text{adj}_{\vec Y}=\prod_{(\alpha, \beta, s)} \left(2\I \sin\left[\pi \left(\fa_\alpha- \fa_\beta+\frac{1}{2}(A_{Y_\alpha}(s)-L_{Y_\beta}(s))\lambda
\pm \fm \right)\right]\right)}
\end{equation}
where the products are over the same triples $(\alpha, \beta, s)$ as given in \eref{alpha-beta-s-cond}.\\

Contribution of fundamental hypers to the instanton partition function is given by:
\be
z^\text{fund}_{\vec Y,\, {\rm Nek.}}= \prod_{(\alpha,s)} 2\sinh\Big({a_\al - m_f + \epsilon_+ + \epsilon_1(i_s-1) + \epsilon_2(j_s-1)}\Big)~,
\ee
where the product is over the pairs $(\alpha,s)$ with $s \in Y_\alpha$.
Under the $U(1)_K$-action \eref{K-fixedpoints}, the argument of the $\sinh$ function transforms as:
\be
\begin{split}
&(a_\al - m_f + \epsilon_+ + \epsilon_1(i_s-1) + \epsilon_2(j_s-1) ) \\
&\to (a_\al - m_f + \epsilon_+ + \epsilon_1(i_s-1) + \epsilon_2(j_s-1) ) + (\mathbf{v}_\al - i_s + j_s)\nu~.
\end{split}
\ee
Invariance under the $U(1)_K$-action requires restricting the product over the pairs $(\alpha,s)$ with $s \in Y_\alpha$, such that
\be\label{al-s-constraint}
\mathbf{v}_\al - i_s + j_s~.
\ee

Therefore, proceeding as before and defining $m_f =2\pi \I \fm_f$, the contribution of the fundamental hyper to $Z_{\rm mono}$ is given as
\begin{equation}\label{mono-fund}
\boxed{z_{\vec Y}^\text{fund}(\fa,m_f,\lambda;B,v)=\prod_{(\alpha,s\in Y_\al)} 2\I \sin\left[\pi\left(\fa_\alpha -\fm_f + \frac{1}{2}\left( i_s+j_s-1\right)\lambda\right)\right]}
\end{equation}
where the product is over the pairs $(\alpha,s)$ satisfying \eref{al-s-constraint}.

\subsection{One-Loop contribution to the 't Hooft defect vev}
For a monopole bubbling sector with effective 't Hooft charge $\mathbf{v} = {\rm diag}(0,\ldots,0)$, we have
$$Z_\text{1-loop}(\fa, \fm_f,\lambda; \mathbf{v}=0)=1~.$$
For a non-zero $\mathbf{v}$, the one-loop contribution to the 't Hooft defect expectation value was explicitly computed in \cite{Ito:2011ea}
and can be written as,
\begin{align}
Z_\text{1-loop}(\fa,\fm_f,\lambda; \mathbf{v}) := Z_\text{1-loop}^\text{vm}(\fa,\lambda;\mathbf{v})Z_\text{1-loop}^\text{hm}(\fa,\fm_f,\lambda;\mathbf{v})~,
\end{align}
where the contribution of the vector multiplet is
\begin{align}
Z_\text{1-loop}^\text{vm}(\fa,\lambda; \mathbf{v})& = \prod_{n\in \mathbb Z}\prod_{\alpha} \prod_{k=0}^{|\alpha \cdot \mathbf{v}|-1}\left[n \varepsilon+\frac 1 2 \lambda
+\alpha \cdot \fa +\left(\frac{|\alpha\cdot \mathbf{v}|-1}2-k\right)\lambda\right]^{-1/2} \nonumber \\
&=\prod_{\alpha>0} \prod_{k=0}^{|\alpha\cdot \mathbf{v}|-1} \prod_\pm \sin^{-1/2}\left[\pi\left( \alpha\cdot \fa \pm\left(\frac{|\alpha\cdot \mathbf{v}|}2-k\right)\lambda\right)\right]~,
\end{align}
and the contribution of the hypermultiplets are
\begin{align}
Z_\text{1-loop}^\text{hm}(\fa,&\fm_f,\lambda;v) = \prod_{n\in \mathbb Z}\prod_{f=1}^{N_\text{F}}\prod_{w\in R}\prod_{k=0}^{|w\cdot \mathbf{v}|-1}
\left[n \varepsilon+ w\cdot \fa-\fm_f +\left(\frac{|w\cdot \mathbf{v}|-1}2-k\right)\lambda\right]^{1/2}\nonumber \\
&= \prod_{f=1}^{N_\text{F}}\prod_{w\in R}\prod_{k=0}^{|w\cdot \mathbf{v}|-1}\sin^{1/2}\left[\pi\left(w\cdot \fa-\fm_f +\left(\frac{|w\cdot \mathbf{v}|-1}2 -k\right)\lambda\right)\right]~,
\end{align}
where $w$ represents a weight of the representation $R$ of the gauge group in which the hypermultiplet transforms.\\

The one-loop contribution can also be derived from the one-loop factor of a five-dimensional supersymmetric index -- we refer the reader
to \cite{Mekareeya:2013ija} for details.

\subsection{IOT formula: $\IP{L_{p,0}}$ in $\N=2^*, SU(2)$ SYM}
For $\N=2^*$ $SU(2)$ SYM, $B$ and $\bf v$ can be parametrized as:
\begin{align}
B= \half {\rm diag} (p, \, -p)\quad, \qquad {\bf v}= \half {\rm diag} (v, \, -v)~,
\end{align}
where $p$ is a positive integer, and $v=p, p-2, p-4,\ldots, -p$. To illustrate the IOT prescription, let us compute the monopole bubbling contribution to $\IP{L_{2,0}}$.
In this case, we have $B=\half \text{diag}(2,-2)$, and the possible values of $\mathbf{v}$ are $\half\text{diag}(2,-2), -\half\text{diag}(2,-2)$ and $\text{diag}(0,0)$.
From \eqref{char-eqn-main}, it is clear that $K$ has no solution (for generic $\nu$) for $\mathbf{v}= \pm \half{\rm diag} (2,-2)$ which implies that there are no monopole bubbling contributions in these cases. For $\mathbf{v}=\text{diag}(0,0)$, there is a solution for $K$ -- a $1 \times 1$ matrix with entry 0. The fixed points therefore correspond to doublets of Young diagrams with total number of boxes equal to one:
\begin{align}\begin{split}
& (1): Y_1=\Yvcentermath1 \begin{Young}   \cr \end{Young}\;, Y_2 = \emptyset~,\\
& (2): Y_1 = \emptyset\;, Y_2=\Yvcentermath1 \begin{Young} \cr \end{Young}~.
\end{split}\end{align}
In the first case, for the only box $s=(1,1) \in Y_1$: $A_{Y_1}(s)=0,L_{Y_1}(s)=0, A_{Y_2}(s)=-1, L_{Y_2}(s)=-1$. The triple $(1,2,s\in Y_1)$ satisfies \eqref{alpha-beta-s-cond} and therefore using \eqref{mono-vec} and \eqref{mono-adj}
\begin{align}\begin{split}
&z^\text{vec}_{\vec Y}(1)=\Big(-\sin{\pi (2 \fa)} \sin{\pi (2 \fa + \lambda)}\Big)^{-1}\quad,\qquad
z^\text{adj}_{\vec Y}(1)=-\sin \pi \left(2\fa+\frac{1}{2}\lambda\pm \fm \right).
\end{split}\end{align}
In the second case, for the only box $s=(1,1) \in Y_2$: $A_{Y_1}(s)=-1,L_{Y_1}(s)=-1, A_{Y_2}(s)=0, L_{Y_2}(s)=0$. The triple $(2,1,s\in Y_2)$ satisfies \eqref{alpha-beta-s-cond} and therefore using \eqref{mono-vec} and \eqref{mono-adj}
\begin{align}\begin{split}
&z^\text{vec}_{\vec Y}(2)=\Big(-\sin{\pi (2 \fa)} \sin{\pi (2 \fa - \lambda)}\Big)^{-1},\\
&z^\text{adj}_{\vec Y}(2)=-\sin \pi \left(2\fa - \frac{1}{2}\lambda\pm \fm \right).
\end{split}\end{align}
Putting together (1) and (2), we have
 \begin{align}\begin{split}
 &Z_\text{mono}(\fa,\fm,\lambda;p=2,v=0) \\
 &=z_{\vec Y}^\text{vec}(1) z^\text{adj}_{\vec Y}(1) +z_{\vec Y}^\text{vec}(2) z^\text{adj}_{\vec Y}(2) \\
 &=\frac{\sin \pi \left(2\fa+\frac{1}{2}\lambda\pm \fm \right)}{\Big(\sin{\pi (2 \fa)} \sin{\pi (2 \fa + \lambda)}\Big)} +\frac{\sin \pi \left(2\fa - \frac{1}{2}\lambda\pm \fm \right)}{\Big(\sin{\pi (2 \fa)} \sin{\pi (2 \fa - \lambda)}\Big)}~.
\end{split}\end{align}
The configurations $\mathbf{v}=\pm \half{\rm diag} (2,-2)$ receive classical and one-loop contributions. Putting those together with $Z_{\rm mono}$ computed above, we obtain the final answer for $\IP{L_{2,0}}$.
\begin{equation}
\begin{split}
\IP{L_{2,0}} = &\Big(e^{4\pi \I \fb} + e^{-4\pi \I \fb}\Big) \frac{\prod_{s_1,s_2=\pm} \sin{\pi(2\fa+ s_1 \fm+\frac{s_2}{2} \lambda)}}{\sin{\pi(2\fa +\half \lambda)}\sin{\pi(2\fa -\half \lambda)}\sin{2\pi \fa}}\\
+ & \frac{\sin \pi \left(2\fa+\frac{1}{2}\lambda\pm \fm \right)}{\Big(\sin{\pi (2 \fa)} \sin{\pi (2 \fa + \lambda)}\Big)} +\frac{\sin \pi \left(2\fa - \frac{1}{2}\lambda\pm \fm \right)}{\Big(\sin{\pi (2 \fa)} \sin{\pi (2 \fa - \lambda)}\Big)}~.
\end{split}
\end{equation}

\printbibliography

\end{document}